\documentclass[twocolumn]{aastex63}
\received{5 April 2022; Accepted 18 June 2022}
\usepackage{amssymb,amsmath,graphicx,longtable,subfigure,wrapfig}
\usepackage{rotating}
\usepackage{float}
\usepackage{color}
\newcommand{\beginappendixA}{%
        \setcounter{table}{0}
        \renewcommand{\thetable}{A\arabic{table}}%
        \setcounter{figure}{0}
        \renewcommand{\thefigure}{A\arabic{figure}}%
        \setcounter{equation}{0}
        \renewcommand{\theequation}{A\arabic{equation}}%
     }

\shorttitle{Low frequency view of Coma cluster of galaxies}
\shortauthors{Lal et~al.}

\begin{document}
%\linenumbers

\title{High-resolution, High-sensitivity, Low-frequency uGMRT View of Coma Cluster of Galaxies}

\correspondingauthor{D. V. Lal}
\email{dharam@ncra.tifr.res.in}

\author[0000-0001-5470-305X]{D. V. Lal}
\affiliation{National Centre for Radio Astrophysics - Tata Institute of Fundamental Research Post Box 3, Ganeshkhind P.O., Pune 411007, India}

\author[0000-0003-4917-7803]{N. Lyskova}
%\affiliation{National Research University Higher School of Economics, Myasnitskaya str 20, Moscow 101000, Russia}
\affiliation{Space Research Institute (IKI), Profsoyuznaya 84/32, Moscow 117997, Russia}
\affiliation{ASC of P.N.Lebedev Physical Institute, Leninskiy prospect 53, Moscow 119991, Russia}
%\affiliation{Max-Planck-Institut f\"ur Astrophysik, Karl-Schwarzschild-Strasse 1, 85741 Garching, Germany}

\author[0000-0001-5888-7052]{C. Zhang}
\affiliation{Department of Astronomy and Astrophysics, The University of Chicago, Chicago, IL 60637, USA}

\author[0000-0002-8476-6307]{T. Venturi}
\affiliation{INAF - Istituto di Radioastronomia, via Gobetti 101, I-40129 Bologna, Italy}

\author[0000-0002-9478-1682]{W. R. Forman}
\affiliation{Harvard-Smithsonian Center for Astrophysics, 60 Garden Street, Cambridge, Massachusetts 02138, USA}

\author[0000-0003-2206-4243]{C. Jones}
\affiliation{Harvard-Smithsonian Center for Astrophysics, 60 Garden Street, Cambridge, Massachusetts 02138, USA}

\author[0000-0002-0322-884X]{E. M. Churazov}
\affiliation{Max-Planck-Institut f\"ur Astrophysik, Karl-Schwarzschild-Strasse 1, 85741 Garching, Germany}
\affiliation{Space Research Institute (IKI), Profsoyuznaya 84/32, Moscow 117997, Russia}

\author[0000-0002-0587-1660]{R. J. van~Weeren}
\affiliation{Leiden Observatory, Leiden University, PO Box 9513, 2300 RA Leiden, The Netherlands}

\author[0000-0002-5068-4581]{A. Bonafede}
\affiliation{DIFA - Universit\'a di Bologna, via Gobetti 93/2, I-40129 Bologna, Italy}
\affiliation{INAF - Istituto di Radioastronomia, via Gobetti 101, I-40129 Bologna, Italy}
\affiliation{Universit\"at Hamburg, Hamburger Sternwarte, Gojenbergsweg 112, D-21029, Hamburg, Germany}

\author[0000-0003-1076-7558]{N. A. Miller}
\affiliation{Stevenson University, Department of Mathematics and Physics, 1525 Greenspring Valley Road, Stevenson, MD, 21153, USA}

\author[0000-0002-0692-0911]{I. D. Roberts}
\affiliation{Leiden Observatory, Leiden University, PO Box 9513, 2300 RA Leiden, The Netherlands}

\author[0000-0003-0037-2288]{A. M. Bykov}
\affiliation{Ioffe Institute, Politekhnicheskaya st. 26, Saint Petersburg 194021, Russia}

\author[0000-0003-3586-4485]{L. Di Mascolo}
\affiliation{Dipartimento di Fisica dell’ Universit\'a di Trieste, Sezione di Astronomia, via Tiepolo 11, I-34131 Trieste, Italy}

\author[0000-0002-3369-7735]{M. Br\"uggen}
\affiliation{Universit\"at Hamburg, Hamburger Sternwarte, Gojenbergsweg 112, D-21029, Hamburg, Germany}

\author[0000-0003-4195-8613]{G. Brunetti}
\affiliation{INAF - Istituto di Radioastronomia, via Gobetti 101, I-40129 Bologna, Italy}

%\author[]{Add additional author(s)}
%\affiliation{If any}

\begin{abstract}
We present high-resolution, high-sensitivity upgraded Giant Metrewave Radio Telescope observations of the Coma cluster (A1656) at 250-500 MHz and 550-850 MHz.  At 250-500 MHz, 135 sources have extensions $>0\farcm45$ (with peak-to-local-noise ratio $> 4$). Of these, 24 sources are associated with Coma-member galaxies.  In addition, we supplement this sample of 24 galaxies with 20 ram pressure stripped (RPS) galaxies from \citet[][eight are included in the original extended radio source sample]{Chenetal} and an additional five are detected and extended.  We present radio morphologies, radio spectra, spectral index maps, and equipartition properties for these two samples. In general, we find the equipartition properties lie within a narrow range (e.g., $P_{\rm min}$ = 1--3 dynes~cm$^{-2}$).  Only NGC\,4874, one of the two brightest central Coma cluster galaxies, has a central energy density and pressure about five times higher and a radio source age about 50\% lower than that of the other Coma galaxies.  We find a diffuse tail of radio emission trailing the dominant galaxy of the merging NGC\,4839 group that coincides with the \textit{slingshot} tail, seen in X-rays. The southwestern radio relic, B1253$+$275, has a large extent $\approx$ 32$^\prime$ $\times$ 10$^\prime$ ($\simeq$ 1.08 $\times$ 0.34 Mpc$^2$). For NGC\,4789, whose long radio tails merge into the relic and may be a source of its relativistic seed electrons, and we find a transverse radio spectral gradient, a steepening from southwest to northeast across the width of the radio source.  Finally, radio morphologies of the extended and RPS samples suggest that these galaxies are on their first infall into Coma on (predominantly) radial orbits.
\end{abstract}


\section{Introduction}
\label{sec:intro}

The rich galaxy cluster Abell 1656 \citep[A1656;][]{Abell1958,Abell1989} in the constellation Coma Berenices is one of a trio (with the Perseus and Virgo clusters) of the most intensively studied clusters.  Coma is nearby \citep[z = 0.0232;][]{Wagneretal}, massive \citep[6 $\times$ 10$^{14}$ $M_{\odot}$;][]{Planck-collaboration}, regular, dominated by E and S0 galaxies, and lies near the Galactic pole ($b^{II} = 88^{\circ}$), away from any significant obscuration, which make it an ideal target across the wavelength band.  Unlike the other members of the cluster trio, Coma appears relatively spherical, and therefore, more amenable to many types of studies that have dramatically advanced our understanding \citep[e.g.,][]{Zwicky1933}.  The concentration of nebulae in Coma was first noted by \citet{Herschel1785}.  The first ADS listing for Coma is the study of \citet{Curtis1918}, who remarked that the region ``contains the most remarkable aggregation of closely packed small nebulae''.  Over the past 100 yr, Coma has been observed by nearly every major telescope across the electromagnetic spectrum, and it is now considered a cluster where merging processes are currently taking place \citep[see][for a recent overview]{Churazovetal2021}.
Among the newest probes of the Coma cluster are those from the Fermi-LAT. Based on 12 yr of Fermi-LAT data, \citet{Adametal} and \citet{Baghmanyanetal} have recently reported emission at $\gamma$-ray energies from the direction of the Coma cluster.  This emission could arise from interactions of the thermal gas with relativistic protons.  Models of cosmic-ray acceleration by multiple large-scale shocks, which heat the intracluster gas \citep[see also][for review]{bykov2019}, predict cosmic-ray proton distribution power-law indices consistent with the reported values. While secondary leptons produced by cosmic-ray protons may produce radio synchrotron emission, their contribution and the relative contribution of point sources to this emission is an open issue \citep[see also][for a detailed discussion]{BL2011,BL2016,Pinzkeetal2017}.

\tabletypesize{\scriptsize}
\def\arraystretch{0.5}
\begin{table}
\tablewidth{0pt}
\caption{Summary of recent wide-area radio observations of the Coma cluster.}
\label{tab:survey-sum}
\begin{center}
\begin{tabular}{lccc}
\hline\hline
Frequency & Area      &  Resolution       & Sensitivity  \\
      & (deg$^2$) &  (${\prime\prime}$) & ($\mu$Jy~beam$^{-1}$)  \\
 \multicolumn{1}{c}{(1)} &  (2) & (3) & (4) \\
\hline\noalign{\smallskip}
 \multicolumn{2}{c}{JVLA}   &   &  \\
 2051–3947 MHz$^\dagger$  & $\sim$0.2 & 15.1 & 151  \\
 1400 MHz$^\ddagger$  & $\sim$0.5 & 4.4 & 22  \\
\hline\noalign{\smallskip}
 \multicolumn{2}{c}{uGMRT}   &   & \\
 250--500 MHz & $\sim$4.0 & 6.0 & 31.1  \\
 550--850 MHz & $\sim$3.8 & 3.5 & 16.4  \\
\hline\noalign{\smallskip}
 \multicolumn{2}{c}{Lofar}   &  &  \\
 144 MHz$^\digamma$      & $\sim$25 & 20 & 150 \\
\hline
\end{tabular}
\end{center}
\tablecomments{See also Sec.~\ref{sec:intro} for more details. \\
The uGMRT observations are presented in this paper. \\
References; $\dagger$: \citet{Ozawaetal2014}; $\ddagger$: \citet{NHM2009}; $\digamma$ \citet{Bonafedeetal2022}. }
\end{table}

Previous radio continuum studies of Coma were both pioneering, uncovering new phenomena, and very comprehensive (see Table~\ref{tab:survey-sum} for a summary of previous large-area surveys that are most recent and comparable to our uGMRT data presented here).  Coma is the first cluster where diffuse steep spectrum megaparsec-scale low-surface brightness emission in the form of a radio halo and a relic was discovered \citep{Willson1970}; \citep[see also][for recent reviews on this topic]{BrunettiJones,vanWeerenetal2019}. It is also the first cluster, where a bridge of emission connecting the halo and the relic was found \citep{Kimetal1990}.  Both the radio halo and relic have been extensively studied with all radio interferometers \citep[i.e.,][]{1990AJ.....99.1381V,Giovanninietal1990,1993ApJ...406..399G,BrownRudnick2011,Bonafedeetal2021} and are considered the prototypes for these classes of cluster radio sources. 
Coma further hosts a number of head-tail and wide-angle-tail radio galaxies \citep[see][]{FerettiVenturi}, which have been studied in the context of the interaction between the radio-emitting plasma of the jets and lobes and the surrounding environment \citep[i.e.,][]{Venturietal1989,Ferettietal1990,Kimetal1994,Lal2020a,Lal2020b}.
Finally, radio continuum tails of ram pressure stripping origin have been recently reported for a number of late-type galaxies \citep{NHM2009,Chenetal,Robertsetal2021}.

\begin{figure*}[ht]
\begin{center}
\begin{tabular}{c}
\includegraphics[width=17.0cm]{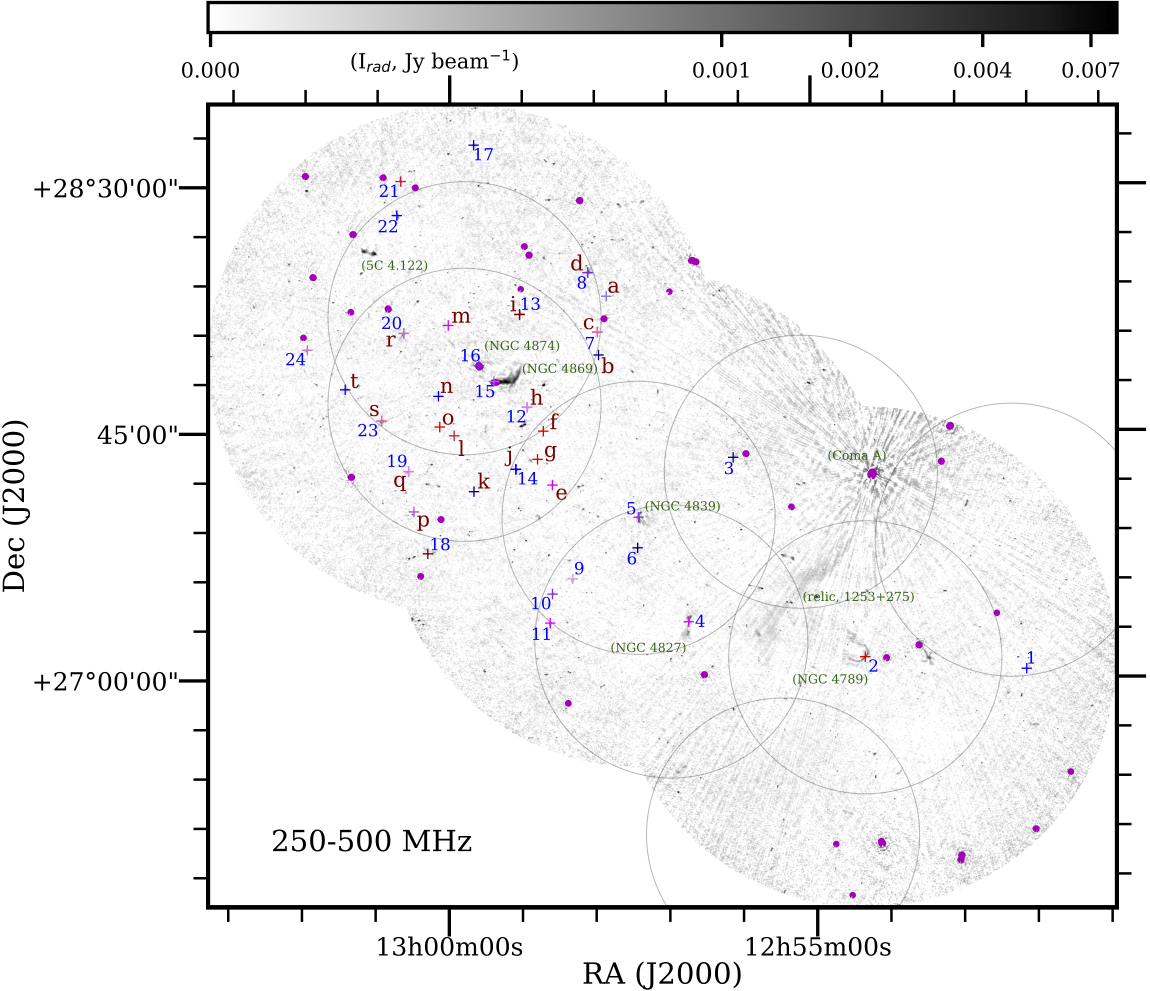}
\end{tabular}
\end{center}
\caption{Mosaic image of three pointings (see Table~\ref{tab:obs-log}) at an angular resolution of $\sim$6\farcs1 in the uGMRT 250--500 MHz band.  Source positions are marked by the `$+$' sign and color coded according to their redshifts, where blue marks the lowest redshift (= 0.01523) galaxy, red marks the highest redshift (= 0.03707) galaxy, and the intervening galaxies are marked with increasing gradient from smaller (blue) to higher (red; see also Fig.~\ref{fig:xmm-24-12}).  The source\_IDs are also labeled adjacent to their positions, extended sources (in blue) and the RPS galaxies (in dark-red) that are members of Coma are listed in Tables~\ref{tab:tab2-ex} and ~\ref{tab:tab3-rps}, respectively.  The magenta points mark the peak positions of dominant (peak $S_\nu$ $\gtrsim$ 30 mJy~beam$^{-1}$ at 250--500 MHz band) radio emission.  The grayscale image is displayed in logarithmic scales to emphasize the extended, low-surface brightness diffuse radio emission and radio emission associated with the radio galaxies.  The pointing containing Coma\,A has higher \textsc{rms} noise ($\approx$36 $\mu$Jy~beam$^{-1}$) than the other pointings (see Table~\ref{tab:obs-log} for \textsc{rms} noise levels).  Note the starry-pattern artifact centered on Coma\,A, limiting the dynamic range that could be achieved for the NGC\,4789 data. Eight circles correspond to the eight pointings in the uGMRT 550--850 MHz band (see Table~\ref{tab:obs-log}).}
\label{fig:f1-b3}
\end{figure*}

In this work, we present observations of the Coma cluster at low frequencies obtained with the upgraded Giant Metrewave Radio Telescope \citep[uGMRT;][]{Swarupetal1991,Guptaetal2017}.  Our observations are made in the frequency bands 250--500 MHz and 550--850 MHz at angular resolutions of $\sim$6\farcs1 and $\sim$3\farcs7, respectively, and cover approximately $\approx$7.5 deg$^2$ using both new and archival pointings (see Table~\ref{tab:obs-log} and Fig.~\ref{fig:f1-b3}).  As Table~\ref{tab:survey-sum} shows, the uGMRT survey we present has better sensitivity and angular resolution than the public LOFAR survey and better sensitivity, larger solid angle, and better resolution than the JVLA Coma survey. For these reasons, we concentrate on the uGMRT data for this paper.

\tabletypesize{\scriptsize}
\def\arraystretch{0.5}
\begin{table*}
\tablewidth{0pt}
\caption{The uGMRT observations.}
\label{tab:obs-log}
\begin{center}
\begin{tabular}{llccccccc}
\hline\hline
Field & Obs\_ID  &  Obs. Date & $\nu$ & $\Delta\nu$ & Ch./Ch.-width & t$_{\rm int.}$ & FWHM & \textsc{rms} \\
   & &       &  (MHz) &   (MHz) & (no. / kHz) & (hr) & ($^{\prime\prime}\times^{\prime\prime}, ^{\circ}$)& ($\mu$Jy~beam$^{-1}$) \\
 \multicolumn{1}{c}{(1)} &  \multicolumn{1}{c}{(2)} & (3) & (4) & (5) & (6) & (7) & (8) & (9) \\
\hline\noalign{\smallskip}
 \multicolumn{2}{c}{250--500\,MHz}   &     &     &              &     &                          & \\
Coma C        & ddtB270 & 2017-04-28 & 400 & 200 & 4096 / 48.8  & 1.8 &~6.65$\times$5.90, ~76.39 & ~21.1 \\
NGC 4839      & 36\_033 & 2019-05-12 & 400 & 200 & 8192 / 24.4  & 3.9 &~6.31$\times$5.73, ~50.61 & ~31.1 \\
NGC 4789      & 36\_033 & 2019-05-23 & 400 & 200 & 8192 / 24.4  & 4.7 &~6.51$\times$5.66, ~41.27 & ~36.1 \\
\hline\noalign{\smallskip}
 \multicolumn{2}{c}{550--850\,MHz}   &     &     &              &     &                          & \\
NGC 4839      & 36\_033 & 2019-06-14 & 700 & 300 & 8192 / 48.8  & 3.4 &~4.21$\times$3.30, ~71.82 & ~19.8 \\
NGC 4789      & 36\_033 & 2019-06-14 & 700 & 300 & 8192 / 48.8  & 3.0 &~4.04$\times$3.35, ~64.58 & ~15.5 \\
B1253$+$275-N & 36\_033 & 2019-06-13 & 700 & 300 & 8192 / 48.8  & 2.3 &~3.89$\times$3.58, ~37.69 & ~42.4 \\
B1253$+$275-S & 36\_033 & 2019-05-31 & 700 & 300 & 8192 / 48.8  & 1.8 &~4.45$\times$3.44, ~45.55 & ~13.7 \\
B1253$+$275-NW& 36\_033 & 2019-05-31 & 700 & 300 & 8192 / 48.8  & 1.7 &~4.30$\times$3.42, ~44.01 & ~18.6 \\
B1253$+$275-SE& 36\_033 & 2019-06-13 & 700 & 300 & 8192 / 48.8  & 1.8 &~3.96$\times$3.44, ~66.10 & ~16.5 \\
Coma H        & 35\_005 & 2018-01-01 & 700 & 300 & 2048 / 195.3 & 2.2 &~3.74$\times$3.16, ~44.79 & ~12.8 \\
Coma L        & 35\_005 & 2018-01-01 & 700 & 300 & 2048 / 195.3 & 1.3 &~3.69$\times$3.08, ~40.73 & ~16.4 \\
\hline
\end{tabular}
\end{center}
\tablecomments{Column~8: The position angle (P.A.) is measured from the north and counterclockwise. \\
Column~9: \textsc{rms} noise at the half power point; see also Sec~\ref{sec:radio-data} and \citet{Lal2020a} for a discussion.}
\end{table*}

We have used the deepest radio images of the Coma cluster in the 250-500 MHz and 550-850 MHz bands to image the extended cluster radio galaxies and study, for the first time, using a radio-selected sample, the connection between the orientation and intrinsic properties of the tails and the cluster environment, as did \citet{RobertsParker} for an optically selected sample of ram pressure stripped (RPS) galaxies (see also \citet{Robertsetal2021} who studied a Sloan Digital Sky Survey (SDSS)-high star formation rate sample of \textit{jellyfish} galaxies). As Table 1 shows, the uGMRT observations provide the highest sensitivity observations over a large field which themselves are adequate for our analysis. Future papers focusing on individual radio sources and structures will exploit radio observations at other wavelengths. In this paper, we investigate whether the extended radio galaxies in the cluster bear information on the cluster dynamics and accretion processes that are taking place, making use of head-tail and wide-angle-tail radio sources, associated with elliptical galaxies and RPS sources, associated with late-type galaxies. To this aim, we use both the radio continuum images as well as spectral imaging in the frequency range covered by our observations. 

The paper is organized as follows:
Our data are summarized in Sec.~\ref{sec:data-red}.
Sec.~\ref{sec:ex-rps-rad} discusses our 24 extended radio sources and 20 RPS galaxies that are associated with the Coma cluster.
We present the analyses of our data, i.e., radio morphologies and spectral structure in Sec.~\ref{sec:ex-morph} and in Sec.~\ref{sec:ex-spectra} for extended radio sources, and in Sec.~\ref{sec:RPS-morph} and Sec.~\ref{sec:RPS-spectra} for RPS galaxies, respectively.
Descriptions of the radio morphologies and optical properties of these extended sources, along with notes from the literature, are given in the Appendices (Sec.~\ref{sec:app-notes-ex} and \ref{sec:app-notes-rps}).  We also provide notes on the spectral structure of four large angular-sized sources in the Appendix (Sec.~\ref{sec:app-notes-spec-ex}).
In Sec.~\ref{sec:discuss}, we examine the nature of radio morphologies of cluster-member galaxies, including equipartition parameters and radiative ages of all extended and detected RPS galaxies (Sec.~\ref{sec:discuss-eq-param}), their relative motions with respect to the cluster center (Sec.~\ref{sec:orient-vector}), and their X-ray properties (Sec.~\ref{sec:X-ray-prop}).
Sec.~\ref{sec:conclusions} summarizes our conclusions.

Throughout this paper, we adopt a $\Lambda$ cold dark matter cosmology with $H_0$ = 70 km s$^{-1}$ Mpc$^{-1}$, $\Omega_{\rm m}$ = 0.27, and $\Omega_{\Lambda}$ = 0.73.
At the mean redshift of the Coma cluster, $z$ = 0.02316, 1~arcsec corresponds to 467~pc at the luminosity distance of 100.7 Mpc.
We define the spectral index, $\alpha$, as $S_\nu$ $\propto$ $\nu^\alpha$; where $S_\nu$ is the flux density at frequency, $\nu$.
Throughout, positions are given in J2000 coordinates.

\section{Observations and Data Reduction}
\label{sec:data-red}

\subsection{Radio Data}
\label{sec:radio-data}

The GMRT \citep{Swarupetal1991} has been upgraded with a completely new set of receivers at
frequencies $<$~1.5 GHz.  The uGMRT has (nearly) seamless frequency coverage in the 0.050--1.50 GHz range \citep{Guptaetal2017}.
The data reported here include archival data (Obs\_IDs" DDT-ddtb270 and 35\_005) and new observations (Obs\_ID 36\_033) detailed in the observation log (Table~\ref{tab:obs-log}).
As far as possible, two pointings were observed alternately for one observing run, each integration lasting $\sim$20~minutes.  Before and after two scans, corresponding to two pointings, a calibration source, 3C\,286 was observed for $\sim$5 minutes.  In this way, a fairly regular and complete sampling of the ($u,v$) plane was obtained.

\begin{figure*}[ht]
\begin{center}
\begin{tabular}{c}
\includegraphics[width=17.0cm]{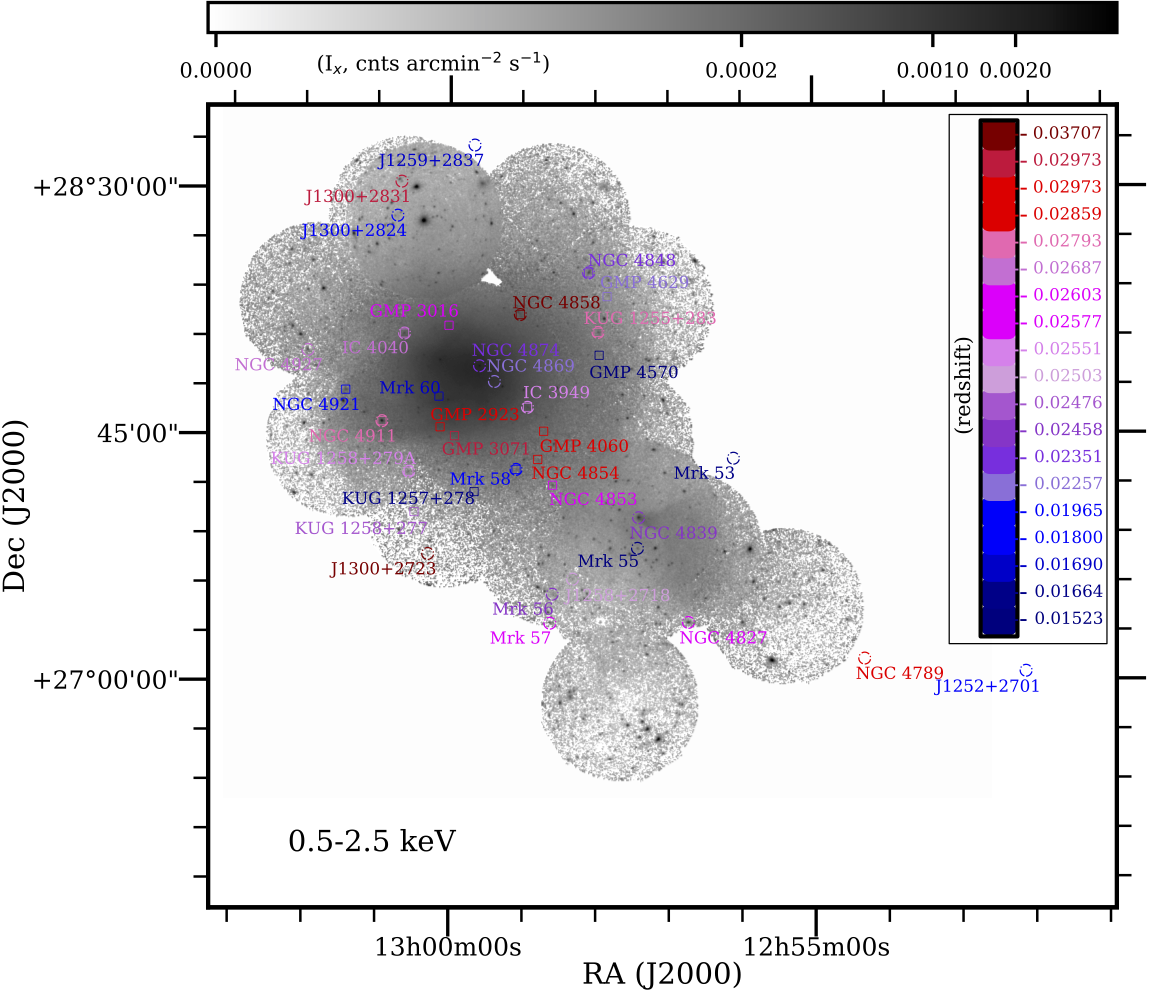} 
\end{tabular}
\end{center}
\caption{The background-subtracted, exposure, and vignetting-corrected {\it XMM-Newton} image of the Coma cluster in the 0.5--2.5 keV energy band.  The positions of the extended radio sources (Table~\ref{tab:tab2-ex}) and the RPS galaxies (Table~\ref{tab:tab3-rps}) are marked with circles and boxes, respectively.  The source name and their positions are color coded according to redshifts, where blue marks the lowest redshift (= 0.01523) galaxy, red marks the highest redshift (= 0.03707) galaxy; the intervening galaxies are marked with increasing gradient from smaller (blue) to higher (red) as depicted in the inset redshift scale bar.}
\label{fig:xmm-24-12}
\end{figure*}

Both the archival and new uGMRT data were analysed in \textsc{aips} and \textsc{casa}, using standard imaging procedures \citep[see also][for a complete and detailed description of the data reduction methodology]{Lal2020a}.
Briefly, after bad data were edited, phases and amplitudes were calibrated using the calibration source observations.  The flux density calibrator, 3C\,286 was used as the secondary phase calibrator during the observations.  We used the calibration source to correct for the bandpass shape and to set the flux density scale \citep{PerleyButler}.  Bad data and data affected due to RFI were identified and flagged.
After the initial calibration, the 200 MHz wide data of the 250--500 MHz band and the 300 MHz wide data of the 550--850 MHz band were split into five 40~MHz and six 50~MHz sub-bands, respectively.  For each pointing, a region of size $\approx$ 3$^\circ$ $\times$ 3$^\circ$ and $\approx$ 1.8$^\circ$ $\times$ 1.8$^\circ$ was imaged, at  250--500 MHz  and  550--850 MHz, respectively, extending far beyond the null of the GMRT primary beam.
In addition to the standard self-calibration procedure, the method of \textit{peeling} was also performed, in \textsc{aips}, on each sub-band to correct for direction-dependent errors.  The calibrated data for five 40~MHz and six 50~MHz sub-bands were further stitched to form full 200 and 300~MHz calibrated visibility data at the two uGMRT bands, 250--500 MHz and 550--850 MHz, respectively.  Next, these combined visibilities were imaged using \textsc{tclean} in \textsc{casa}.  We used 3D imaging (gridder = `widefield'), two Taylor coefficients (nterms = 2), and Briggs weighting (robust = 0.5) in \textsc{tclean} to ensure image fidelity over the full field of view and wide band.
A final amplitude and phase self-calibration with a solution interval equal to the length of observation was then applied,
and the two polarizations, RR and LL, were combined to obtain the final image.  The final image was corrected for the primary beam shape of the GMRT antennas.
The uncertainty in the estimated flux density, both due to calibration and due to systematics is $\lesssim$5 per cent \citep[see also][]{Lal2020a}.

These final images have a noise level (from a portion of an image at the half power points) of 21.1--36.6 $\mu$Jy~beam$^{-1}$ and 12.8--42.4 $\mu$Jy~beam$^{-1}$ at the angular resolutions of $\sim$6\farcs1 and $\sim$3\farcs7 in the 250--500 MHz and 550--850 MHz bands, respectively (see Table~\ref{tab:obs-log}).  These are the deepest images so far reported at these bands covering a total area of $\sim$4~deg$^2$, despite the modest integration time ($\approx$2.5 hr on sky for a typical pointing).  The (conservative) uncertainty in the estimated flux density, both due to calibration and due to systematics, is $\lesssim$4\% at both 250--500 MHz and 550--850 MHz bands.  The \textsc{rms} noise is a factor of $\approx$2 and 3 higher close to the dominant radio sources, NGC\,4874 (one of the two core BCGs) and Coma\,A, respectively.
Based on the range for the brightest pixels (in the three 250--500 MHz band pointings and the eight 550--850 MHz band pointings), i.e., 0.71--0.61 Jy~beam$^{-1}$ and 0.23--0.02 Jy~beam$^{-1}$, we reach dynamic ranges of $\sim$20,000:1--29,000:1 and $\sim$5,500:1--15,700:1 in the 250--500 MHz and 550--850 MHz bands, respectively, considering that a pointing contained the very bright Coma\,A radio source.  The three corrected primary beam pointings at 250--500 MHz band were knitted together and are presented in Fig.~\ref{fig:f1-b3}.  Also shown are eight circles that correspond to the eight pointings at 550--850 MHz.  Note that the two mosaics were not affected by an offset with respect to the phase center, since the images revealed (i) no systematic differences besides typical differences in flux density calibration, and (ii) no positional offsets that are larger than the size of a pixel ($\approx$0.05~arcsec and $\approx$0.03~arcsec at 250--500 MHz and 550--850 MHz bands, respectively).

\subsection{X-Ray Data}

\def\arraystretch{0.5}
\setlength\LTcapwidth{\textwidth}
\begin{table*}
\tablewidth{0pt}
\caption{Radio Properties of the 24 Extended Coma Cluster-member Sources Detected using the uGMRT.}
\label{tab:tab2-ex}
\begin{center}
\scriptsize
\begin{tabular}{lllcccrrrl}
\hline \hline
 Source\_ID & \multicolumn{2}{c}{Optical Position} & \multicolumn{1}{c}{$z$} &\multicolumn{2}{c}{\textsc{rms}} & \multicolumn{1}{c}{$S_{\rm 400 MHz}$} & \multicolumn{1}{c}{$S_{\rm 700 MHz}$} & & \multicolumn{1}{l}{Radio} \\
\cline {1-1}
 & \multicolumn{1}{c}{R.A.} & \multicolumn{1}{c}{Decl.} &  & \multicolumn{1}{l}{B-3} & \multicolumn{1}{c}{B-4} & & & \multicolumn{1}{c}{$\alpha_{\rm 400 MHz}^{\rm 700 MHz}$} & \multicolumn{1}{l}{Morphology} \\
    &  \multicolumn{2}{c}{(J2000)} &   & \multicolumn{2}{c}{($\mu$Jy~b$^{-1}$)} & \multicolumn{2}{c}{(mJy)} & & \\
  \multicolumn{1}{c}{(1)} & \multicolumn{1}{c}{(2)} & \multicolumn{1}{c}{(3)} & (4) & \multicolumn{1}{c}{(5)} & (6) & \multicolumn{1}{c}{(7)} & \multicolumn{1}{c}{(8)} & \multicolumn{1}{c}{(9)} & \multicolumn{1}{c}{(10)} \\
\hline\noalign{\smallskip}
  ~1 2MASX J12520684$+$2701352          & 12:52:06.84 & $+$27:01:35.2 & 0.02135 & 12 & 24 &   3.3 $\pm$0.2 &   1.7 $\pm$0.1 & $-$1.23 $\pm$0.13 & A \\ 
  ~2 NGC 4789                           & 12:54:19.02 & $+$27:04:04.9 & 0.02790 & 40 & 24 & 156.8 $\pm$1.2 & 119.5 $\pm$0.5 & $-$0.49 $\pm$0.02 & HT, NAT \\ 
  ~3 Mrk 53                             & 12:56:06.10 & $+$27:40:41.2 & 0.01657 & 24 & 24 &  14.5 $\pm$0.4 &   9.5 $\pm$0.2 & $-$0.75 $\pm$0.07 & T \\ 
  ~4 NGC 4827                           & 12:56:43.52 & $+$27:10:43.7 & 0.02545 & 24 & 18 & 132.9 $\pm$0.5 & 101.2 $\pm$0.3 & $-$0.49 $\pm$0.01 & CD, FR\,II \\ 
  ~5 NGC 4839                           & 12:57:24.35 & $+$27:29:51.8 & 0.02456 & 36 & 18 & 196.9 $\pm$1.6 & 118.9 $\pm$1.0 & $-$0.90 $\pm$0.02 & HT, WAT, cD \\ 
  ~6 Mrk 55                             & 12:57:25.25 & $+$27:24:16.5 & 0.01620 & 24 & 12 &   7.9 $\pm$0.5 &   5.2 $\pm$0.3 & $-$0.78 $\pm$0.14 & Tr \\ 
  ~7 KUG 1255$+$283$^\dagger$           & 12:57:57.75 & $+$28:03:41.5 & 0.02768 & 24 & 18 &  17.3 $\pm$0.7 &  13.4 $\pm$0.4 & $-$0.45 $\pm$0.08 & T \\ 
  ~8 NGC 4848$^\dagger$                 & 12:58:05.58 & $+$28:14:33.3 & 0.02351 & 24 & 24 &  60.7 $\pm$2.1 &  43.4 $\pm$0.8 & $-$0.60 $\pm$0.07 & T \\ 
  ~9 2MASX J12581865$+$2718387          & 12:58:18.62 & $+$27:18:38.9 & 0.02484 & 24 & 18 &  10.7 $\pm$0.5 &   9.4 $\pm$0.2 & $-$0.22 $\pm$0.10 & T \\ 
  10 Mrk 56                             & 12:58:35.34 & $+$27:15:52.9 & 0.02458 & 12 & 12 &   7.3 $\pm$0.3 &   7.8 $\pm$0.1 &    0.11 $\pm$0.08 & T \\ 
  11 Mrk 57                             & 12:58:37.28 & $+$27:10:35.8 & 0.02556 & 18 & 24 &  14.4 $\pm$0.4 &   8.9 $\pm$0.2 & $-$0.86 $\pm$0.05 & T \\ 
  12 IC 3949$^\dagger$                  & 12:58:55.95 & $+$27:50:00.4 & 0.02525 & 18 & 12 &   7.5 $\pm$0.3 &   5.3 $\pm$0.2 & $-$0.61 $\pm$0.10 & A, T \\ 
  13 NGC 4858$^\dagger$                 & 12:59:02.07 & $+$28:06:56.4 & 0.03141 & 36 & 12 &  22.1 $\pm$0.7 &  12.1 $\pm$0.5 & $-$1.07 $\pm$0.10 & T \\ 
  14 Mrk 58$^\dagger$                   & 12:59:05.30 & $+$27:38:39.9 & 0.01853 & 24 & 24 &  12.2 $\pm$0.6 &   7.9 $\pm$0.3 & $-$0.76 $\pm$0.11 & T \\ 
  15 NGC 4869                           & 12:59:23.33 & $+$27:54:41.8 & 0.02288 & 72 & 36 &1381.0 $\pm$9.7 & 735.9 $\pm$2.1 & $-$1.12 $\pm$0.01 & HT, NAT \\ 
  16 NGC 4874                           & 12:59:35.71 & $+$27:57:33.8 & 0.02394 &384 & 96 & 476.6 $\pm$2.3 & 305.0 $\pm$2.4 & $-$0.80 $\pm$0.02 & HT, NAT, cD \\ 
  17 2MASX J12594009$+$2837507$^\digamma$ & 12:59:40.10 & $+$28:37:50.8 & 0.01787 & 12 &$-$ &   4.5 $\pm$0.1 &\multicolumn{1}{c}{$-$}&\multicolumn{1}{c}{$-$}& A \\
  18 2MASX J13001780$+$2723152$^\digamma$ & 13:00:17.81 & $+$27:23:15.1 & 0.03707 & 24 &$-$ &  16.3 $\pm$0.1 &\multicolumn{1}{c}{$-$}&\multicolumn{1}{c}{$-$}& A, D \\
  19 KUG 1258$+$279A$^\dagger$          & 13:00:33.73 & $+$27:38:15.6 & 0.02494 & 24 & 12 &  10.5 $\pm$0.6 &   7.0 $\pm$0.2 & $-$0.74 $\pm$0.11 & T \\ 
  20 IC 4040$^\dagger$                  & 13:00:37.86 & $+$28:03:28.7 & 0.02615 & 24 & 12 &  53.3 $\pm$1.5 &  29.6 $\pm$0.3 & $-$1.05 $\pm$0.05 & T \\ 
  21 2MASX J13004067$+$2831116$^\digamma$ & 13:00:40.68 & $+$28:31:11.6 & 0.02969 & 24 &$-$ &  18.2 $\pm$0.5 &\multicolumn{1}{c}{$-$}&\multicolumn{1}{c}{$-$}& T \\
  22 2MASX J13004385$+$2824586          & 13:00:43.88 & $+$28:24:58.8 & 0.02095 & 24 & 24 &  31.6 $\pm$0.6 &  16.4 $\pm$0.2 & $-$1.17 $\pm$0.04 & HT, WAT \\ 
  23 NGC 4911$^\dagger$                 & 13:00:56.06 & $+$27:47:27.2 & 0.02663 & 24 & 12 &  50.6 $\pm$1.0 &  29.2 $\pm$0.2 & $-$0.98 $\pm$0.04 & A \\ 
  24 NGC 4927$^\digamma$                  & 13:01:57.59 & $+$28:00:21.0 & 0.02590 & 18 &$-$ &   9.9 $\pm$0.7 &\multicolumn{1}{c}{$-$}&\multicolumn{1}{c}{$-$}& WAT \\
\hline
\end{tabular}
\end{center}
\tablecomments{Column~1: source name as identified in the NED; $\dagger$: RPS galaxies (see also Table~\ref{tab:tab3-rps} and Sec.~\ref{sec:RPS-morph}; $\digamma$: sources not covered in our 550--850 MHz band pointings. \\
Columns~2 and 3: The right ascension and declination (J2000) of the optical host associated with the probable radio core; in increasing right ascension. \\
Column~4: spectroscopic redshift. \\
Columns~5 and 6: The local \textsc{rms} noise (in $\mu$Jy~beam$^{-1}$) near the source for 250--500 MHz (band 3) and 550--850 MHz (band 4). \\
Columns~7 and 8: The flux densities of the RPS galaxies at 250--500 MHz (band 3) and 550--850 MHz (band 4) at angular resolutions of $\sim$6\farcs1 and $\sim$3\farcs7, respectively. \\
Column~9: The spectral index at 250--500 MHz (band 3) and 550--850 MHz (band 4) at these angular resolutions (see also discussion in Sec.~\ref{sec:ex-spectra}). \\
Column~10: The radio source morphology classification following \citet{Rudnick2021}: A = amorphous; HT = head tail; WAT = wide-angle tailed; T = tailed source; CD = classical double; FR\,II: FR type II source; cD = compact dominant; Tr: Triple; NAT: narrow-angle tailed; and D: diffuse.}
\end{table*}

The Coma cluster has been observed multiple times with \textit{Chandra} and \textit{XMM}-\textit{Newton}.
\citet{Lyskovaetal2018} presented an anaylsis of \textit{XMM}-\textit{Newton} data from the EPIC/MOS detector, covering a field of more than 1~deg$^2$ (see Fig.~\ref{fig:xmm-24-12}).  The data were analysed by removing background flares using the light curve of the detected events and renormalizing the `blank fields' background to match the observed count rate in the 11--12 keV band \citep[shown in Fig.~\ref{fig:xmm-24-12},][]{Churazovetal2003}.

To investigate the interaction of the radio lobes of one of two core bright cluster galaxies (BCGs), NGC\,4874 with its environment, we used \textit{Chandra} observations (Obs\_IDs 13993 13994, 13995, 13996, 14406, 14410, 14411, and 14415).
These data were selected since they are nearly contemporaneous (within one month in 2012) and do not suffer from varying ACIS responses.  The accumulated data is about 400~ks.  The data were analysed following the steps described in \citet{Vikhlininetal2005}, which include filtering of high background, use of the latest calibration corrections, and determination of the background for each observation.
The image in the 0.5--4.0~keV band was generated along with the exposure and background maps \citep[see][for details]{Churazovetal2012}.

\section{Extended Radio Sources and RPS Galaxies}
\label{sec:ex-rps-rad}

The Coma cluster has been studied over the years, providing a vast amount of literature.
Using an extensive survey, \citet{Huchra1990} and \citet{KentandGunn} compiled a list of redshifts between blue band magnitude 10.94 and 21.63 in the Coma cluster and its surrounding environs to a radius of six degrees from NGC\,4874, one of the two core BCGs of the Coma cluster, for galaxies whose radial velocities are in the range betweeb 4000 and 9400~km~s$^{-1}$, considered to be associated with the Coma cluster.   We have used this catalog to identify cluster-member galaxies.

A search of the 250--500 MHz image was made to find the position and flux densities of all strong (point and extended) sources.  These were then ``blanked" from the map to enable searches for weaker sources.
Preliminary inspection suggests a myriad ($\simeq$ 2000) of extended and point-like discrete radio sources is seen in the image, most of them being background active galactic nuclei (AGNs).  The analysis of these detected (point and extended) sources will be undertaken as separate projects.

\begin{figure*}[ht]
\begin{center}
\begin{tabular}{c}
\includegraphics[width=15.5cm]{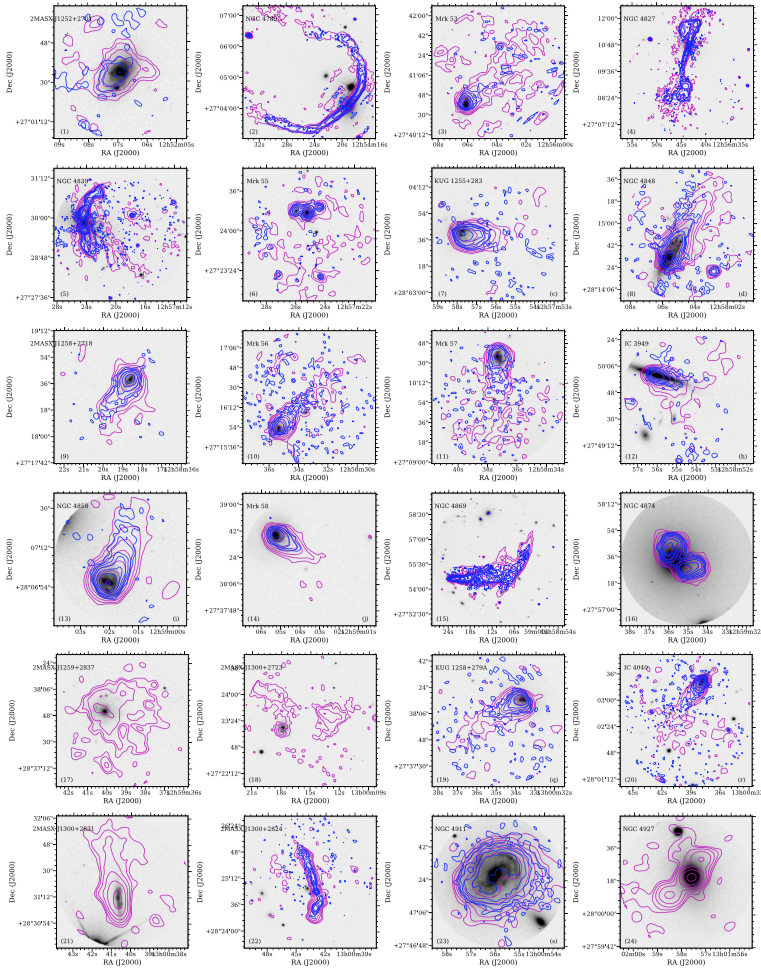}
\end{tabular}
\end{center}
\caption{Images of extended ($\gtrsim$12.6~kpc) confirmed Coma cluster-member radio sources.  These radio sources are in the order presented in Table~\ref{tab:tab2-ex}, from left to right and top to bottom.  The source\_ID, as tabulated in the Table~\ref{tab:tab2-ex}, is labeled in the lower-left corner of each image.  The label at the lower-right corner of each image is the source\_ID for (eight) RPS galaxies as tabulated in the Table~\ref{tab:tab3-rps}.  The magenta and blue radio contours correspond to 250--500 MHz (band 3) and 550--850 MHz (band 4) at angular resolutions of 6\farcs1 and 3\farcs7, respectively, of the uGMRT that are are overlaid on the grayscale SDSS (DR12) $r$-band image.  The lowest radio contour is three times the local \textsc{rms} noise and subsequent contours increase by factors of 2.  The local \textsc{rms} noise is tabulated in Tables~\ref{tab:tab2-ex} and \ref{tab:tab3-rps} for 250--500 MHz (Column~5) and 550--850 MHz (Column~6) of the uGMRT.  The grayscale images are displayed in logarithmic scales to emphasize the optical hosts associated with the respective radio galaxies.}
\label{fig:fig-ex}
\end{figure*}

The definition and identification of multiple systems, e.g., double lobes, is a common problem in radio astronomy and is especially difficult for deep radio observations. We therefore looked at images to find all obvious faint and/or extended radio sources that are at least 4--5 beams across, i.e., sources showing projected radio extent $\gtrsim$0\farcm45 ($\approx$ 12.6~kpc) at the mean redshift of the Coma cluster down to a peak-to-local-noise ratio = 4 at an angular resolution of $\sim$6\farcs1 in the 250--500 MHz band image.
We have identified a total of 135 sources in the 250--500 MHz band.  Of these, 31 sources were not covered in the 550--850 MHz band pointings. 
Furthermore, of the 135 extended radio sources, half ($\approx$ 55~per cent) had redshift measurements (as reported in the NED) and only 24 ($\approx$ 18~per cent) are confirmed members of the Coma cluster.
Note that we have not included the historical Coma\,A (3C\,277.3, $z$ = 0.08536) radio galaxy in this list. 
The confirmed members of the Coma cluster have a redshift range from $z$ = 0.0162 (Mrk~55) to $z$ = 0.02969 (2MASX J13004067$+$2831116) around the mean Coma cluster redshift $z$ = 0.02310.

All 20 RPS galaxies selected by \citet{Chenetal} from the combined galaxy samples of \citet[][UV trails]{Smith2010}, \citet[][12 H$\alpha$]{Yagi2010}, \citet{Kenney2015}, and \citet{Gavazzi1989}, that are confirmed members of the Coma cluster, are covered by our low-frequency radio observations. 
Eight of our extended radio sources are also galaxies with RPS tails of gas and embedded young stars (listed in both Tables~\ref{tab:tab2-ex} and \ref{tab:tab3-rps}).  Of the remaining 12 RPS galaxies, five are detected and seven are undetected in our 250--500 MHz band.  Similarly, in the 550--850 MHz band, of these same 12 RPS galaxies, three are detected and seven are undetected, and two are not covered in our 550--850 MHz pointings.  These are listed together in Table~\ref{tab:tab3-rps}.

Tables~\ref{tab:tab2-ex} and \ref{tab:tab3-rps} list these 24 extended radio sources and 20 RPS galaxies, respectively, along with their source parameters.
Since almost all the extended radio sources have complex morphologies, we simply integrated the signal within interactively defined boundaries and determined integrated flux densities.
Fig.~\ref{fig:f1-b3} depicts the positions of the extended sources and RPS galaxies, marked with `$+$' signs, and color coded accordingly to their redshifts (see Fig.~\ref{fig:xmm-24-12}).
The source\_IDs are also shown, (blue) numeric and (red) alphabetic characters for extended sources and RPS galaxies, respectively, adjacent to their positions.
In Appendix (Sec.~\ref{app:other}) we present the remaining 111 extended radio sources.  Table~\ref{tab:app-tab1} lists these 111 extended radio sources and Fig.~\ref{fig:b3-app-f1} depicts source\_IDs and positions (magenta `$+$' marks) that are either foreground/background galaxies or do not have redshift data.
The radio images of these extended radio sources are presented in Fig.~\ref{fig:app1}, in the order listed in Table~\ref{tab:app-tab1}.
Below we discuss, using our 250--500 MHz and 550--850 MHz band continuum (with deeper sensitivity) data on extended sources and RPS galaxies in the Coma cluster.

\section{Results and Analysis}

\subsection{Radio Morphologies of Extended Sources}
\label{sec:ex-morph}

In Table~\ref{tab:tab2-ex}, we list 24 extended ($\gtrsim$12.6~kpc) radio sources that are confirmed members of the Coma cluster.  Individual columns are source\_ID, source position, spectroscopic redshift, \textsc{rms} noise at 250--500 MHz and 550--850 MHz in the near vicinity of the source, integrated flux densities at 250--500 MHz and 550--850 MHz, and the spectral index of the source using the 250--500 MHz and 550--850 MHz band flux densities.  Uncertainties on the spectral index are determined using uncertainties on the flux calibrations ($\lesssim$4\%, see also Sec.~\ref{sec:radio-data}) and uncertainties on the flux densities, where the latter correspond to the \textsc{rms} noise in the near vicinity of the source times the square root of the number of beams covering the source in each observing band.
The source\_IDs of the extended sources (in blue `$+$' marks) are labeled adjacent to their positions in Fig.~\ref{fig:f1-b3} listed in Table~\ref{tab:tab2-ex}.

\begin{figure*}[ht]
\begin{center}
\begin{tabular}{c}
\includegraphics[width=16.5cm]{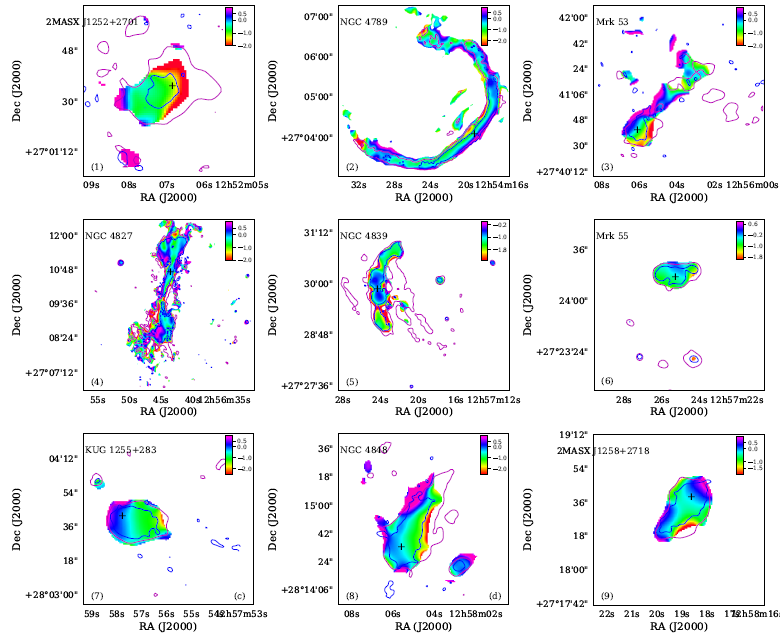}
\end{tabular}
\end{center}
\caption{Spectral index images of extended ($\gtrsim$12.6~kpc) radio sources in the Coma cluster field that are imaged using 250--500 MHz band and 550--850 MHz band data at matched angular resolutions (= 8$^{\prime\prime}$).  The source\_ID, as tabulated in Tables~\ref{tab:tab2-ex} and \ref{tab:tab3-rps}, is labeled in the lower-left and lower-right corners of each image.  The optical host positions are marked by a `$+$' sign.  These radio sources are in the order presented in Table~\ref{tab:tab2-ex}, from left to right and top to bottom.
The images are displayed in linear scales, the scale bar gives the spectral index range, and the typical spectral index uncertainty is 0.03.
The magenta and blue radio contours (= five times the local \textsc{rms} noise) are also overlaid, which correspond to the 250--500 MHz (band 3) and 550--850 MHz (band 4) data at the angular resolutions of 6\farcs1 and 3\farcs7, respectively.}
\label{fig:spec-in}
\end{figure*}

The radio images of these extended cluster-member radio sources are presented in Fig.~\ref{fig:fig-ex}, in the order listed in Table~\ref{tab:tab2-ex}.
The source name, as tabulated in the Table~\ref{tab:tab2-ex}, is labeled in the lower-left corner of each image.  The radio contours, in magenta and blue, correspond respectively to the 250--500 MHz and the 550--850 MHz band images that are overlaid on the grayscale SDSS (DR12) $r$-band image.  Descriptions of the radio morphologies and their optical properties of these extended sources, along with notes from the literature, are given in the Appendix (Sec.~\ref{sec:app-ex-rps}).  To study the plasma properties of these extended sources, we also elaborate below on their radio morphologies (Sec.~\ref{sec:app-notes-ex}) and spectral properties (Sec.~\ref{sec:app-notes-spec-ex}) at low frequencies.  For RPS galaxies detected as extended sources, their alphabetic designation (as in Table~\ref{tab:tab3-rps}) is included along with the extended source serial number from Table~\ref{tab:tab2-ex} (see also Sec.~\ref{sec:RPS-morph}).

\subsubsection{Spectral Structure of Extended Sources}
\label{sec:ex-spectra}

\setcounter{figure}{3}
\begin{figure*}[ht]
\begin{center}
\begin{tabular}{c}
\includegraphics[width=16.5cm]{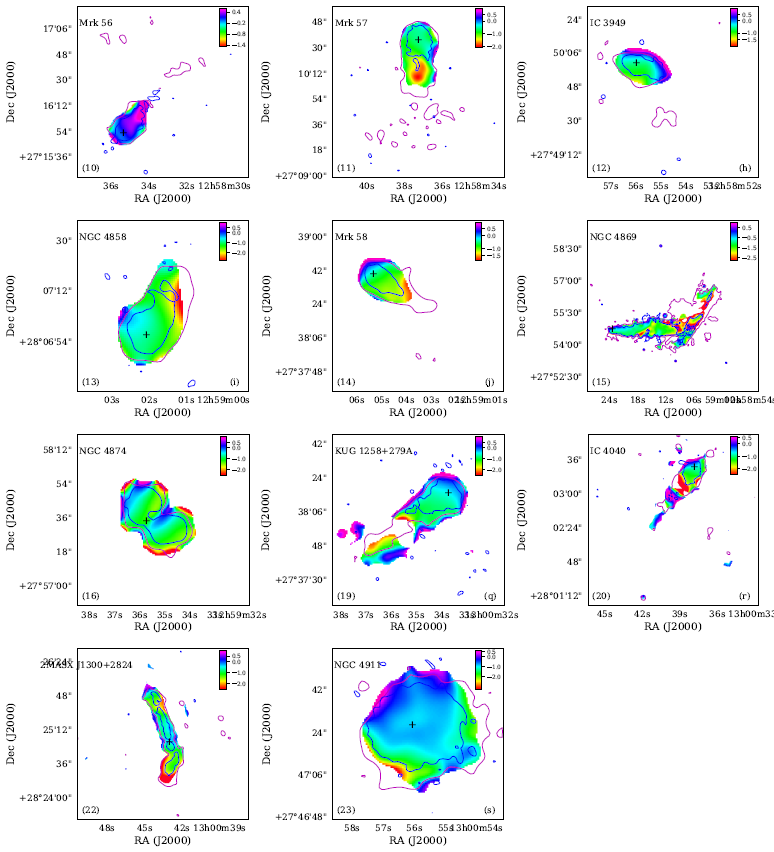}
\end{tabular}
\end{center}
\caption{Continued}
\end{figure*}

Twenty of 24 extended sources were also covered in the 550--850 MHz band pointings, which allows us to investigate the spectral index distribution of these sources.  Therefore, the final calibrated ($u,v$)-data at 550--850 MHz were mapped using a ($u,v$)-taper of 0--36~k$\lambda$, which is similar to that of the 250--500 MHz data and then restored, using the restoring beam corresponding to the 250--500 MHz band image.  We also made images at matched lower angular resolutions to emphasize low-surface brightness diffuse structures for both bands.  The restored and matched images at the 250--500 MHz and 550--850 MHz bands were further used to construct spectral index images for 20 of the 24 extended radio sources.
The standard direct method, i.e., the ratio of
${\rm log}\left[S_{\rm 400~MHz} (\rm x,y)/S_{\rm 700~MHz} (\rm x,y)\right]$
and
${\rm log}\left[{\rm 400~MHz}/{\rm 700~MHz}\right]$,
where
$S_{\rm 400~MHz}(\rm x,y)$ and $S_{\rm 700~MHz}(\rm x,y)$
are the images at two bands, 250--500 MHz and 550--850 MHz, respectively, was used to determine the spectral index distribution.

Fig.~\ref{fig:spec-in} shows spectral index images for 20 of the 24 extended radio sources that are associated with the Coma cluster (in the order presented in Table~\ref{tab:tab2-ex}).  The source\_ID, as tabulated in Table~\ref{tab:tab2-ex}, is labeled in the lower-left corner of each image.
All tailed radio sources are characterized by elongated morphologies found in both bands of the uGMRT.
As usual, sources that have a high brightness head close to the optical galaxy and a narrow low brightness tail show spectral structure that gradually steepens from the flat spectrum head with increasing distance from the head along the tail. The prevalence of flat spectrum radio cores in almost all extended radio sources implies that they may still be active.  These spectral trends are rather similar when the uncertainties are taken into account for the majority of the extended radio sources shown in Fig.~\ref{fig:spec-in}.

\begin{figure}[ht]
\begin{center}
\begin{tabular}{c}
\includegraphics[width=7.8cm]{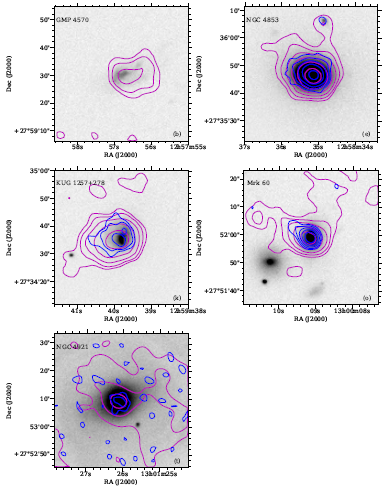}
\end{tabular}
\end{center}
\caption{Images of the five (of 20) RPS galaxies not detected as extended radio sources.  Eight RPS galaxies detected as extended radio sources
are presented in Fig.~\ref{fig:fig-ex}.  The remaining seven RPS galaxies are not detected in our low-frequency radio data.  These radio sources are in the order presented in Table~\ref{tab:tab3-rps}, from left to right and top to bottom.  The source\_ID as tabulated in the Table~\ref{tab:tab3-rps} (as alphabetical letters) is labeled in the lower-right corner of each image.  The magenta and blue radio contours correspond to 250--500 MHz (band 3) and 550--850 MHz (band 4) at angular resolutions of 6\farcs1 and 3\farcs7, respectively, which are overlaid on the grayscale SDSS (DR12) $r$-band image.  The lowest radio contour plotted is three times the local \textsc{rms} noise and increasing by factors of 2. The local \textsc{rms} noise is tabulated in Table~\ref{tab:tab3-rps} for the 250--500 MHz band (Column~5) and 550--850 MHz band (Column~6) of the uGMRT.  The grayscale images are displayed in logarithmic scales to emphasize the optical hosts associated with the respective radio galaxies.}
\label{fig:fig-rps}
\end{figure}

The projected source size at lower frequencies, 250--500 MHz, is larger than the source size at higher frequencies, the 550-850 MHz band, for all extended sources presented here, probably due to synchrotron ageing.
Note that the integrated flux densities given in Table~\ref{tab:tab2-ex} for the 250--500 MHz and 550-850 MHz band images are at angular resolutions of $\sim$6\farcs1 and $\sim$3\farcs7, respectively.  These measurements would neither affect the small-angular sized radio sources nor affect the sources showing large angular extent that have associated low-surface brightness diffuse radio emission, namely, NGC\,4789 and the associated relic radio source, 1253$+$275 in our case, and NGC\,4869.  This is because our observations are full synthesis observing runs at both bands, along with their large fractional bandwidths, which result in good ($u,v$)-coverage at the short as well as the long spacings that are critical to map $\approx$ 10$^{\prime}$-sized structures.
The spectral structure of the large ($\gtrsim$ 1\farcm85) angular-sized sources, namely, NGC\,4789, NGC\,4827, NGC\,4839, NGC\,4869 and 2MASX J13004385$+$2824586 are elaborated further in the Appendix (Sec.~\ref{sec:app-notes-spec-ex}).

\subsection{Radio Morphologies of RPS Galaxies}
\label{sec:RPS-morph}

\def\arraystretch{0.5}
\begin{table*}
\tablewidth{0pt}
\caption{A Summary of Radio Properties of the RPS Galaxies in the Coma Cluster Field Observed Using the uGMRT.}
\label{tab:tab3-rps}
\begin{center}
\scriptsize
\begin{tabular}{lllcccrrrcl}
\hline \hline
 Source\_ID & \multicolumn{2}{c}{Optical Position} & \multicolumn{1}{c}{Vel.} &\multicolumn{2}{c}{\textsc{rms}} & \multicolumn{1}{c}{$S_{\rm 400 MHz}$} & \multicolumn{1}{c}{$S_{\rm 700 MHz}$} & & \multicolumn{1}{c}{$S_{\rm 1.4 GHz}^\amalg$} & \multicolumn{1}{l}{Radio} \\
\cline {1-1}
 & \multicolumn{1}{c}{R.A.} & \multicolumn{1}{c}{Decl.} &  & \multicolumn{1}{l}{B-3} & \multicolumn{1}{c}{B-4} & & & \multicolumn{1}{c}{$\alpha_{\rm 400 MHz}^{\rm 700 MHz}$} & & \multicolumn{1}{l}{Morphology} \\
          & \multicolumn{2}{c}{(J2000)}  &   & \multicolumn{2}{c}{($\mu$Jy~b$^{-1}$)} & \multicolumn{2}{c}{(mJy)} & & \multicolumn{1}{c}{(mJy)} &  \\
   \multicolumn{1}{c}{(1)} & \multicolumn{1}{c}{(2)} & \multicolumn{1}{c}{(3)} & (4) & \multicolumn{1}{c}{(5)} & (6) & \multicolumn{1}{c}{(7)} & \multicolumn{1}{c}{(8)} & \multicolumn{1}{c}{(9)} & \multicolumn{1}{c}{(10)} & \multicolumn{1}{c}{(11)} \\
\hline\noalign{\smallskip}
  (a) GMP 4629$^\digamma$      & 12:57:50.27 & 28:10:13.7 & 6918 & 22 &$-$ &  $\lesssim$0.80 &  $\lesssim$0.33$^\Gamma$ &\multicolumn{1}{c}{$-$}&  0.03 $\pm$0.01 & $-$ \\
  (b) GMP 4570$^\digamma$      & 12:57:56.81 & 27:59:30.6 & 4565 & 18 &$-$ &  1.04 $\pm$0.22 &  $\lesssim$0.60$^\Gamma$ &$\lesssim$ $-$0.98 &  0.55 $\pm$0.06 & T \\
  (c) KUG 1255$+$283$^\dagger$ & 12:57:57.75 & 28:03:41.5 & 8136 & 24 & 18 & 17.29 $\pm$0.69 & 13.43 $\pm$0.36 & $-$0.45 $\pm$0.08 &  6.76 $\pm$0.08 & T \\
  (d) NGC 4848$^\dagger$       & 12:58:05.58 & 28:14:33.3 & 7184 & 24 & 24 & 60.66 $\pm$2.12 & 43.41 $\pm$0.80 & $-$0.60 $\pm$0.07 & 23.85 $\pm$0.11 & T \\
  (e) NGC 4853                 & 12:58:35.20 & 27:35:47.1 & 7688 & 24 & 24 &  6.79 $\pm$0.41 &  4.43 $\pm$0.12 & $-$0.77 $\pm$0.12 &  1.35 $\pm$0.15 & A, T \\
  (f) GMP 4060                 & 12:58:42.60 & 27:45:38.0 & 8686 & 22 & 14 &  $\lesssim$0.40 &  $\lesssim$0.31 &\multicolumn{1}{c}{$-$}&  $<$0.083 & $-$ \\
  (g) NGC 4854                 & 12:58:47.44 & 27:40:29.3 & 8383 & 22 & 12 &  $\lesssim$0.52 &  $\lesssim$0.37 &\multicolumn{1}{c}{$-$}&  $<$0.098 & $-$ \\
  (h) IC 3949$^\dagger$        & 12:58:55.95 & 27:50:00.4 & 7526 & 18 & 12 &  7.48 $\pm$0.34 &  5.29 $\pm$0.19 & $-$0.61 $\pm$0.10 &  2.37 $\pm$0.09 & T \\
  (i) NGC 4858$^\dagger$       & 12:59:02.07 & 28:06:56.4 & 9416 & 36 & 12 & 22.11 $\pm$0.68 & 12.13 $\pm$0.48 & $-$1.07 $\pm$0.10 &  8.73 $\pm$0.13 & T \\
  (j) Mrk 58$^\dagger$         & 12:59:05.30 & 27:38:39.9 & 5419 & 24 & 24 & 12.15 $\pm$0.61 &  7.93 $\pm$0.27 & $-$0.76 $\pm$0.11 &  3.29 $\pm$0.15 & T \\
  (k) KUG 1257$+$278           & 12:59:39.82 & 27:34:35.9 & 5011 & 12 & 18 &  2.28 $\pm$0.21 &  1.41 $\pm$0.19 & $-$0.86 $\pm$0.29 &  0.51 $\pm$0.07 & T \\
  (l) GMP 3071                 & 12:59:56.15 & 27:44:47.3 & 8920 & 22 & 14 &  $\lesssim$0.24 &  $\lesssim$0.11 &\multicolumn{1}{c}{$-$}&  0.05 $\pm$0.02 & $-$ \\
  (m) GMP3016                  & 13:00:01.08 & 28:04:56.2 & 7765 & 22 & 12 &  $\lesssim$0.29 &  $\lesssim$0.23 &\multicolumn{1}{c}{$-$}&  $<$0.043 & $-$ \\
  (n) GMP 2923                 & 13:00:08.07 & 27:46:24.0 & 8672 & 22 & 14 &  $\lesssim$0.30 &  $\lesssim$0.21 &\multicolumn{1}{c}{$-$}&  $<$0.034 & $-$ \\
  (o) Mrk 60                   & 13:00:09.14 & 27:51:59.3 & 5316 & 24 & 12 &  2.29 $\pm$0.52 &  1.36 $\pm$0.18 & $-$0.93 $\pm$0.46 &  1.12 $\pm$0.03 & A, T \\
  (p) KUG 1258$+$277           & 13:00:29.23 & 27:30:53.7 & 7395 & 22 & 12 &  $\lesssim$0.19 &  $\lesssim$0.12 &\multicolumn{1}{c}{$-$}&  $<$0.094 & $-$ \\
  (q) KUG 1258$+$279A$^\dagger$& 13:00:33.73 & 27:38:15.6 & 7485 & 24 & 12 & 10.52 $\pm$0.56 &  6.96 $\pm$0.20 & $-$0.74 $\pm$0.11 &  3.70 $\pm$0.09 & T \\
  (r) IC 4040$^\dagger$        & 13:00:37.86 & 28:03:28.7 & 7675 & 24 & 13 & 53.33 $\pm$1.51 & 29.62 $\pm$0.26 & $-$1.05 $\pm$0.05 & 16.31 $\pm$0.10 & T \\
  (s) NGC 4911$^\dagger$       & 13:00:56.06 & 27:47:27.2 & 7985 & 24 & 12 & 50.58 $\pm$0.99 & 29.24 $\pm$0.16 & $-$0.98 $\pm$0.04 & 15.08 $\pm$0.23 & A \\
  (t) NGC 4921                 & 13:01:26.15 & 27:53:09.5 & 5470 & 32 & 12 &  1.89 $\pm$0.48 &  $\lesssim$0.65 &$\lesssim$ $-$2.05 &  0.44 $\pm$0.05 & A \\
\hline
\end{tabular}
\end{center}
\tablecomments{Column~1: source name as identified in NED; $\dagger$: extended radio source (see Table~\ref{tab:tab2-ex} and Sec.~\ref{sec:RPS-morph} for a discussion); $\digamma$: sources not covered in our 550--850 MHz band pointings. \\
Columns~2 and 3: the right ascension and declination (J2000) of the optical host; in increasing right ascension. \\
Column~4: optical velocity from \citet{Chenetal}. \\
Columns~5 and 6: the local \textsc{rms} noise (in $\mu$Jy~beam$^{-1}$) in near vicinity of the source for 250--500 MHz (band 3) and 550--850 MHz (band 4) images. \\
Columns~7 and 8: the flux densities of the RPS galaxy at 250--500 MHz (band 3) and 550--850 MHz (band 4) at angular resolutions of $\sim$6\farcs1 and $\sim$3\farcs7, respectively. An upper limit of five times the local \textsc{rms} is quoted for non-detected RPS galaxies; $\Gamma$: measurements (upper limits) are from the nearest locations covered in our 250--500 MHz band pointings. \\
Column~9: the spectral index at 250--500 MHz (band 3) and 550--850 MHz (band 4) of the source at these angular resolutions. \\
Column~10: the flux densities of the RPS galaxy at 1.4 GHz; $\amalg$: taken from \citet{Chenetal}. \\
Column~11: the radio source morphology classification following \citet[][see also Table~\ref{tab:tab2-ex}]{Rudnick2021}.}
\end{table*}

Eight of our extended radio sources are also galaxies with RPS tails of gas and embedded young stars.  Below, we discuss using our 250--500 MHz and 550--850 MHz bands continuum (with deeper sensitivity) data, the RPS galaxies in the Coma cluster.
Their radio properties, measured from our data along with measurements at 1.4\,GHz \citep{Chenetal}, are listed in Table~\ref{tab:tab3-rps}, and Fig.~\ref{fig:f1-b3} depicts source\_IDs (alphabetic characters) and positions (`$+$' marks).
As mentioned in Sec.~\ref{sec:ex-rps-rad}, the radio emission associated with the ultra-diffuse galaxies that is found in our data will be undertaken as a separate study.

RPS galaxies are characterized by gas being stripped, from the affected galaxy, by the intracluster medium (ICM).  This ram pressure, especially at the early interaction stage with compression of the interstellar medium (ISM), can trigger star formation.  Subsequently, as the cold ISM is depleted, the star formation is quenched \citep[e.g.,][]{GunnGott,Nulsen1982,KoopmannKenney2004,Crowletal,Bosellietal2016}.
Thus, ram pressure stripping is an important process affecting galaxy evolution in rich environments. 

In Table~\ref{tab:tab3-rps}, we present the radio results for 20 RPS galaxies from
\citet{Chenetal} that are confirmed members of the Coma cluster, ordered in increasing right ascension.  Individual columns depict source\_ID, source position, spectroscopic redshift, \textsc{rms} noises at the 250--500 MHz and 550--850 MHz bands in the near vicinity of the source, integrated flux densities at the 250--500 MHz and 550--850 MHz bands, and the spectral index $\alpha$(400--700 MHz) of the source.
%, and the largest linear size of the source at 250--500 MHz (band 3) of the uGMRT.
The source\_IDs of the RPS galaxies (in dark-red `$+$' signs) are labeled adjacent to their positions in Fig.~\ref{fig:f1-b3}, which are listed in Table~\ref{tab:tab3-rps}.

Eight of the 20 RPS galaxies \citep{Chenetal} are also extended ($\gtrsim$ 12.6~kpc) radio sources, and are marked by `$\dagger$' alongside their source\_IDs in Table~\ref{tab:tab2-ex}.  
In addition to the eight RPS galaxies, we detected five more galaxies in our low radio frequency images (IDs = b, e, a, o, and t; see Table~\ref{tab:tab3-rps}).  Figs.~\ref{fig:fig-ex} and \ref{fig:fig-rps} show images of these eight and five (total 13) RPS galaxies, respectively, detected in our low frequency images.  The source\_ID as tabulated in Table~\ref{tab:tab3-rps} is labeled in the lower-right corner of each image in Figs.~\ref{fig:fig-ex} and \ref{fig:fig-rps}.
The radio contours, in magenta and blue correspond, respectively, to the 250--500 MHz and 550--850 MHz band images that are overlaid on the grayscale SDSS (DR12) $r$-band image.
Two sources, GMP\,4629 and GMP\,4570, i.e., source\_IDs (a) and (b) respectively, were not covered in our 550--850 MHz band pointings.
Seven RPS galaxies that are not detected in either of the two bands are GMP\, 4629, GMP\,4060, NGC\,4854 (a.k.a. GMP\,4017), GMP\,3071, GMP\,3016, GMP\,2923 and KUG\,1258$+$277 (a.k.a. GMP\,2640), i.e., source\_IDs, (a), (f), (g), (l), (m), (n), and (p), respectively.
Two sources, GMP\,4629 and GMP\,3071, from these seven sources, were marginally detected at 1.4 GHz \citep{Chenetal}, and
assuming a spectral index $\alpha$ = $-$0.7 for these two sources, the non-detections are consistent in our 250--500 MHz and 550--850 MHz band images within the uncertainties.

\citet{Grishinetal} observed nine low-mass post-starburst galaxies in the Coma cluster using the multiobject Binospec spectrograph on board the 6.5~m MMT.  They reported that two RPS galaxies (GMP\,4060 and GMP\,2923) exhibit spectacular tails of material stripped by the ram pressure of the hot ICM with H$\alpha$ emission suggesting ongoing star formation, although their disks are not star forming \citep[see also][]{Venturietal2022}.
The lack of star formation in their disks suggests that they could be classified as \textit{post-jellyfish} galaxies; similar but typically more massive, \textit{jellyfish} systems have active star formation in their disks and tails \citep{Smith2010,Yagi2010}.
In Appendix (Sec.~\ref{sec:app-notes-rps}), we describe the radio morphologies of the additional five RPS galaxies that are detected in our low radio frequency images, namely, source\_IDs, (b), (e), (k), (o) and (t).  We also provide salient spectral features, using the formulation discussed above, for the three of these five RPS sources that are detected in both bands of uGMRT in Sec.~\ref{sec:RPS-spectra} below, and images of the remaining three of five RPS sources in Fig.~\ref{fig:fig-rps-spix}.

\subsubsection{Spectral Structure of RPS Galaxies}
\label{sec:RPS-spectra}

\begin{figure}[t]
\begin{center}
\begin{tabular}{c}
\includegraphics[width=7.8cm]{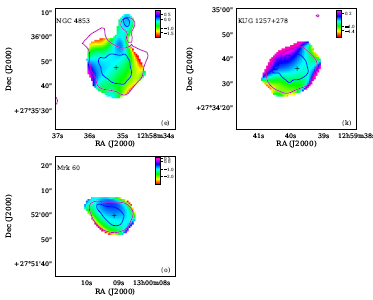}
\end{tabular}
\end{center}
\caption{Spectral index images of the remaining three of five RPS galaxies in the Coma cluster field that are imaged using 250--500 MHz band and 550--850 MHz band data at matched angular resolutions (= 8$^{\prime\prime}$).
The source\_ID, as tabulated in Table~\ref{tab:tab3-rps}, is labeled in the lower-right corner of each image.
The optical host positions are marked by a `$+$' sign.
The images are displayed in linear scales and the scale bar depicts the spectral index range.
The magenta and blue radio contours (= five times the local \textsc{rms} noise) are also overlaid, which correspond to 250--500 MHz (band 3) and 550--850 MHz (band 4) at the angular resolutions of 6\farcs1 and 3\farcs7, respectively.  See also Fig.~\ref{fig:spec-in} for more details.}
\label{fig:fig-rps-spix}
\end{figure}

Our measurements of the integrated spectral index for all RPS galaxies at the 250--500 MHz and 550--850 MHz bands agree, within uncertainties, with measurements at 1.4 GHz using the VLA \citep{Chenetal}.
Fig.~\ref{fig:fig-rps-spix} shows spectral index images of three of five RPS galaxies that are detected in both bands, 250--500 MHz and 550--850 MHz images (in the order presented in Table~\ref{tab:tab3-rps} with the source\_ID labeled in the lower-right  corner of each image).  Similar to the spectral structure of all tailed radio sources, these remaining sources also show spectral structures that gradually steepen from the flat spectrum head with increasing distance from the head along the tail.  The spectral trends of all RPS galaxies are rather similar.  Our results hint that the radio tail has a steeper spectrum than the radio core associated with the host galaxy (see below for a quantitative analysis).

The majority, 13 out of 20 RPS galaxies, show signatures of ram pressure stripping, based on our low radio frequency, 250--500 MHz and 550--850 MHz band images.
The presence of radio continuum tails in 13 out of 20 RPS galaxies is reported for the first time at the 250--500 MHz and 550--850 MHz bands of uGMRT, which reveal the presence of relativistic electrons and magnetic fields in the stripped tails.  Of these 13, radio continuum tails in three RPS galaxies, (namely, NGC\,4853, IC\,3949, and NGC\,4911) were not detected by \citet{Chenetal}. However, radio tails in NGC\,4853 and IC\,3949 were detected by \citep{Robertsetal2021} and we detect these, as well as the new tail for NGC\,4911.
%Of these 13, the new detection of radio continuum tails in three RPS galaxies (namely NGC\,4853, IC\,3949 and NGC\,4911), compared to the earlier study by \citet{Chenetal}, are reported here for the first time.}
%Radio tails in two galaxies (NGC\,4853 and IC\,3949) of these three were also reported at 144 MHz using LOFAR \citep{Robertsetal2021}. 
The larger extent of the 250--500 MHz band image as compared to the extent of the 550-850 MHz band image along with our spectral index images unambiguously suggest that the tail has a steeper spectrum than the radio emission associated with the host galaxy.
Furthermore, the spectral index between 250--500 MHz and 550--850 MHz shows hints of spectral steepening as a function of distance from the head to the tail.   This is typical of head-tail radio sources, suggesting ageing of synchrotron electrons \citep{LalandRao2004}.
\citet{Chenetal} also detected significant H$\alpha$ or star formation tails in half of the RPS galaxies.  Finally, the radio continuum tails in our observations for all 13 RPS galaxies are spatially coincident with the H$\alpha$ or star formation tails, consistent with \citet{BravoAlfaro} and \citet{Chenetal}.

\section{Discussion}
\label{sec:discuss}

We have used the deepest radio images of the Coma cluster in the 250-500 MHz and 550-850 MHz bands to image the tailed cluster radio galaxies and study, for the first time using a radio-selected sample, the connections between the orientation and intrinsic properties of the tails and the cluster environment.
With the primary novel goal of studying the interplay of numerous tailed radio galaxies in the Coma cluster and the merger and accretion history of the Coma cluster, we have presented high-resolution, high-sensitivity radio morphologies of 24 extended ($\gtrsim$12.6~kpc) radio sources associated with Coma cluster galaxies and spectral index distributions for 20 of the 24 extended radio sources using the uGMRT.  We have also presented high-resolution, high-sensitivity radio morphologies for 13 (of 20) detected RPS galaxies associated with the Coma cluster and spectral index distributions for 11 of the 13 RPS galaxies using the uGMRT.
Barring a few sources, e.g., NGC\,4827, the surface brightness of almost all extended sources is a maximum at the location of the optical host and decreases away from the center as expected for radio sources residing in a cluster environment.
Similarly, the surface brightness of RPS galaxies is also a maximum at the location of the optical host and decreases away from the center along the tail.

\subsection{Spectral Properties}
\label{sec:discuss-spec-prop}

\begin{figure}[hb]
\begin{center}
\begin{tabular}{c}
\hspace*{-0.2cm}\includegraphics[width=8.4cm]{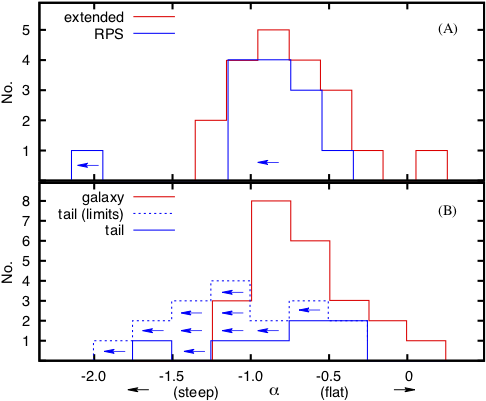}
\end{tabular}
\end{center}
\caption{Distributions of $\alpha$, spectral index between 250--500 MHz and 550--850 MHz (i) for the 20 (of 24) extended radio sources (in red; Column 9 of Table~\ref{tab:tab2-ex}), and 13 (of 20) RPS galaxies (in blue; Column 9 of Table~\ref{tab:tab3-rps}), shown in the upper panel (A), and (ii) for the host galaxy (in red) and radio tails (in blue) of the 23 (20 extended radio sources including 8 RPS galaxies and additional 3 RPS galaxies) sources, shown in the lower panel (B).  In both panels, the sources in the left bins, $\alpha$ $\lesssim$ 0.75, are steep spectrum sources and sources in the right bins, $\alpha$ $\gtrsim$ 0.75 are flat spectrum sources.  Blue arrows denote RPS galaxies that are detected at 250--500 MHz, but undetected at 550--850 MHz.
Note that more than half of the extended radio sources are steep spectrum sources, and a significant number of RPS galaxies also have steep spectra (see upper panel, A).  The tail emission has steeper spectra than the emission at the location of the host galaxy in a majority of sources (see lower panel, B).}
\label{fig:ex-rps-spin}
\end{figure}

Fig.~\ref{fig:ex-rps-spin} (upper panel, A) shows the distributions of $\alpha$, the spectral index between 250--500 MHz and 550--850 MHz for the 20 (of 24) extended radio sources (in red; Column 9 of Table~\ref{tab:tab2-ex}), and 13 (of 20) RPS galaxies (in blue; Column 9 of Table~\ref{tab:tab3-rps}).
A majority of the extended radio sources (55\%) and RPS galaxies (69\%) have steep spectra.
In addition, the tail region spectra appear steeper than the cores.  Apart from the radiative cooling of relativistic electrons that produce a break/steepening in the spectrum of particles, the electrons' adiabatic losses, and the decrease of the magnetic fields in radio-emitting plasma that outflows from the cores can further shift the break in the observed radio spectrum to lower frequencies, contributing to the spectral steepening.

Fig.~\ref{fig:ex-rps-spin} (lower panel, B) shows the distributions of $\alpha$, spectral index between 250--500 MHz and 550--850 MHz for the host galaxy (in red) and radio tails (in blue) of the 23 sources (20 extended radio sources, which includes eight RPS galaxies and three additional RPS only galaxies).  We determined emission from the host galaxy (size = D$_{25}$), and wherever possible, emission from the tail, whose size is equal to the host galaxy.
This suggests that the radio properties of the host galaxy are in general agreement with those of star-forming galaxies, with a spectral index $\alpha$ $\approx$ $-$0.7, while the tail appears significantly different.  The tail spectral index is found to be steeper compared to the host galaxy, reaching a mean spectral index of $\alpha$ $\approx$ $-$1.4, indicating a significant ageing of electrons and/or adiabatic expansion through the tail.
Although we used only uGMRT data to derive the spectral results presented in this paper, we note that our integrated spectra are consistent with earlier work, e.g., RPS galaxies: \citet{Chenetal} and NGC4869: \citet{Lal2020b}.
Furthermore, the radio continuum in RPS galaxies is found to be colocated with H$\alpha$, not only within the host galaxy, but also within the extended tails in several galaxies \citep{Chenetal}.
Although the nature of the interplay between gas stripping and star formation activity might vary from galaxy to galaxy depending on their size, location, and orbit, it seems that gas removal is associated with the quenching of star formation \citep[see also][]{Bosellietal}.  A detailed analysis of these properties for these sources, along with fainter and detected (unresolved and extended) sources, will be undertaken as a separate project.

\subsection{Equipartition Parameters and Radiative Ages}
\label{sec:discuss-eq-param}

\tabletypesize{\scriptsize}
\def\arraystretch{0.5}
\begin{table*}
\caption{Radio Luminosity and Equipartition Parameters Along with Radiative Age for 16 Extended Cluster-member Sources, Eight Extended/RPS Sources, and Five RPS galaxies; see also Sect.~\ref{sec:discuss-eq-param}.}
\label{eq-para-ex-rps}
\begin{center}
\tablewidth{0pt}
\begin{tabular}{rclcccccc}
\hline \hline
 & & Source\_ID & \multicolumn{1}{c}{LLS} & \multicolumn{1}{c}{L$_{\rm 400 MHz}$}     & $B_{\rm min}$ & \multicolumn{1}{c}{$U_{\rm min}$}   & \multicolumn{1}{c}{$P_{\rm min}$}    & \multicolumn{1}{c}{Age} \\
 &      &  &  \multicolumn{1}{c}{(arcmin)} &  \multicolumn{1}{c}{(erg~s$^{-1}$~Hz$^{-1}$)} & ($\mu$G)      & \multicolumn{2}{c}{(erg~cm$^{-3}$) (dyne~cm$^{-2}$)} & \multicolumn{1}{c}{(yr)} \\
 &      &     &  &  ($\times10^{20}$)                  &               & ($\times$ $10^{-13}$)               & ($\times$ $10^{-13}$)                 & ($\times$ $10^{8}$) \\
 \multicolumn{3}{c}{(1)} & \multicolumn{1}{c}{(2)} & \multicolumn{1}{c}{(3)} & \multicolumn{1}{c}{(4)} & \multicolumn{1}{c}{(5)} & \multicolumn{1}{c}{(6)} & \multicolumn{1}{c}{(7} \\
\hline\noalign{\smallskip}
   1 &     & 2MASX J12520684$+$2701352 & 0\farcm45 &    0.4 &  2.4 &   5.8 &   1.9  & 1.7   \\
   2 &     & NGC 4789                  & 7\farcm37 &   28.1 &  2.5 &   5.8 &   1.9  & 1.7   \\
   3 &     & Mrk 53                    & 1\farcm23 &    0.9 &  3.1 &   9.1 &   3.0  & 1.6   \\
   4 &     & NGC 4827                  & 4\farcm66 &   19.7 &  3.0 &   8.7 &   2.9  & 1.6   \\
   5 &     & NGC 4839                  & 2\farcm47 &   27.2 &  3.1 &   9.3 &   3.1  & 1.6   \\
   6 &     & Mrk 55                    & 0\farcm57 &    0.5 &  3.0 &   8.5 &   2.8  & 1.6   \\
    & (b) & GMP 4570$^\ddagger$        & 0\farcm25 &    0.1 &  2.7 &   7.2 &   2.4  & 1.7   \\
   7 & (c) & KUG 1255$+$283$^\dagger$  & 0\farcm67  &    3.0 &  3.5 &  11.9 &   4.0  & 1.5   \\
   8 & (d) & NGC 4848$^\dagger$        & 1\farcm23 &    7.7 &  3.5 &  11.5 &   3.8  & 1.5   \\
   9 &     & 2MASX J12581865$+$2718387 & 0\farcm71 &    1.5 &  2.5 &   6.3 &   2.1  & 1.7   \\
    & (e) & NGC 4853                   & 0\farcm26 &    1.0 &  3.4 &  10.9 &   3.6  & 1.5   \\
  10 &     & Mrk 56                    & 1\farcm21 &    1.0 &  2.0 &   4.0 &   1.3  & 1.8   \\
  11 &     & Mrk 57                    & 0\farcm90 &    2.2 &  2.8 &   7.6 &   2.5  & 1.7   \\
  12 & (h) & IC 3949$^\dagger$         & 0\farcm50 &    1.1 &  2.8 &   7.7 &   2.6  & 1.7   \\
  13 & (i) & NGC 4858$^\dagger$        & 0\farcm79 &    2.8 &  3.1 &   9.4 &   3.1  & 1.6   \\
  14 & (j) & Mrk 58$^\dagger$          & 0\farcm74 &    0.9 &  3.0 &   8.6 &   2.9  & 1.6   \\
  15 &     & NGC 4869                  & 6\farcm63 &  165.1 &  2.9 &   8.3 &   2.8  & 1.6   \\
  16 &     & NGC 4874                  & 0\farcm50 &   62.5 &  7.0 &  46.6 &  15.5  & 0.8   \\
    & (k) & KUG 1257$+$278             & 0\farcm38 &    0.1 &  2.4 &   5.4 &   1.8  & 1.8   \\
  17 &     & 2MASX J12594009$+$2837507 & 0\farcm72 &    0.3 &  1.3 &   1.6 &   0.5  & 1.7   \\
    & (o) & Mrk 60                     & 0\farcm41 &    0.2 &  2.9 &   8.4 &   2.8  & 1.7   \\
  18 &     & 2MASX J13001780$+$2723152 & 1\farcm79 &    5.2 &  1.5 &   2.1 &   0.7  & 1.7   \\
  19 & (q) & KUG 1258$+$279A$^\dagger$ & 1\farcm01 &    1.5 &  2.9 &   8.4 &   2.8  & 1.6   \\
  20 & (r) & IC 4040$^\dagger$         & 2\farcm18 &    8.4 &  3.2 &  10.2 &   3.4  & 1.6   \\
  21 &     & 2MASX J13004067$+$2831116 & 0\farcm96 &    3.7 &  2.8 &   7.8 &   2.6  & 1.6   \\
  22 &     & 2MASX J13004385$+$2824586 & 1\farcm85 &    3.2 &  3.0 &   8.7 &   2.9  & 1.6   \\
  23 & (s) & NGC 4911$^\dagger$        & 0\farcm73 &    8.2 &  3.0 &   8.7 &   2.9  & 1.6   \\
    & (t) & NGC 4921$^\ddagger$        & 0\farcm47 &    0.1 &  3.2 &   9.7 &   3.2  & 1.6   \\
  24 &     & NGC 4927                  & 0\farcm66 &    1.5 &  3.5 &  12.0 &   4.0  & 1.5   \\
\hline
\end{tabular}
\end{center}
\tablecomments{Column~1: source name as identified in NED; all 29 sources are ordered in increasing right ascension; $\dagger$: extended radio source that is also a RPS galaxy (see Table~\ref{tab:tab2-ex} and Sec.~\ref{sec:RPS-morph} for a discussion); $\ddagger$: We assume $\alpha$ = $-$0.7 for sources that are not detected at 550--850 MHz (band 4) of uGMRT. \\
Column~2: the largest linear size at 250--500 MHz (band 3). \\
Column~3: monochromatic radio luminosity at 400 MHz using 250--500 MHz band data. \\
Columns~4--6: equipartition magnetic field, minimum energy density and minimum pressure. \\
Column~7: maximum radiative lifetime.}
\end{table*}

Due to a lack of polarization information, for both the extended and RPS galaxies in the Coma cluster, the estimates of the magnetic field are performed using the minimum energy conditions.
We assume (i) a cylindrical or a spherical geometry for the source, (ii) an electron population radiating at frequencies from 100 MHz to 10 GHz, (iii) the energy carried by the electrons is half the total energy of all particles, (iv) the filling factor of the emitting region is unity, and (v) the angle between the uniform magnetic field and the line of sight is 90$^\circ$ \citep{Miley1980}.
We note that this method calculates the synchrotron luminosity using a fixed high- and low-frequency cutoffs, instead, low and high energy cut-offs for the particle distribution should be used \citep[][and references therein]{BSC1997,2005AN....326..414B,vanWeerenetal2009}.  Taking the latter into account, the corrected equipartition parameters do not change significantly.  Furthermore, taking a ratio of 100 instead of 2 for the energy in relativistic protons to that in electrons results in equipartition magnetic field strengths about a factor of 3 higher.
We measure the integrated flux densities from both bands, the 250--500 MHz and 550--8500 MHz images, and determine spectral indices, a simple power law, and estimate the equipartition magnetic field, the minimum energy density, and the minimum pressure exerted by the relativistic gas in the radio source.  We further assume $\alpha$ = $-$0.7 for sources that are not detected in the 550--850 MHz band of uGMRT because the non-detections are consistent in our 250--500 MHz and 550--850 MHz band images within the uncertainties.

We also derive the radiative lifetime \citep{JaffePerola1973} using
$$
{t} = 1590 \times \frac{\sqrt{B_{\rm eq}}}{(B_{\rm eq}^2 + B_{\rm IC}^2)} \times \frac{1}{\sqrt{(1 + z)~\nu_{\rm br}}}~{\rm Myr}.
$$
Here, ${B_{\rm eq}}$ is the equipartition magnetic field (in $\mu$G), ${B_{\rm IC}}$ is the inverse Compton equivalent magnetic field (= 3.25~(1 + $z)^2$ $\mu$G), and the frequency where the radio spectrum changes by 0.5, called the break frequency, $\nu_{\rm br}$ is expressed in GHz \citep{Miley1980}.
Compared to earlier studies at high frequencies ($\gtrsim$ 1.4\,GHz), here, we reported three new detections for the RPS galaxies for the first time at the 250--500 MHz and 550--850 MHz bands of uGMRT.  We thus believe that $\nu_{\rm br}$ should be below 1.4\,GHz, and therefore, assume $\nu_{\rm br} =$ 0.7 GHz.

\begin{figure}[t]
\begin{center}
\begin{tabular}{c}
\includegraphics[height=13.0cm]{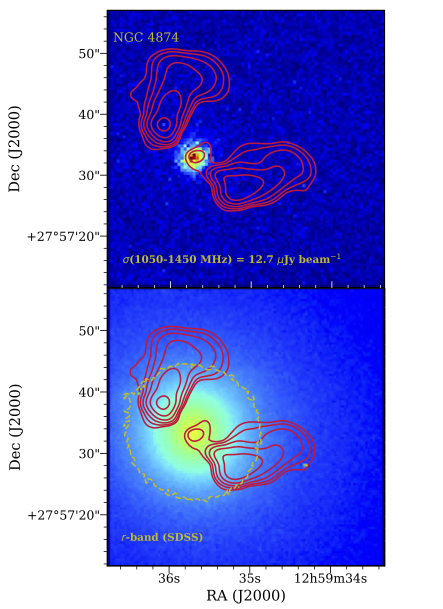}
\end{tabular}
\end{center}
\caption{(Top) NGC\,4874 \textit{Chandra} X-ray image (bin = 1$^{\prime\prime}$, convolved with $\sigma$ = 3$^{\prime\prime}$ Gaussian) in the 0.5--4.0~keV energy band with uGMRT 1050--1450 MHz band contours (red; 2\farcs1 resolution from \citet{Lal2020b}). uGMRT provides the combination of best sensitivity and higher angular resolution image (see also Table\ref{tab:survey-sum}).  The lowest radio contour plotted is three times the local \textsc{rms} noise (labeled in the lower-left corner) and subsequent contours increase by factors of 2. (bottom) SDSS $r$-band image with the same uGMRT contours as in the top panel. The yellow SDSS $r$-band surface brightness contour is plotted at 0.1 $\times$ the peak surface brightness (= 8.73 photons~s$^{-1}$~arcsec$^{-2}$).
The low-frequency radio lobes bend and flare up outside the X-ray mini-corona, but are completely embedded within the optical host (D$_{25}$ = 144$^{\prime\prime}$) galaxy (for previous studies of the radio lobes around NGC\,4874, see \citet{Sunetal2005} and \citet{Sandersetal}). Comparing these higher frequency radio observations to those in Fig.~3 (250-500 MHz and 550-850 MHz), we note there is no evidence of additional extended radio structures (e.g. at larger distances from NGC\,4874) that might have arisen from an earlier/older outburst from the central SMBH.}
\label{fig:n4874}
\end{figure}

Table~\ref{eq-para-ex-rps} lists the largest linear size of the source at 250--500 MHz (band 3) of the uGMRT, and these equipartition parameters, namely, monochromatic luminosity at 400 MHz using 250--500 MHz band data, equipartition magnetic field, minimum energy density, minimum pressure, and the radiative age.
These radiative age estimates are consistent with the hypothesis that the radio plasma ages due to synchrotron cooling and inverse Compton losses.

As Table~\ref{eq-para-ex-rps} shows, the minimum pressure values for all the cluster-member galaxies are comparable (a few dyne~cm$^{-2}$), with one exception, NGC\,4874 (see Fig.~\ref{fig:n4874}).  NGC\,4874 is one of the two BCGs at the cluster core (the other being NGC\,4889, not detected in the radio band, see also Fig.~\ref{fig:sdss-24-12}).
The projected linear extent of NGC\,4874 (in the 250--500 MHz band) is 0\farcm81 ($\approx$ 22.7~kpc) and the spectral index is $-$0.67 $\pm$0.08 \citep{Lal2020a}, which suggests that it could be a compact steep spectrum source \citep[see also][]{Fantietal1990,ODea1998}.  Its age (Table~\ref{eq-para-ex-rps}) is consistent with this possibility.
It hosts a mini-corona of $\sim1$~keV gas \citep[see][]{Vikhlininetal,Sunetal2005,Sandersetal}.
\citet{Sunetal2007} showed that such systems were quite common around massive galaxies in clusters.  For NGC\,4874, the radio jet appears to be confined by the 1~keV mini-corona \citep{Sunetal2007,Sandersetal} and only \textit{blossoms} into radio lobes after the jet has transitioned from the corona to the hot (nearly 10~keV) ICM.  Indeed, Fig.~\ref{fig:n4874} shows the prominent wide-angle tail becoming prominent outside the central corona.  With its high pressure environment, at the bottom of the potential well of Coma, it should not be surprising that the minimum pressure, derived from the radio, is significantly higher than that of the other noncentral Coma galaxies.

\subsection{Orientation of Trailing Features}
\label{sec:orient-vector}

\begin{figure*}[ht]
\begin{center}
\begin{tabular}{c}
\includegraphics[width=16.0cm]{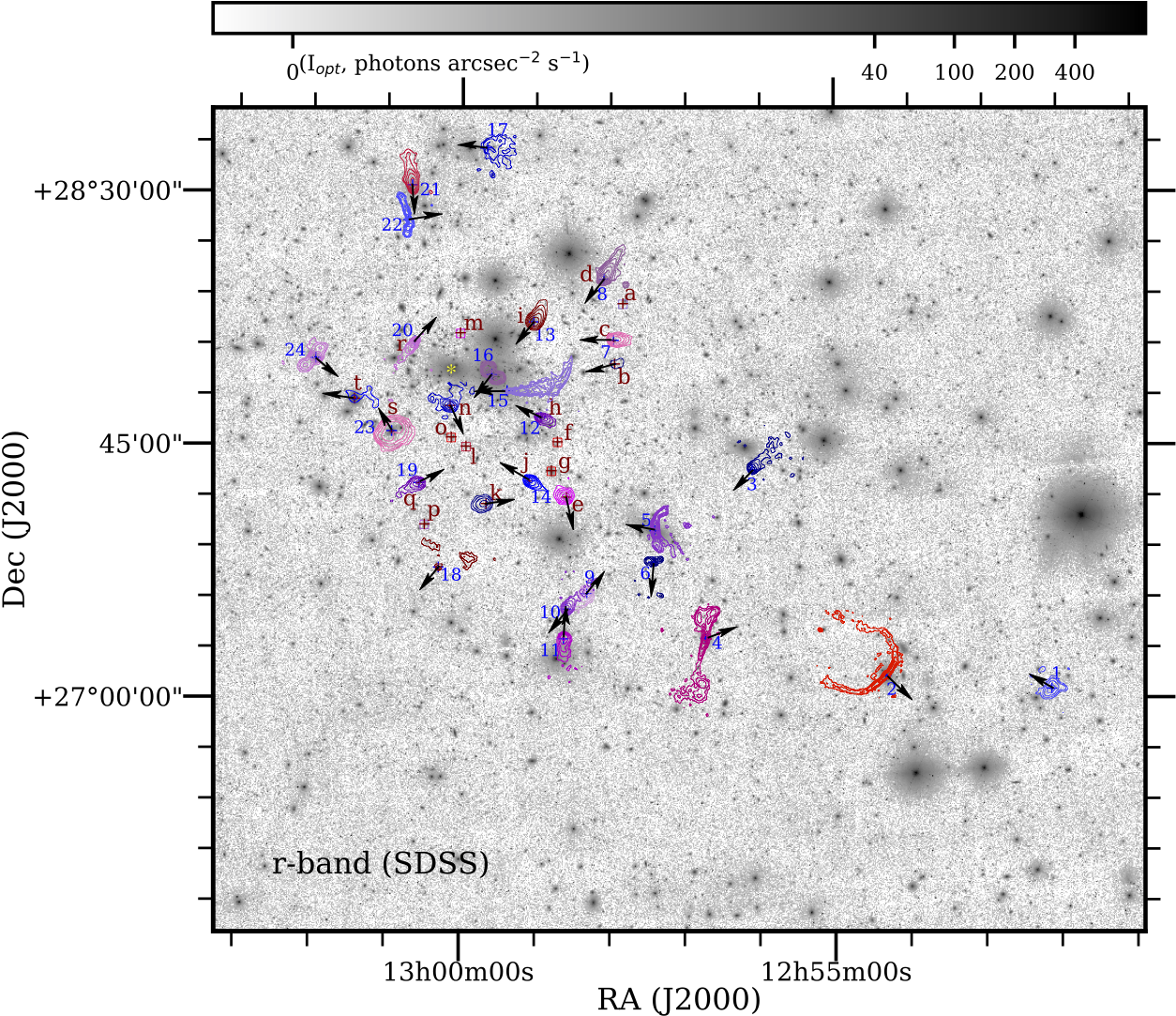}
\end{tabular}
\end{center}
\caption{Composite SDSS (DR12) $r$-band image of the Coma cluster with superposed uGMRT (250--500 MHz band) contours.  Two BCGs, NGC\,4874 (source\_ID 16) and NGC\,4889 (marked with `$*$' and is 4\farcm58 east of NGC\,4874), are also shown.  The positions of the extended radio sources (Table~\ref{tab:tab2-ex}) and the RPS galaxies (Table~\ref{tab:tab3-rps}) are marked with magenta `$+$' and blue `$+$', respectively.  The radio contours corresponding to the 250--500 MHz band (displayed in thumbnail images presented in Figs.~\ref{fig:fig-ex} and ~\ref{fig:fig-rps}) are displayed in logarithmic scale (= logarithm of 10 times the angular size of the source in arcminutes) and overlaid on the SDSS (DR12) $r$-band image at their proper position to emphasize the optical hosts associated with the respective radio galaxies.  The radio contours are color coded accordingly to redshifts, where blue marks the lowest redshift (= 0.01523) galaxy, red marks the highest redshift (= 0.03707) galaxy; the intervening galaxies are colored with increasing gradient from smaller (blue) to higher (red; see also Fig.~\ref{fig:xmm-24-12} for the redshift scale bar).  The lowest radio contour is six times the local \textsc{rms} noise, for clarity and subsequent contours increase by factors of 2. The local \textsc{rms} noise is tabulated in Tables~\ref{tab:tab2-ex} and \ref{tab:tab3-rps} for 250--500 MHz (Column~5) of the uGMRT.  Seven RPS galaxies that are neither detected in the 250--500 MHz band nor in the 550--850 MHz band of uGMRT are marked with boxes (also color coded accordingly to redshifts).  The vectors depict projected motion of these galaxies in the ICM as inferred from their radio morphologies.}
\label{fig:sdss-24-12}
\end{figure*}

A characteristic trait of an RPS galaxy is the tail of gas trailing behind its host galaxy, opposite to its direction of motion.
The slender curved morphology suggests an orbit through the cluster, along which a trail of radio-emitting material that has been left behind.
For example, we find an ensemble of seven sources, namely, KUG 1255$+$283, GMP\,4570, NGC\,4848, IC\,3949, NGC\,4858, Mrk\,58, and NGC\,4869, all within $\sim$14$^\prime$ ($\approx$ 390~kpc) of each other, located $\sim$15\farcm2 (= 445~kpc) east of NGC\,4874, which seem to be moving into the cluster center, though seen in projection.
We visually estimated the direction of the stripping features of all extended radio sources and RPS galaxies relative to one of the two core BCGs of the Coma cluster, NGC\,4874.
To estimate the projected direction of motion of the host galaxy, we assume that tails of radio sources are collimated jets as they trail away from the host galaxy.  In our high-resolution, high-sensitivity deep images, the radio tails have been detected in almost all extended sources and detected RPS galaxies.  It is not, however, possible to produce 3D curvature because we lack knowledge of both the tangential velocities and the complete orbits.

\begin{figure}[ht]
\begin{center}
\begin{tabular}{c}
\includegraphics[width=7.0cm]{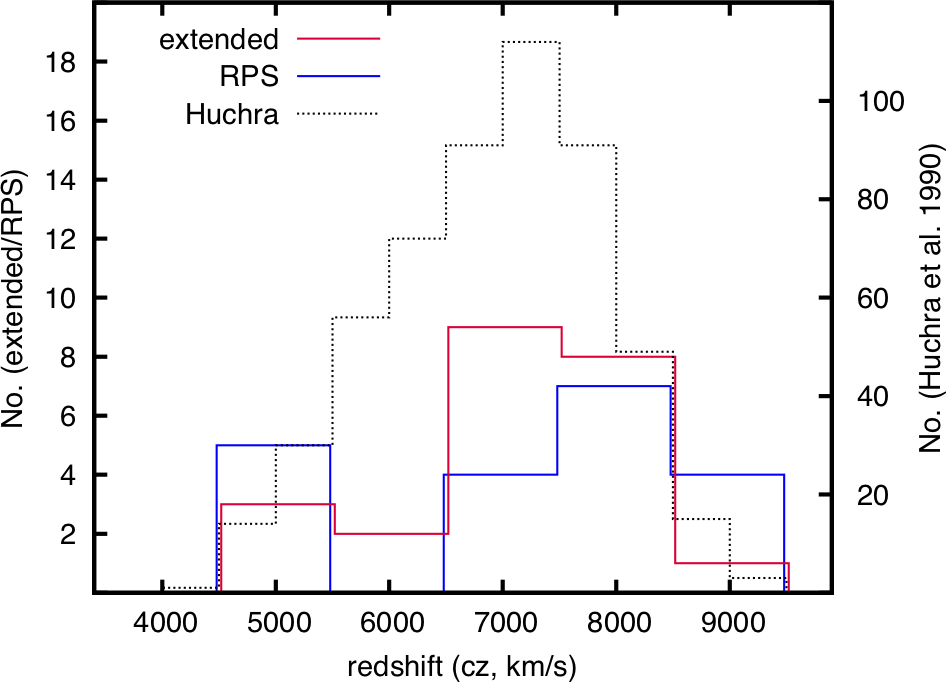} \\ [0.2cm]
\includegraphics[width=7.0cm]{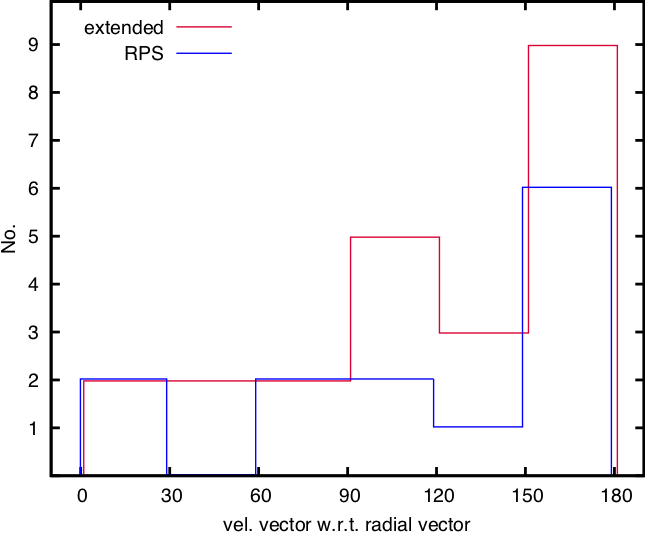}
\end{tabular}
\end{center}
\caption{Histograms showing distributions of redshifts (upper panel) and orientation of the projected motion of the host galaxy with respect to its radial vector, $\theta$ (lower panel), in projection.  Eight of our extended radio sources that are also galaxies with RPS tails of gas and embedded young stars are included in both distributions. The histograms (slightly shifted with respect to one another for clarity) are for the extended radio sources (in red) and for RPS galaxies (in blue). The histogram showing the distribution of all galaxies within the mapped ($\sim$7.5 deg$^2$) area and with 4000 $<$ $cz$ $<$ 9400~km~s$^{-1}$ corresponds to the Y-axis on the right side \citep[from][]{Huchra1990}.  Note that there is weak evidence for more extended radio sources and RPS galaxies with $cz$ larger than the mean Coma value $\sim 6900\,{\rm km\,s^{-1}}$,
which suggests that the extended radio sources and RPS galaxies samples do not follow the given \citet{Huchra1990} distribution.
The radial vector is the line joining one of the two core BCGs of the Coma cluster, NGC\,4874, with each host galaxy.  The position angles of the projected motions of the host galaxies are shown in Fig.~\ref{fig:sdss-24-12}.  The sources with large $\theta$ $\gtrsim$ 90$^\circ$ are moving toward the cluster center and the sources with small $\theta$ $\lesssim$ 90$^\circ$ are moving away from the cluster center.
There is a clear excess of sources with a large angle with respect to the radial vector connecting one of the two core BCGs, NGC\,4874, to their position.  Hence, there is an excess of infalling galaxies, suggesting first infall.  While not of high significance, the excess is seen for both the RPS and extended radio source samples.}
\label{fig:re.vector}
\end{figure}

The spatial orientations of the projected motions, shown as vectors, of the host galaxies are shown in Fig.~\ref{fig:sdss-24-12}.  These projected motions are consistent with the spectral index distribution between the 250--500 MHz band and 550--850 MHz band, i.e., the spectra are flattest toward the optical hosts, in the vicinity of the radio cores and they gradually steepen along the radio tails.  In the background, we show an SDSS (DR12) $r$-band image of the Coma cluster with 250--500 MHz band contours overlaid (from the thumbnail images presented in Figs.~\ref{fig:fig-ex} and ~\ref{fig:fig-rps}).  Seven RPS galaxies that are undetected at 250--500 MHz and 550--850 MHz with uGMRT are marked with boxes.
The subsamples of 24 extended radio sources (see Table~\ref{tab:tab2-ex}) and 20 RPS galaxies (see Table~\ref{tab:tab3-rps}) are not homogeneous, and the sample of \citet{Huchra1990} is apparent magnitude limited.  Here, we are aiming to characterize the morphologies of extended radio sources and RPS galaxies in light of the recent (ongoing) merger of the Coma cluster with the NGC\,4839 group. Hence, we believe that inhomogeneities in the subsamples are not an issue, although their small sizes are limiting.

Following the analysis in \citet{RobertsParker}, in Fig.~\ref{fig:re.vector}, we show the distributions of the redshifts (upper panel) and the distributions of the orientation of the projected motion of the host galaxy with respect to its radial vector, $\theta$ (lower panel).
An angle, $\theta$ = 180$^\circ$ corresponds to the host galaxy moving directly toward the cluster center, and an angle $\theta$ = 0$^\circ$ corresponds to the host galaxy moving away from the cluster center.
Note that there is evidence for more extended radio sources and RPS galaxies with $cz$ larger than the mean Coma value $\sim 6900\,{\rm km\,s^{-1}}$, even after excluding the 2MASX J13001780$+$2723152 (source\_ID 18, $cz$ = 11121 km~s$^{-1}$) extended radio source. The $p$-value of the Kolmogorov-Smirnov test is 0.0145, which suggests that the extended radio sources and RPS galaxies samples do not follow the given \citet{Huchra1990} distribution.
Equivalently, there is a suggestion that the mean (= 7367~km~s$^{-1}$ for extended radio sources and 7025 ~km~s$^{-1}$ for RPS galaxies), as well as the median (= 7467~km~s$^{-1}$ for extended radio sources and 7526~km~s$^{-1}$ for RPS galaxies) velocity of sources 
%moving radially towards the cluster center 
is higher than the mean velocities of Coma cluster galaxies.
%The radial motions suggest that there is a preference for RPS galaxies and those with extended radio emission to be in the early phases of merging with the Coma cluster.  
This difference suggests that RPS galaxies and those with extended radio emission are in the early phases of merging with the Coma cluster.\footnote{We note that all the effects are seen in projection and, hence, clearer correlations with orientation and velocity will be complicated by unknown projection effects and the location of galaxies along the line of sight through the cluster.}
Furthermore, a majority of galaxies, both extended radio sources as well as the RPS galaxies, show angles between 90$^\circ$ and 180$^\circ$, i.e., preferentially moving toward the cluster center.  Given that the timescales for stripping  can be short \citep[see][for review and references therein]{Bosellietal}, especially for galaxies entering the dense core of the cluster, such skewness can arise naturally if the majority of objects recognized as RPS galaxies and extended radio sources are infalling deep into the cluster potential well for the first time. The fact that RPS galaxies and extended radio sources share the same trend is also suggestive that stripping makes an impact on the activity of the central radio source. 

Our result that a majority of galaxies with tails are moving toward the cluster center, i.e., the radio trails point away from the cluster center, is fully consistent with the findings of \citet{Smith2010}, \citet{RobertsParker} and \citet{Robertsetal2021}. We note that the tail orientations of the small post-starburst galaxy sample in \citet{Grishinetal} are less clear and that their selection criteria may favor objects with more tangential orbits.
\citet{RobertsParker} noted that the motion of the majority of stripping galaxies was pointing away from the cluster center, i.e., these galaxies are moving toward the cluster center.  Here, note that a majority of our extended sources and RPS galaxies are within 1.4~deg$^2$, whereas \citet{RobertsParker} extend to much larger radii covering $\sim$9~deg$^2$ with the bulk of the galaxies lying between 1.4~deg$^2$ and $\sim$9~deg$^2$.  Thus, not only the galaxies in the near vicinity of the cluster center, but also galaxies, with tails and signs of ram pressure stripping, located much farther away, as seen in projection, all seem to be moving toward the cluster center. 

\subsection{X-Ray Properties}
\label{sec:X-ray-prop}

There has been an increasing effort to obtain multiwavelength data for these extended radio sources and RPS galaxies.  Fig.~\ref{fig:xmm-24-12} shows the \textit{XMM-Newton} 0.5--2.5~keV image of the Coma cluster \citep{Lyskovaetal2018}, with positions of the extended and the RPS galaxies marked with circles and boxes, respectively.  The names of the sources and their positions are color coded accordingly to redshifts.
The X-ray emission from the bright core and extended X-ray emission from several extended radio sources has been detected (see Fig.~\ref{fig:xmm-24-12}).
Similarly, the X-ray emission from the core and the tails for the RPS galaxies have been detected.  For example, \citet{Sandersetal} reported a narrow X-ray tail in Mrk\,60 using \textit{Chandra} \citep[see also][]{Sunetal2021}.

Barring 2MASX J12520684$+$2701352, Mrk\,53, and 2MASX J12594009$+$2837507, extended radio sources that do not have X-ray data (see Fig.~\ref{fig:xmm-24-12}), all extended sources, as well as the RPS galaxies, are detected in the X-ray band.  The radio core and the associated radio jets of the wide-angle-tail radio source are not covered in the \textit{XMM}-\textit{Newton} image, but the extended radio relic source, 1253$+$275 is partly covered in the \textit{XMM}-\textit{Newton} image (see Fig.~\ref{fig:xmm-24-12}).
\citet{Lal2020b} reported in detail, the radio morphology and the hot gas environment of NGC\,4869, and below we discuss the interaction of NGC\,4789 and NGC\,4839 with their environments.

\begin{figure}[t]
\begin{center}
\begin{tabular}{c}
\hspace*{-0.3cm}\includegraphics[height=7.6cm]{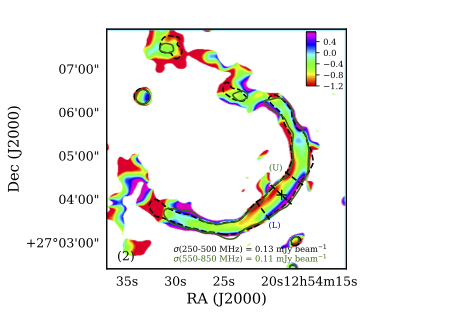} \\
\hspace*{-2.6cm}\includegraphics[width=8.4cm]{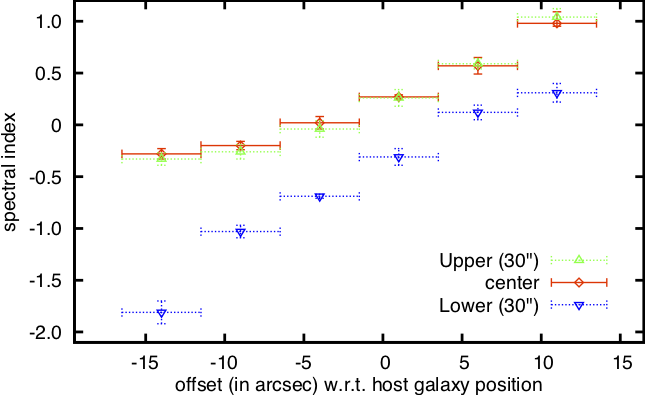}
\end{tabular}
\end{center}
\caption{uGMRT low-angular resolution (= 12$^{\prime\prime}$) spectral index image of NGC\,4789 (upper panel). The optical host position is marked with a `$+$' sign.  The image is displayed in linear scale (the scale bar depicts the spectral index range).
The black (dashed) and green radio contours (= five times the local \textsc{rms} noise) are also overlaid, which correspond to 250--500 MHz (band 3) and 550--850 MHz (band 4), respectively.
The lower panel shows (over-sampled) transverse spectral structure, a gradual steepening from southwest to northeast, across the width of the radio source (along the direction of motion).  The profiles are determined along three \textit{slices} marked in the upper panel, at the location of the host galaxy, and 30$^{\prime\prime}$ to the north/upper (marked `U') and 30$^{\prime\prime}$ to the south/lower (marked `L').}
\label{fig:n4789-spin.profile}
\end{figure}

\subsubsection{Interaction of Radio Sources with the Surrounding ICM}
\label{sec:src-interact}
\citet[][and references therein]{Churazovetal2021} discussed the ongoing merger of the NGC\,4839 group with Coma. They argue that the group has completed its first passage through the cluster core \citep[see also][]{Burns1990,Lyskovaetal2018,Zhangetal}.  In this model, the Coma core has experienced two shocks, first through the bow shock driven by NGC\,4839 during its first passage through the cluster, and subsequently, through a \textit{secondary shock} associated with the gas settling back to quasi-hydrostatic equilibrium in the core.
The bow shock associated with the first infall of NGC\,4839 has detached from the group and propagated southwest to the (current) location of 1253$+$275, the relic radio. Interior to the secondary shock (a.k.a. \textit{mini-accretion} shock), is a more diffuse and extended radio halo.
In this context, we discuss, below, the salient features, i.e., the interplay of the radio synchrotron emission and its hot gas environment of two (NGC\,4789 and the 1253$+$275 relic, and NGC\,4839) of the largest, $\gtrsim$ 2\farcm5 ($\gtrsim$ 70~kpc) radio sources in our sample.

\paragraph{NGC\,4789 and the 1253$+$275 relic}
The upper panel of Fig.~\ref{fig:n4789-spin.profile} shows the uGMRT low-angular resolution (= 12$^{\prime\prime}$) spectral index image of NGC\,4789.  The head located close to the optical host has a flat spectrum between 250--500 MHz and 550--850 MHz. The spectrum between these two bands gradually steepens with increasing distance from the head along the two radio jets, typical of head-tail radio galaxies.  In addition, we notice transverse spectral structure, a gradual steepening from southwest to northeast across the width of the radio tail, close to the head, before the jets bend toward the northwestern direction, possibly due to the ballistic motion of the host galaxy through a dynamic ICM.
In the lower panel of Fig.~\ref{fig:n4789-spin.profile}, we show profiles of the spectral index to quantify the transverse spectral structure at three locations (marked in the upper panel, Fig.~\ref{fig:n4789-spin.profile}), at the location of the host galaxy, and 30$^{\prime\prime}$ to the north and 30$^{\prime\prime}$ to the south.  The north/upper and the center (via the radio core) profiles are similar within uncertainties, while the south/lower profile is relatively steeper.

\begin{figure}[t]
\begin{center}
\begin{tabular}{c}
\hspace*{-0.3cm}\includegraphics[height=7.2cm]{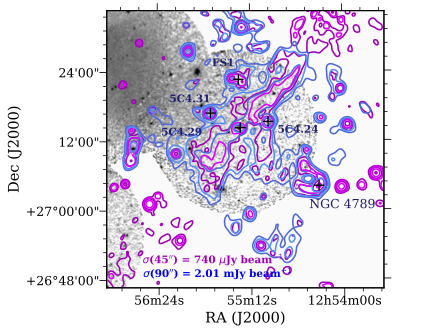}
\end{tabular}
\end{center}
\caption{Low-angular resolution image of NGC\,4789 and radio relic, 1253$+$275, at 45$^{\prime\prime}$ (in light red) and 75$^{\prime\prime}$ (in light-blue) in the 250--500 MHz band of uGMRT; the radio contours are overlaid on the grayscale background-subtracted, exposure corrected \textit{XMM}-\textit{Newton} image (0.5--2.5~keV).
The lowest radio contour is three times the local \textsc{rms} noise (see lower-left corner) and subsequent contours increase by factors of 3.
The compact radio sources marked with the `$+$' sign (are labeled) were subtracted to determine the integrated flux density of the radio relic source.  Note the extended radio relic source shows diffuse emission (in 75$^{\prime\prime}$, low-angular resolution image) with a central ridge, a bridge of emission connected with the narrow-angle-tailed galaxy NGC\,4789.}
\label{fig:n4789-xray}
\end{figure}

Fig.~\ref{fig:n4789-xray} displays the low-angular resolution (= 45$^{\prime\prime}$) image of the wide-angle radio source NGC\,4789 and the extended radio relic source in the 250--500 MHz band.  We clearly detect the extended source, 1253$+$275 which was otherwise resolved in our high-angular resolution image.  It is not detected in our 550--850 MHz band image, possibly due to (i) the presence of the compact, bright Coma\,A source, $\sim$33$^\prime$ to the north of NGC\,4789, (ii) the ($u,v$) coverage and (iii) the largest linear structure ($\approx$ 17$^{\prime}$ at $\sim$5$^{\prime\prime}$ angular resolution) that can be imaged in this band.

The radio relic source, 1253$+$275, plausibly associated with the \textit{runaway} merger shock \citep[see also][]{Zhangetal}
has a large extent $\approx$ 32$^\prime$ $\times$ 10$^\prime$ ($\simeq$ 1.08 $\times$ 0.34 Mpc$^2$), which is broadly consistent with \citet{Giovanninietal1990} \citep[see also][Figs.~1 and 4, respectively]{Bonafedeetal2021,Bonafedeetal2022}. The sharpness of the surface brightness contours suggests that the relic consists of a high-surface-brightness southwestern part and a fainter, diffuse low-surface-brightness northwestern part.
Although in the XMM-Newton X-ray image (Fig.~\ref{fig:n4789-xray}) a shock is not readily seen, the presence of a discontinuity in X-ray and Sunyaev-Zel'dovich data at the location of the radio relic source is confirmed by several studies \citep[see, e.g.][]{2013ApJ...775....4S,2016A&A...591A.142B}.
The radio relic is located at the cluster periphery, $\sim$70$^\prime$ ($\simeq$ 2~Mpc) from the Coma cluster center, in the habitable zone of \textit{runaway} shocks, in the southwest direction, consistent with the predictions of \citet{Zhangetal} \citep[see also][for more detailed discussions of the relic as originating from a runaway shock]{Lyskovaetal2018,Churazovetal2021}.  Its flux density in the 250--500 MHz band is $S_{\rm 400\,MHz}$ = 0.98 $\pm$0.06 Jy, not including contributions from NGC\,4789, 5C\,4.24, 5C\,4.29, 5C\,4.31, and FS1 radio sources.  Our measurement for the radio relic source is consistent with the fairly straight spectrum, $\alpha$ = $-$1.18 $\pm$0.06 found by \citet{Giovanninietal1990} using WSRT 326 and 608 MHz data.

\begin{figure}[t]
\begin{center}
\begin{tabular}{c}
\hspace*{-0.3cm}\includegraphics[height=13.0cm]{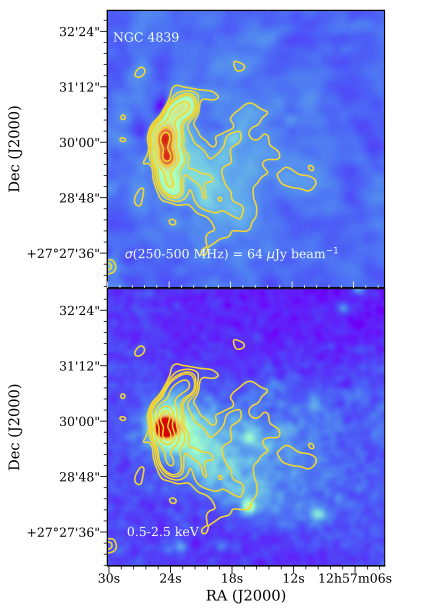}
\end{tabular}
\end{center}
\caption{(Top) The uGMRT 250--500 MHz  (15$^{\prime\prime}$ resolution) image of NGC\,4839 with  radio contours (starting at three times the local \textsc{rms} noise; labeled in the lower-left corner;  and increasing by factors of 2). (bottom) The 0.5--2.5~keV
background-subtracted, exposure corrected \textit{XMM}-\textit{Newton} image with the same uGMRT contours as in the top panel.  
%Note the tail-like X-ray structure (in light green) is co-spatial with the radio tail. 
Note the tail-like X-ray emission (in light green), whose spectrum is well described by thermal plasma with the temperature $\sim 2-4$ keV \citep[e.g.][]{Lyskovaetal2018}
is co-spatial with the radio tail. This suggests that both thermal and nonthermal plasmas are affected by the same velocity field. The detailed surface brightness of the radio tail appears to match that in the X-ray, e.g., a spine of emission extends along the southeast side of the tail.
Spectral steepening can be seen in the spectral index map (Fig.~\ref{fig:spec-in}) between the pair of compact inner lobes (top panel red) and the more extended (light green) outer radio structures.}
\label{fig:n4839-xray}
\end{figure}

The tailed radio source to the southwest of the relic is associated with  NGC\,4789, seen clearly in our high-resolution images (Fig.~\ref{fig:fig-ex} and see also Fig.~\ref{fig:n4789-spin.profile}).
The extended radio relic shows diffuse emission with a central ridge, a bridge of emission connected to the narrow-angle tailed galaxy NGC\,4789 (see Fig.~\ref{fig:n4789-xray}), which suggests that this tailed radio galaxy could be the source of the seed particles feeding the radio relic, 1253$+$275, providing a direct connection between a radio relic source and the radio galaxy, NGC\,4789 \citep{Giovanninietal1990,vanWeeren2017,Bonafedeetal2021,Bonafedeetal2022}.

\paragraph{NGC\,4839}
Fig.~\ref{fig:n4839-xray} displays the  0.5--2.5 keV XMM-Newton image with 15$^{\prime\prime}$ resolution radio contours overlaid.  We see indications of a bow shock and of ram pressure stripping around NGC\,4839 in both radio and X-ray data.
The hot sheath, seen in the X-ray data surrounds the optical host NGC\,4839 and coincides with the low-surface brightness diffuse radio emission.
We also note the tail-like X-ray structure, $\simeq$ 2.5$^{\prime}$ ($\sim70$~kpc), pointing away from the Coma cluster core. Co-spatial with the X-ray tail is radio emission whose detailed surface brightness appears to match that in the X-ray, e.g., a spine of emission extending along the southeast side of the tail. Integrating the band 3 and band 4 emission over the region, which is at least five times the beam size, and above their 3 $\times$ \textsc{rms} contour to reduce statistical errors, in the tail, we derive a spectral index $-$1.72 $\pm0.31$, significantly steeper than that of the inner radio lobes (see the spectral index map of NGC\,4839 in Fig.~\ref{fig:spec-in}).  The X-ray emission, corresponding to the NGC\,4839 group, seems somewhat disconnected from the diffuse, extended X-ray emission from the Coma cluster.
The two-sided radio jet with slight distortions emerging in P.A. $\approx$0$^\circ$, with a bow-shock-like feature with indications of ram pressure stripping (see also Fig.~\ref{fig:fig-ex}) was used to suggest that the NGC\,4839 group was on its initial infall into the Coma cluster \citep{2001A&A...365L..74N}.
However,  subsequent analysis showed that the detailed X-ray morphology of the tail (Fig.~\ref{fig:n4839-xray}), and the location of the radio relic were best explained as arising from a \textit{slingshot} effect, as NGC\,4839, initially infalling into Coma from the northeast, slows and crosses the apocenter, and reverses its radial velocity \citep[see][]{Lyskovaetal2018,Sheardown2019,Zhangetal}.

\section{Conclusions}
\label{sec:conclusions}

A study of a rich and nearby cluster like Coma can establish a local reference for studies of clusters at higher redshifts.
Hence, we obtained new high-resolution, high-sensitivity, deep uGMRT images of a $\sim7.5$ deg$^2$ region centered on the Coma cluster in two bands, 250--500 MHz band and 550--850 MHz, with angular resolutions of $\sim$6\farcs1 and $\sim$3\farcs7, and \textsc{rms} noise levels of 21.1–-36.6 $\mu$Jy and 12.8–-42.4 $\mu$Jy, respectively.  We detected 135 sources in the 250–500 MHz band that have clear extensions (minimum size of $0\farcm45$, 12.6 kpc at the distance of the Coma cluster) with peak-to-local-noise ratio $> 4$.  We culled a complete sample of 24 radio sources that are associated with the Coma cluster galaxies, whose radial velocities are in the range 4000 to 9400 km~s$^{-1}$.  We also supplement this radio-selected sample with 20 RPS galaxies from \citep{Chenetal}, of which eight are included in the extended radio source sample and an additional five are detected and extended (but with either peak-to-local-noise ratio $< 4$ or extent $< 0\farcm45$).

We presented radio morphologies, radio spectra (from 250-–500 MHz and 550--850 MHz), spectral maps, and equipartition properties for these two, extended radio source and RPS galaxy, samples.  Our results are summarized as follows:

\begin{itemize}
\item[(i)] Radio continuum tails, as expected from the effects of ram pressure, are found in almost all the sources (19 of 24 extended radio sources and nine of 13 detected RPS galaxies) in both bands.
\item[(ii)] The radio extent at two bands of the radio continuum tails is consistent with the length being limited by synchrotron cooling time.
\item[(iii)] The derived spectral structure in all tailed radio sources suggests that the radio spectra steepen from the heads to the outer parts of the tails, as is typically seen in tailed radio sources.
\item[(iv)] The majority of extended radio sources (55\%) and RPS galaxies (69\%) are steep spectrum sources.  In addition to
radiative (synchrotron and inverse Compton) cooling, these sources may also undergo adiabatic cooling and the decrease of the magnetic fields in radio-emitting plasma that outflows from the cores can further shift the break in the observed radio spectrum to lower frequencies, contributing to the spectrum steepening.
\item[(v)] The equipartition properties lie within a narrow range (e.g., $P_{\rm min}$ = 1--3 dynes~cm$^{-2}$), with the exception of NGC\,4874, one of the two core BCGs, whose central energy density and pressure are about five times higher than all the other Coma-member galaxies, and whose radio source age is about 50\% that of the other Coma galaxies.  It is plausible that selection effects may be responsible for the narrow range.
\item[(vi)] The radio jet of NGC\,4874 appears to be confined by the 1~keV mini-corona, and they \textit{blossom} into radio lobes after the jet has transitioned from the corona to the hot ICM \citep[see also][]{Sunetal2007,Sandersetal}. A comparison of the lower and higher radio frequency observations shows no evidence, at lower frequencies, for a population of relativistic particles from any earlier (older) ourbursts from the central SMBH. In addition, the NGC\,4874 radio source is relatively young - its age is at least $\sim2$ times smaller than all the other extended sources (see Table~5). The activity of the central AGN in the radio band can be viewed as one of the arguments in favor of NGC\,4874, rather than NGC\,4889, being the true center of the Coma cluster potential well in the near past.
\item[(vii)] We find an ensemble of seven sources, within $\sim$14$^\prime$ ($\approx$ 390~kpc), located $\sim$ 15\farcm2 (= 445~kpc) east of the cluster center, and all moving toward the cluster center, though seen in projection.
\item[(viii)] We show that in both the extended radio source sample and the RPS galaxy sample, the radio morphologies and their spectra suggest that these galaxies are moving radially toward the cluster center, in agreement with \citet{Smith2010},
\citet{RobertsParker} and \citet{Robertsetal2021} who used differently selected samples.  This implies that these galaxies are (predominantly) recent newcomers to Coma and are infalling into Coma for the first time. We emphasize that this refers to objects located in the Coma core (at least in projection).
\item[(ix)] We find a suggestion that the mean recession velocity of sources moving radially toward the cluster center is higher than the mean velocity of Coma cluster galaxies. Coupled with the argument that tailed sources in the core are on their first infall, this suggests that most of them are infalling from the near side of the Coma cluster.
\item[(x)] The diffuse tail of emission trailing the tailed radio galaxy associated with NGC\,4789 is detected in both uGMRT bands.  Moreover, we report transverse spectral structure, a gradual steepening from southwest to northeast across the width of the radio source.
\item[(xi)] The radio relic, B1253$+$275, plausibly associated with a \textit{runaway} merger shock \citep[see also][]{Zhangetal} has a large extent $\approx$ 32$^\prime$ $\times$ 10$^\prime$ ($\simeq$ 1.08 $\times$ 0.34 Mpc$^2$) and the low-frequency flux density $S_{\rm 400\,MHz}$ = 0.98 $\pm$0.06 Jy.
This extended radio relic shows diffuse emission with a central ridge, a bridge of emission connected with the narrow-angle-tailed galaxy NGC\,4789, which suggests that this tailed radio galaxy could be the source of the seed particles feeding the radio relic, 1253$+$275 \citep[see also][]{Giovanninietal1990,Bonafedeetal2021,Bonafedeetal2022}.
\item[(xii)] The uGMRT observations show a long, broad tail of radio emission ($\simeq$ 2.5$^{\prime}$ = 70~kpc; see Fig.~13) co-spatial with the X-ray slingshot tail of the wide-angle radio galaxy NGC\,4839. The X-ray and radio tails extend to the southwest, away from the Coma core. The detailed surface brightness of the radio tail appears to match that in the X-ray, e.g., a spine of emission extends along the southeast side of the tail.  Spectral steepening between the compact inner lobes and the outer tailed radio structures is observed.
\end{itemize}

\section*{Acknowledgments}

We thank the anonymous referee for his/her comments that improved this paper.
D.V.L. acknowledges the support of the Department of Atomic Energy, Government of India, under project no. 12-R\&D-TFR-5.02-0700.
W.F. and C.J. acknowledge support from the Smithsonian Institution, the \textit{Chandra} High Resolution Camera Project through NASA contract NAS8-03060, and NASA Grants 80NSSC19K0116, GO1-22132X, and GO9-20109X.
N.L. acknowledges partial support by the RSF grant 19-12-00369.
R.J.vW. and I.D.R. acknowledge support from the ERC Starting Grant ClusterWeb 804208
We thank the staff of the GMRT who made these observations possible. The GMRT is run by the National Centre for Radio Astrophysics of the Tata Institute of Fundamental Research.
This research has made use of the NED, which is operated by the Jet Propulsion Laboratory, Caltech, under contract with the NASA, and NASA's Astrophysics Data System.
This research has made use of the NASA/IPAC Infrared Science Archive, which is funded by the NASA and operated by the California Institute of Technology.
This research has made extensive use of SAOImage DS9, in addition to software provided by the CXC in the application packages CIAO.

Facilities: {GMRT, \textit{Chandra}, Sloan Digital Sky Survey}

Software: {AIPS \citep{van_moorsel_aips_1996}, CASA \citep{mcmullin_casa_2007}, DS9 \citep{joye_new_2003}, Astropy \citep{robitaille_astropy_2013}}

%\facilities{GMRT, \textit{Chandra}, Sloan Digital Sky Survey}

%\software{AIPS \citep{van_moorsel_aips_1996}, CASA \citep{mcmullin_casa_2007}, DS9 \citep{joye_new_2003}, Astropy \citep{robitaille_astropy_2013,the_astropy_collaboration_astropy_2018}, NumPy \citep{harris_array_2020}, Matplotlib \citep{hunter_matplotlib_2007}, pyregion (\url{https://github.com/astropy/pyregion}), Adobe Photoshop 2020 (\url{https://www.adobe.com/products/photoshop.html})}

\beginappendixA

\appendix

\section{Extended Radio Sources and RPS Galaxies}
\label{sec:app-ex-rps}

In Tables~\ref{tab:tab2-ex} and ~\ref{tab:tab3-rps}, we list 24 extended ($\gtrsim$12.6~kpc) radio sources and 20 RPS galaxies, respectively, that are confirmed members of the Coma cluster, ordered in increasing right ascension.
The radio images of these extended cluster-member radio sources and RPS galaxies are presented in Figs.~\ref{fig:fig-ex} and \ref{fig:fig-rps}, in the order listed in Tables~\ref{tab:tab2-ex} and \ref{tab:tab3-rps}, respectively.
The radio contours, in magenta and blue, correspond, respectively, to the 250--500 MHz and the 550--850 MHz band images that are overlaid on the grayscale SDSS (DR12) $r$-band image.

\subsection{Notes on the Radio Morphology of Individual Extended Sources}
\label{sec:app-notes-ex}

Descriptions of the radio morphologies and their optical properties of these extended sources along with notes from the literature are given below.

\paragraph{1: 2MASX J12520684$+$2701352} (a.k.a. KUG 1249$+$272B)
An emission-line galaxy \citep{KentandGunn}.
The radio morphologies in the 250--500 MHz and the 550--850 MHz bands are amorphous.  The low-frequency, 250--500 MHz image has larger extent as compared to the 550--850 MHz image, probably due to synchrotron cooling.
The largest projected linear extent of the source in the 250--500 MHz band is 0\farcm45 ($\approx$ 12.6~kpc).
The integrated spectral index is very steep, $\alpha$(400--700 MHz) = $-$1.23 $\pm$0.13.

\paragraph{2: NGC 4789}
The optical host is type SA0 \citep{KentandGunn}.
The radio morphologies at both bands show the source to be a narrow-angle-tailed radio galaxy.  We barely detect the extended source, 1253$+$275 in our high angular resolution image (see Fig.~\ref{fig:f1-b3}).  Instead, the tailed radio source at the southwest is seen clearly in our low-resolution image (Fig.~\ref{fig:n4789-xray}).
The largest projected linear extent of the NGC\,4789 galaxy in the 250--500 MHz band image (Fig.~\ref{fig:fig-ex}) is 7\farcm37 ($\approx$ 206.5~kpc).
The host galaxy (D$_{25}$ = 114\farcs3) and the tail have spectral index, $\alpha$(400--700 MHz), $-$0.46 $\pm$0.02 and $-$0.51 $\pm$0.03, respectively.
The integrated spectral index is flat, $\alpha$(400--700 MHz) = $-$0.49 $\pm$0.02.

\paragraph{3: Mrk 53} (a.k.a. KUG 1253$+$279)
The optical host is Sa with a bright knot to the south \citep{KentandGunn}.
The radio emission at both 250--500 MHz and 550--850 MHz bands is one-sided showing asymmetric tails toward the northwest, extending beyond the optical host galaxy.
The largest projected linear extent of the source in the 250--500 MHz band is 1\farcm23 ($\approx$ 34.5~kpc).
The host galaxy (D$_{25}$ = 19\farcs0) and the tail have spectral index, $\alpha$(400--700 MHz), $-$0.64 $\pm$0.03 and $-$0.43 $\pm$0.04, respectively.
The integrated spectral index is flat, $\alpha$(400--700 MHz) = $-$0.75 $\pm$0.07.

\paragraph{4: NGC 4827}
The optical host is S0 and is a low-ionization nuclear emission-line region galaxy \citep{KentandGunn}.
The radio morphology at both bands, 250--500 MHz and 550--850 MHz, is nearly identical.  The north and south radio lobes show a disturbed morphology, extending toward the east, suggesting that the motion of the host galaxy in the ICM is possibly responsible for these extensions.  Furthermore, the northern jet is of smaller extent than the southern radio jet, suggesting that projection effects are also at play.
The largest projected linear extent of the source in the 250--500 MHz band is 4\farcm66 ($\approx$ 130.6~kpc).
The host galaxy (D$_{25}$ = 86\farcs7) and the mean of the northern and southern tails (or radio lobes) have spectral index, $\alpha$(400--700 MHz), $-$0.11 $\pm$0.01 and $-$0.39 $\pm$0.02, respectively.
The integrated spectral index is flat, $\alpha$(400--700 MHz) = $-$0.49 $\pm$0.01.

\paragraph{5: NGC 4839}
The optical host is an SA0, a low/average surface brightness disk-dominated galaxy \citep{Oemler1976}, and is the BCG of the NGC\,4839 group \citep{KentandGunn}.  It lies in the cluster outskirts ($\sim$1~Mpc in projection) southwest of the cluster center.  X-ray observations of the Coma cluster have revealed a wealth of substructures and one of the most prominent is the infalling group of galaxies associated with the galaxy NGC\,4839.  The associated radio source has a wide-angle tailed morphology, i.e., the two-sided radio jet with slight distortions emerging in P.A. $\approx$0$^\circ$ with a bow-shock-like  feature with indications of ram pressure stripping, which suggests that the NGC\,4839 group is falling into the Coma cluster \citep{2001A&A...365L..74N}.  The high-resolution image, along with the low-angular resolution image of the source in the 250--500 MHz band (see Fig.~\ref{fig:n4839-xray}), suggests there are two finger-like features emanating toward the southwest.  These images also suggest that there is low-surface brightness diffuse radio emission toward the west, which trails behind the wide-angle-tailed radio galaxy.  The largest projected linear extent of the source in the 250--500 MHz band is 2\farcm47 ($\approx$ 69.2~kpc).
%The integrated spectral index is $\alpha$(400--700 MHz) = $-$0.90 $\pm$0.02 for the source. 

\paragraph{6: Mrk 55} (a.k.a. KUG 1254$+$276) 
The optical host is Sa with no signs of interaction and is classified as a star-forming galaxy showing signatures of low-ionization nuclear emission-line region galaxy \citep{KentandGunn,MillerOwen2002}.
The radio morphology shows emission associated with the optical host and two additional knot-like (lobes) emission aligned along the east-west.  There is also very low-surface brightness diffuse radio emission detected at 250--500 MHz.  Although the radio morphology at the 550--850 MHz band is identical to the 250--500 MHz band, it has smaller extent with no extended very low-surface brightness diffuse radio emission.
The largest projected linear extent of the source in the 250--500 MHz band is 0\farcm57 ($\approx$ 16.0~kpc).
%The integrated spectral index is  $\alpha$(400--700 MHz) = $-$0.78 $\pm$0.14 for the source.

\paragraph{7 (c): KUG 1255$+$283} (a.k.a.  (GMP~4555, MCG\,$+$05$-$31$-$035)
The elliptical optical host is Sb and there are possibly two overlapping galaxies with their nuclei $\sim$3$^{\prime\prime}$ apart \citep{BothunandDressler}.
The radio emission at both bands, 250--500 MHz and 550--850 MHz, is one-sided, showing asymmetric tails toward the west, extending $\sim$19.9~kpc beyond the optical host galaxy.  The radio extent in the 550--850 MHz band image is smaller than in the 250--500 MHz band image.
The source is also classified as an RPS galaxy \citep[][see Sec.~\ref{tab:tab3-rps}]{Chenetal}.
The largest projected linear extent of the source in the 250--500 MHz is 0\farcm67 ($\approx$ 18.8~kpc).
The source has a flat integrated spectral index, $\alpha$(400--700 MHz) = $-$0.45 $\pm$0.08, which is consistent with $S_{\rm 1.4 GHz}$ = 6.76 $\pm$0.08 mJy, within uncertainties \citep{Chenetal}.

\paragraph{8 (d): NGC 4848}
The optical host is a blue disk Scd that shows perturbed gas distributions with most of the neutral hydrogen in the north and a high H\,I gas deficiency \citep{Gavazzi1989}.
The radio emission at both bands, 250--500 MHz and 550--850 MHz, is one-sided with asymmetric tails toward the northwest, extending beyond the optical host galaxy.  The 550--850 MHz image shows smaller extent than the 250--500 MHz image.
The source is classified as an AGN \citep{KentandGunn}, and is listed as a RPS galaxy by \citet[][see Sec.~\ref{tab:tab3-rps}]{Chenetal}.
The largest projected linear extent of the source at 250--500 MHz band is 1\farcm23 ($\approx$ 34.5~kpc).
The integrated spectral index is  $\alpha$(400--700 MHz) = $-$0.60 $\pm$0.07, which agrees with $S_{\rm 1.4 GHz}$ = 23.85 $\pm$0.11 mJy, within uncertainties \citep{Chenetal}.
 
\paragraph{9: 2MASX J12581865$+$2718387} (a.k.a. KUG 1255$+$275)
The optical host is an irregular galaxy showing emission lines.  More specifically, it is a H\,I rich galaxy with regular gas distribution \citep{KentandGunn,Gavazzi1989}.
The radio emission at both bands, 250--500 MHz and 550--850 MHz, is one-sided, showing asymmetric tails toward the southeast, extending beyond the optical host galaxy with the 550--850 MHz image showing smaller extent than the 250--500 MHz image.
The largest projected linear extent of the source in the 250--500 MHz band is 0\farcm71 ($\approx$ 19.9~kpc).
The integrated spectral index is very flat, $\alpha$(400--700 MHz) = $-$0.22 $\pm$0.10.

\paragraph{10: Mrk 56} (KUG 1256$+$275)
The optical host is an S0 with emission-lines \citep{MazzarellaandBalzano}.
The radio emission at both bands, 250--500 MHz and 550--850 MHz, is one-sided showing asymmetric tails towards the northwest, extending beyond the optical host galaxy.  The 550--850 MHz image shows a smaller extent than the 250--500 MHz image.
The largest projected linear extent of the source in the 250--500 MHz band is 1\farcm21 ($\approx$ 33.9~kpc).
The integrated spectral index is very flat, $\alpha$(400--700 MHz) = $-$0.11 $\pm$0.08.

\paragraph{11: Mrk 57}
The optical host is an irregular H\,II galaxy with emission-lines \citep{KentandGunn}.  It is H\,I rich with a H\,I distribution displaying an extension to the north-south \citep{BravoAlfaro}.
The radio emission at both bands, 250--500 MHz and 550--850 MHz, is one-sided, showing asymmetric tails toward the south, extending beyond the optical host galaxy with the 550--850 MHz image showing smaller extent than the 250--500 MHz image.
The largest projected linear extent of the source in the 250--500 MHz band is 0\farcm90 ($\approx$ 25.2~kpc).
The host galaxy (D$_{25}$ = 29\farcs25) and the tail have a spectral index, $\alpha$(400--700 MHz), $-$0.57 $\pm$0.03 and $-$1.12 $\pm$0.05, respectively.
%The integrated spectral index is  $\alpha$(400--700 MHz) = $-$0.86 $\pm$0.05 for the source.

\paragraph{12 (h): IC 3949} (a.k.a. KUG 1256$+$281)
The optical host is SA0, an edge-on spiral galaxy and the radio emission is dominated by star formation in its disk with a more quiescent central bulge \citep{BothunandDressler}.
The source is classified as an AGN by \citet{KentandGunn} and it is also classified as the RPS galaxy \citep[][see Sec.~\ref{tab:tab3-rps}]{Chenetal}.
The radio emission at both 250--500 MHz and 550--850 MHz is one-sided.  Only the 250--500 MHz image shows asymmetric tails toward the southwest, extending slightly beyond the optical host galaxy, whereas the 550--850 MHz image has smaller extent and is coincident with the extent of the optical host.
The largest projected linear extent of the source in the 250--500 MHz is 0\farcm50 ($\approx$ 14.0~kpc).
The integrated spectral index is  $\alpha$(400--700 MHz) = $-$0.61 $\pm$0.10, and agrees with $S_{\rm 1.4 GHz}$ = 2.37 $\pm$0.09 mJy, within uncertainties \citep{Chenetal}.

\paragraph{13 (i): NGC 4858}
The optical host  is an Sb showing interaction with NGC\,4860 (at $z$ = 0.02645), which is located to the northeast \citep{BothunandDressler}.
The radio emission at both bands, 250--500 MHz and 550--850 MHz, is one-sided, showing asymmetric tails toward the northwest, extending beyond the optical host galaxy with the 550--850 MHz image showing smaller extent than the 250--500 MHz image.
The source is also classified as the RPS galaxy \citep[][see Sec.~\ref{tab:tab3-rps}]{Chenetal}.
The largest projected linear extent of the source in the 250--500 MHz band is 0\farcm79 ($\approx$ 22.1~kpc).
The integrated spectral index is steep, $\alpha$(400--700 MHz) = $-$1.07 $\pm$0.10, and it agrees with $S_{\rm 1.4 GHz}$ = 8.73 $\pm$0.13 mJy, within uncertainties \citep{Chenetal}.

\paragraph{14 (j): Mrk 58} (a.k.a. KUG 1256$+$279)
The optical host is a blue disk Sa galaxy \citep{KentandGunn}.  It shows a very asymmetric gas distribution; most of the H\,I gas is lies to the west side, while the east appears depleted \citep{MazzarellaandBalzano}.
The radio emission at both bands, 250--500 MHz and 550--850 MHz, is one-sided showing asymmetric tails toward the southwest, extending beyond the optical host galaxy with the 550--850 MHz image showing smaller extent than 250--500 MHz image.
The source is also classified as RPS galaxy \citep[][see Sec.~\ref{tab:tab3-rps}]{Chenetal}.
The largest projected linear extent of the source in the 250--500 MHz band is 0\farcm74 ($\approx$ 20.7~kpc).
The integrated spectral index is  $\alpha$(400--700 MHz) = $-$0.76 $\pm$0.11, and it agrees with $S_{\rm 1.4 GHz}$ = 3.29 $\pm$0.15 mJy, within uncertainties \citep{Chenetal}.

\paragraph{15: NGC 4869}
A narrow-angle-tailed radio galaxy is hosted by an elliptical galaxy \citep{1990AJ.....99.1381V}.  The radio morphology at both bands 250--500 MHz and 550--850 MHz, is typical of a head-tail radio source with a weak unresolved radio core, two oppositely directed radio jets, and a long-low-surface brightness tail, pointing away from the cluster center.  The radio tail has a conical shape, which initially expands and then recollimates.  The radio jet at $\sim$3\farcm5 ($\approx$ 96.1~kpc) bends by $\sim$70~deg with respect to the initial direction of propagation, after the two radio jets have undergone a twist and wrap of the two tails.  A detailed study of the radio source is presented in \citet{Lal2020b,Ferettietal1990}.
The largest projected linear extent of the source in the 250--500 MHz band is 6\farcm63 ($\approx$ 185.8~kpc).
The host galaxy (D$_{25}$ = 51\farcs24) and the tail have spectral index, $\alpha$(400--700 MHz), $-$0.62 $\pm$0.01 and $-$1.70 $\pm$0.02, respectively.
The integrated spectral index is steep, $\alpha$(400--700 MHz) = $-$1.12 $\pm$0.01.

\paragraph{16: NGC 4874}
One of the two dominant BCGs of the Coma cluster \citep{BaierTiersch1990} is a low-ionization nuclear emission-line region galaxy \citep{KentandGunn}.  The optical host is a regular low brightness elliptical galaxy \citep{2000A&A...362..871C}.
Unlike many relaxed clusters that host a single BCG, Coma has two very bright galaxies in the core - NGC\,4889 and NGC\,4874. While the former is brighter by $\sim 0.5$ magnitude, the latter is rounder and possesses a rich system of satellite galaxies.
NGC\,4874 is also a prominent radio galaxy, which is yet another characteristic feature of  giant ellipticals residing at the bottom of  cluster potential wells \citep{Burns1990}.
The source is located close to the peak of the distributions of galaxies and X-ray gas.
The radio source has a wide-angle tail (WAT) radio morphology and its projected maximum angular extension is $\sim$30 arcsec, corresponding to $\sim$15~kpc.
We detect the radio core and two radio jets forming a wide angle between them in both of our 250--500 MHz and 550--850 MHz band images.
The largest projected linear extent of the source in the 250--500 MHz band is 0\farcm81 ($\approx$ 22.7~kpc).
%The integrated spectral index is  $\alpha$(400--700 MHz) = $-$0.80 $\pm$0.02 for the source.

This one of two core BCGs of the Coma cluster, NGC\,4874 also retains its central X-ray gas corona, i.e., the interstellar gas with a temperature of 1--2~keV is confined by the hot intergalactic medium of the Coma Cluster into a compact ($\simeq$ 3~kpc) cloud.
The radio jets are found to be anti-correlated with the X-ray emission, and there are also small X-ray indentations of the nucleus, where the jets leave the coronal gas (see also Section~\ref{sec:discuss}).  It seems that the supermassive black hole in NGC\,4874 is depositing its energy outside of the corona, and it may be ineffective in preventing the cooling taking place \citep{Vikhlininetal,Sandersetal}.

\paragraph{17: 2MASX J12594009$+$2837507} (a.k.a. KUG 1257$+$288B)
The optical host is Sc type \citep{KentandGunn}.
The 250--500 MHz band image suggests that the source has a wide-angle tail morphology, whose radio emission extends much beyond the extent of the host galaxy.
The largest projected linear extent of the source in the 250--500 MHz band is 0\farcm72 ($\approx$ 20.2~kpc).
The source was not covered in the 550--850 MHz band pointings.

\paragraph{18: 2MASX J13001780$+$2723152} (a.k.a. KUG 1257$+$276)
The optical host is an irregular face-on spiral galaxy \citep{KentandGunn}.
The 250--500 MHz band image suggests that the radio source emerges from the host galaxy and has two radio lobes, one just northeast of the radio core and the other east.  Both radio lobes have similar surface brightnesses, and the radio core is offset from the two symmetrical radio lobes.
The largest projected linear extent of the source in the 250--500 MHz band is 1\farcm79 ($\approx$ 50.2~kpc).
The source was not covered in the 550--850 MHz band pointings.

\paragraph{19 (q): KUG\,1258$+$279A}
The optical host shows emission lines of Sa type and is an asymmetrical blue disk galaxy possibly due to the presence of dust \citep{BothunandDressler}.
The radio emission at both bands, 250--500 MHz and 550--850 MHz, is one-sided showing asymmetric tails to the southeast, extending beyond the optical host galaxy with the 550--850 MHz band image showing smaller extent than the 250--500 MHz band image.
The source is also classified as an RPS galaxy \citep[][see Sec.~\ref{tab:tab3-rps}]{Chenetal}.
The largest projected linear extent of the source in the 250--500 MHz band is 1\farcm01 ($\approx$ 28.3~kpc).
The host galaxy (D$_{25}$ = 39\farcs24) and the tail have spectral index, $\alpha$(400--700 MHz), $-$0.39 $\pm$0.07 and $-$0.56 $\pm$0.12, respectively.
The integrated spectral index is  $\alpha$(400--700 MHz) = $-$0.74 $\pm$0.11, and it agrees with $S_{\rm 1.4 GHz}$ = 3.70 $\pm$0.09 mJy, within uncertainties \citep{Chenetal}.

\paragraph{20 (r): IC 4040} (a.k.a.  GMP~2559, KUG 1258$+$283, CGCG\,160$-$252)
The optical host shows emission lines hosted by a galaxy of spiral or irregular morphological type \citep{KentandGunn}.
The radio emission at both bands, 250--500 MHz and 550--850 MHz, is one-sided showing asymmetric tails toward the southeast, extending beyond the optical host galaxy, with the 550--850 MHz band image showing smaller extent than the 250--500 MHz band image.
The source is also classified as an RPS galaxy \citep[][see Sec.~\ref{tab:tab3-rps}]{Chenetal}.
The largest projected linear extent of the source in the 250--500 MHz band is 2\farcm18 ($\approx$ 61.1~kpc).
The integrated spectral index is steep, $\alpha$(400--700 MHz) = $-$1.05 $\pm$0.05, and it agrees with $S_{\rm 1.4 GHz}$ = 16.31 $\pm$0.10 mJy, within uncertainties \citep{Chenetal}.

\paragraph{21: 2MASX J13004067$+$2831116} (a.k.a. KUG 1258$+$287)
The optical host is an edge-on spiral galaxy \citep{KentandGunn}.
The 250--500 MHz band image suggests that the source has tailed morphology, whose radio emission extends much beyond the extent of the host galaxy.
The largest projected linear extent of the source in the 250--500 MHz band is 0\farcm96 ($\approx$ 26.9~kpc).
The source was not covered in the 550--850 MHz band pointings.

\paragraph{22: 2MASX J13004385$+$2824586}
The optical host is an elliptical galaxy \citep{KentandGunn}.
The radio morphology in both bands, 250--500 MHz and 550--850 MHz, is nearly identical.  The two jets traveling toward north and south show knotty features. In addition, the north and south radio lobes show disturbed morphologies, which are bent toward the east, suggesting that the motion of the host galaxy in the ICM is possibly responsible for these extensions.  Furthermore, the northern jet is of larger extent than the southern radio jet, and the southern jet is also more bent, suggesting that projection effects are likely at play.
The largest projected linear extent of the source in the 250--500 MHz band is 1\farcm85 ($\approx$ 51.8~kpc).
The host galaxy (D$_{25}$ = 86\farcs7) and the tail have spectral indices, $\alpha$(400--700 MHz), $-$0.53 $\pm$0.04 and $-$0.83 $\pm$0.05, respectively.
The integrated spectral index is steep, $\alpha$(400--700 MHz) = $-$1.17 $\pm$0.04.

\paragraph{23 (s): NGC 4911}
The optical host is a low-ionization nuclear emission-line region galaxy of Sb type.  It is one of the two giant spirals in the Coma cluster showing a close interaction with its neighbor DRCG\,27$-$62 \citep{KentandGunn,MillerOwen2002}.
The radio emission at both bands, 250--500 MHz and 550--850 MHz, is symmetrically surrounding the optical host, and radio emission in both bands are of identical extent.
The source is also classified as an RPS galaxy \citep[][see Sec.~\ref{tab:tab3-rps}]{Chenetal}.
The source is nearly circular in shape, and the largest projected linear extent of the source in the 250--500 MHz band is 0\farcm73 ($\approx$ 20.5~kpc).
The integrated spectral index is  $\alpha$(400--700 MHz) = $-$0.98 $\pm$0.04, and it agrees with $S_{\rm 1.4 GHz}$ = 15.08 $\pm$0.23 mJy, within uncertainties \citep{Chenetal}.

\paragraph{24: NGC 4927}
The optical host is a very bright giant early spiral of type Sb \citep{KentandGunn} and is possibly interacting with its neighbor DRCG\,27--62 \citep{BravoAlfaro}.
The radio morphology shows emission associated with the optical host and two additional radio lobes along the northwest and southeast directions.  Both radio lobes are of nearly identical surface brightnesses and are located just outside the optical host galaxy.  The northern radio lobe shows slight extensions of radio emission toward the northeast, suggesting motion of the optical host with respect to the ICM is at play.
The largest projected linear extent of the source in the 250--500 MHz band is 0\farcm66 ($\approx$ 18.5~kpc).
The source was not covered in the 550--850 MHz band pointings.

\subsubsection{Spectral Structure of Extended Sources}
\label{sec:app-notes-spec-ex}

The spectral structure of the large ($\gtrsim$ 1\farcm85) angular-sized sources, namely, NGC\,4789, NGC\,4827, NGC\,4839, NGC\,4869, and 2MASX J13004385$+$2824586 are elaborated further.

\paragraph{2: NGC\,4789}
The spectral structure of the classic wide-angle-tailed radio galaxy NGC\,4789 is presented in Fig.~\ref{fig:spec-in}.  The head, located close to the optical host, has a flat spectrum between 250--500 MHz and 550--850 MHz. The spectrum between these two bands gradually steepens with increasing distance from the head.  In addition, we notice transverse spectral structure, a gradual steepening from southwest to northeast across the width of the radio tail, close to the head, before the jets bend toward the northwestern direction, possibly due to the ballistic motion of the host galaxy through a dynamic ICM.  The faintest, and presumably oldest, far-end part of the tails have felt large scale motions for the longest period of time, and hence, in the far-end part of the tails, we do not see transverse spectral structure.  As discussed above, this wide-angle-tailed radio galaxy is feeding the radio relic source, 1253$+$275.  Unfortunately, the radio galaxy, Coma\,A causes uncertainties in the 550--850 MHz image.  Thus our dynamic range limited map at 550--850 MHz band does not allow us to perform spectral imaging of this radio relic source (see also Sec.~\ref{sec:src-interact}, `NGC\,4789 and the 1253$+$275 relic' for a detailed discussion).

\paragraph{4: NGC\,4827}
The spectral structure of the classic FR\,II type radio galaxy, NGC\,4827 shows a flat spectrum radio core and jets, which form relatively steeper spectrum radio lobes.  The spectral information is similar in the two lobes and the lobes are filled with backflow material, which is older in the center than on the edges of the radio galaxy.  The spectrum steepens from the outer regions toward the center of the source, and also steepens outwards, perpendicular to the major axis of the source, similar to the spectral structure of FR\,II radio galaxies.  The radio lobes are not symmetric across the radio jet axis, i.e., the eastern extent is larger than the western extent, which is possibly due to the motion of the host galaxy.
In the host galaxy (D$_{25}$ = 86\farcs7), the northern and southern tails have spectral indices, $\alpha$(400--700 MHz), $-$0.11 $\pm$0.01, $-$0.35 $\pm$0.02  $-$0.43 $\pm$0.02, respectively.
The integrated spectral index is flat, $\alpha$(400--700 MHz) = $-$0.49 $\pm$0.01.

\paragraph{5: NGC\,4839}
The spectral structure of the BCG of the NGC\,4839 group shows transverse spectral steepening, similar to NGC\,4789, in addition to gradual spectral steepening from the flat spectrum radio core toward the northern and southern radio tails.  The two finger-like features emanating toward the southwest have the steepest spectrum, which is comparable to the low-surface brightness diffuse emission towards the west, that trails behind the wide-angle-tailed radio galaxy.
The integrated spectral index is $\alpha$(400--700 MHz) = $-$0.90 $\pm$0.02 (see also Sec.~\ref{sec:src-interact}, paragraph `NGC\,4839' for a detailed discussion).
%(see also Fig.~\ref{fig:n4789-n4839}, right panel).

\paragraph{15: NGC\,4869}
\citet{Lal2020b} presented a detailed treatment of the head-tail radio galaxy, NGC\,4869.  Its radio spectrum shows progressive steepening with increasing distance from the head because of the effects of radiation losses.  They also reported that the jet bends by $\sim$70$^\circ$ with respect to the initial radio tail direction and there is a clear presence of a ridge, i.e., the flaring of a straight, collimated radio jet as it crosses a surface brightness edge.  The low-frequency radio emission on either side of this surface brightness edge is remarkably different in spectral index, $-$1.21 $\pm$0.07 (inside) and $-$1.58 $\pm$0.11 (outside) using our 250--500 MHz and 550-850 MHz band images (see Fig.~\ref{fig:spec-in}).

\paragraph{22: 2MASX J13004385$+$2824586}
The spectral structure shows spectral steepening from the radio core along the north and south jets and radio lobes.  There are occasional departures from this simple picture of gradual spectral steepening at the locations of jet knots, which are the sites of re-acceleration.  The integrated spectral index is steep, $\alpha$(400--700 MHz) = $-$1.17 $\pm$0.04.

\subsection{Notes on Radio Morphology of Remaining RPS Galaxies}
\label{sec:app-notes-rps}

\paragraph{(b): GMP 4570} The radio emission peaks at the location of the host galaxy and extends toward the west, beyond the extent of the optical host, suggesting it is a tailed radio morphology source.
The largest projected linear extent of the source in the 250--500 MHz band is $\approx$ 7.0~kpc.
The source was not covered in our 550--850 MHz band pointings.  It was detected in the measurement at 1.4 GHz \citep[see Table~\ref{tab:tab3-rps},][]{Chenetal}.
Thus, the integrated spectral index is $\alpha$(400--1400 MHz) = $-$0.51 $\pm$0.11, which agrees, within uncertainties with the upper limit, $\alpha$(400--700 MHz) $\lesssim$ $-$0.98 $\pm$0.52.

\paragraph{(e): NGC 4853} (a.k.a. GMP~4156)
The 250--500 MHz band image shows slight extensions in radio emission toward the north, beyond the extent of the optical host, whereas, the extent in the 550-250 MHz image is similar to the extent of the optical host.
The largest projected linear extent of the source in the 250--500 MHz band is $\approx$ 7.3~kpc.
The integrated spectral index is  $\alpha$(400--700 MHz) = $-$0.77 $\pm$0.12, which agrees, within uncertainties with the measurement at 1.4 GHz \citep[see Table~\ref{tab:tab3-rps},][]{Chenetal}.

\paragraph{(k): KUG 1257$+$278} (a.k.a. GMP~3271)
Both the 250--500 MHz and 550-250 MHz band images show extensions in radio emission toward the east, much beyond the extent of the optical host.
The largest projected linear extent of the source in the 250--500 MHz band is $\approx$ 10.6~kpc.
The integrated spectral index is  $\alpha$(400--700 MHz) = $-$0.86 $\pm$0.29, which agrees, within uncertainties with the measurement at 1.4 GHz \citep[see Table~\ref{tab:tab3-rps},][]{Chenetal}.

\paragraph{(o): Mrk 60} (a.k.a. D100, GMP~2910)
Both, the 250--500 MHz and 550-250 MHz band images show extensions in radio emission toward the northeast, beyond the extent of the optical host.  The 250--500 MHz band image also shows diffuse extensions toward the north, providing it an amorphous-like radio morphology, though the 550-250 MHz band image suggests it has a tailed radio morphology.
The largest projected linear extent of the source in the 250--500 MHz band is $\approx$ 11.5~kpc.
The integrated spectral index is  $\alpha$(400--700 MHz) = $-$0.93 $\pm$0.46, which agrees, within uncertainties with the measurement at 1.4 GHz \citep[see Table~\ref{tab:tab3-rps},][]{Chenetal}.

\paragraph{(t): NGC 4921} (a.k.a. GMP~2059)
The 250--500 MHz band image suggests it is an irregular-shaped radio morphology source.
The largest projected linear extent of the source in the 250--500 MHz band is $\approx$ 13.2~kpc.
We do not detect it in our 550-250 MHz band image, but the source was detected in the measurement at 1.4 GHz \citep[see Table~\ref{tab:tab3-rps},][]{Chenetal}.
Thus, the integrated spectral index is $\alpha$(400--1400 MHz) = $-$1.16 $\pm$0.22, which agrees, within uncertainties with the upper limit, $\alpha$(400--700 MHz) $\lesssim$ $-$1.90 $\pm$0.58.

\section{Extended Radio Sources: Either with foreground/background Galaxies or without Redshift Data}
\label{app:other}

\begin{figure*}[ht]
\begin{center}
\begin{tabular}{c}
\includegraphics[width=17.0cm]{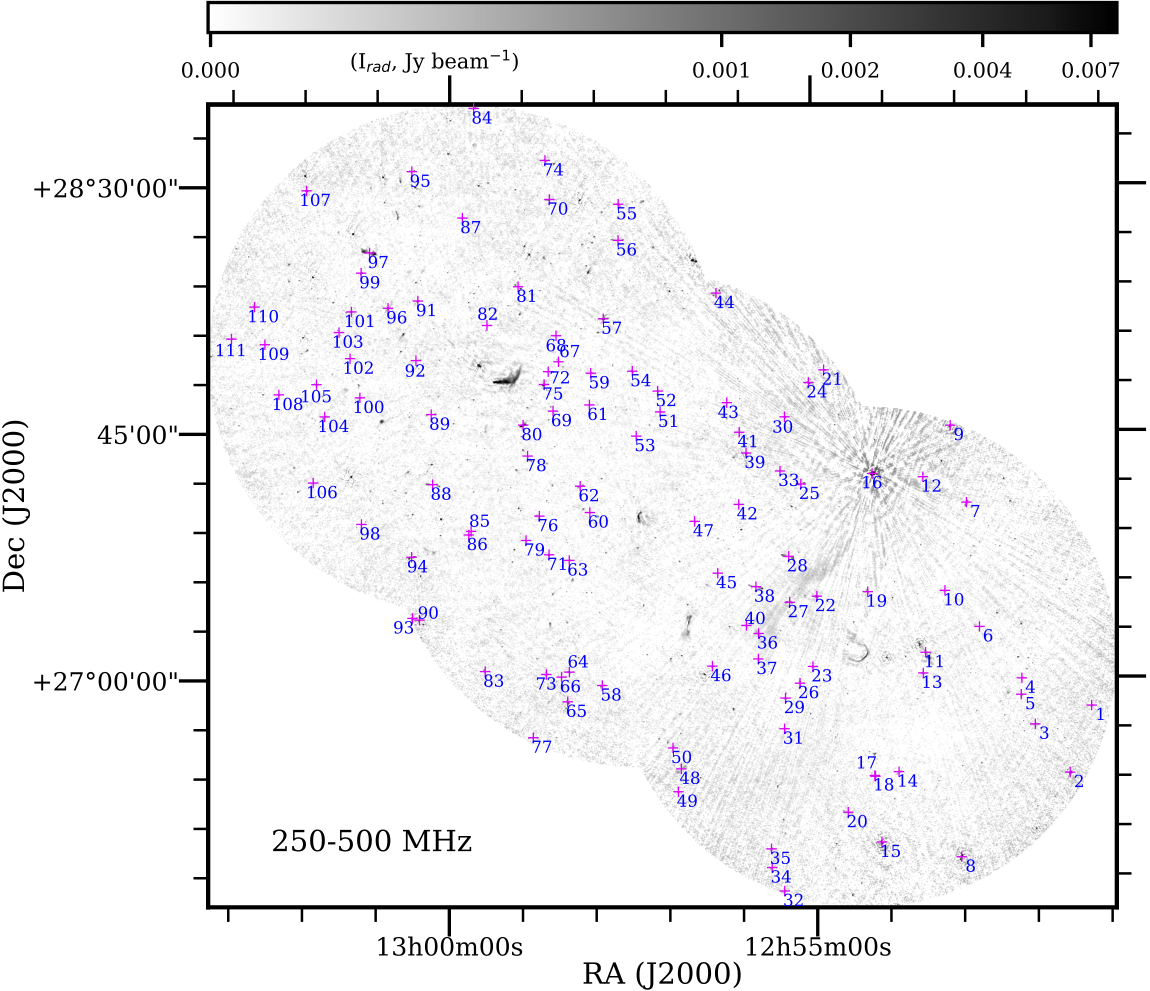}
\end{tabular}
\end{center}
\caption{Mosaic image of three pointings at an angular resolution of $\sim$6\farcs1 at the 250--500 MHz band of uGMRT, described in Table~\ref{tab:obs-log}.  The census of extended sources that are mapped in fields of view of the Coma cluster, listed in Table~\ref{tab:app-tab1}.  The magenta points mark the radio positions and their sources\_IDs are depicted in blue, adjacent to the extended sources.  The grayscale image is displayed in logarithmic scales to emphasize the extended, low-surface brightness diffuse radio emission and radio emission associated with the radio galaxies.}
\label{fig:b3-app-f1}
\end{figure*}

Table~\ref{tab:app-tab1} lists the remaining 111 extended sources, labeled in Fig.~\ref{fig:b3-app-f1} that are either foreground/background galaxies or do not have redshift data.
The radio images of these extended radio sources are presented in Fig.~\ref{fig:app1}, in the order presented in  Table~\ref{tab:app-tab1}.
The columns in Table~\ref{tab:app-tab1} are as follows: (1) source name; (2--3) probable radio core position, right ascension and declination (J2000) at the 250--500 MHz; (4) redshift as reported in NED; (5--6) local \textsc{rms} noise in the near vicinity of the source for the 250--500 MHz and 550--850 MHz images; (7) the radio morphology of the sources shown in the maps presented; and (8) alternate name.
The source\_ID as tabulated in the Table~\ref{tab:app-tab1} is labeled in the upper-right corner of each image in Fig.~\ref{fig:app1}.  The radio contours in magenta and blue correspond, respectively, to the 250--500 MHz and 550--850 MHz band images that are overlaid on the grayscale SDSS (DR12) $r$-band image.
We adopt the definitions of \citet{Rudnick2021} for the classification of radio sources.

\tabletypesize{\scriptsize}
\def\arraystretch{0.5}
\begin{table*}[ht]
\tablewidth{0pt}
\caption{Summary of the Nature of Radio Sources in the Coma Cluster Field Using the uGMRT.}
\label{tab:app-tab1}
\begin{center}
\begin{tabular}{rlllcccll}
\hline \hline
 & Source\_ID & \multicolumn{2}{c}{Radio Core Position} & \multicolumn{1}{c}{$z$} &\multicolumn{2}{c}{\textsc{rms}} & Radio & Alternate Name(s) \\
\cline {1-2}  & &\multicolumn{2}{c}{R.A. (400 MHz) Decl.}             &  & \multicolumn{1}{c}{B-3} & \multicolumn{1}{c}{B-4} & Morphology & \\
  &     & \multicolumn{2}{c}{(J2000)}  &   & \multicolumn{3}{l}{($\mu$Jy~beam$^{-1}$)} & \\
  & (1) & \multicolumn{1}{c}{(2)} & \multicolumn{1}{c}{(3)} & (4) & (5) & \multicolumn{1}{c}{(6)} & \multicolumn{1}{c}{(7)} & \multicolumn{1}{c}{(8)} \\
\hline\noalign{\smallskip}
  1 &                                & 12:51:14.19 & $+$26:54:40.5 &         & 24 &    & A, D  & \\
  2 &                                & 12:51:32.67 & $+$26:42:30.3 &         & 96 &    & CD, FR\,II  & \\
  3 & WISEA J125201.62$+$265126.7    & 12:52:00.51 & $+$26:51:25.1 & 0.40942 & 48 &    & Tr  & \\
  4 &                                & 12:52:10.94 & $+$26:59:50.7 &         & 18 &    & D, Tr  & \\
  5 & WISEA J125211.65$+$265649.1    & 12:52:11.64 & $+$26:56:51.6 & 0.90399 & 24 &    & CD  & \\
  6 & WISEA J125245.04$+$270920.3    & 12:52:45.10 & $+$27:09:20.6 & 0.60126 & 24 & 12 & Tr  & \\
  7 &                                & 12:52:54.36 & $+$27:32:03.5 &         & 24 & 24 & CD, FR\,II  & \\
  8 & WISEA J125302.24$+$262722.3    & 12:53:02.23 & $+$26:27:21.9 & 1.26153 & 96 &    & CD, FR\,II  & \\
  9 &                                & 12:53:06.86 & $+$27:46:08.3 &         &192 & 96 & FD, FR\,II  & \\
 10 &                                & 12:53:13.02 & $+$27:16:00.9 &         & 24 & 24 & CD, FR\,II  & \\
 11 &                                & 12:53:29.57 & $+$27:04:45.5 &         & 48 & 56 & Tr  & \\
 12 &                                & 12:53:29.94 & $+$27:36:48.3 &         & 48 & 36 & A, D  & \\
 13 & WISEA J125332.57$+$270050.0    & 12:53:31.64 & $+$27:00:59.3 & 2.04431 & 24 & 12 & CD, FR\,II  & \\
 14 &                                & 12:53:52.46 & $+$26:43:02.0 &         & 24 & 12 & Tr, FR\,II  & \\
 15 & WISEA J125405.51$+$263028.2    & 12:54:07.09 & $+$26:30:14.1 & 0.16842 &192 & 96 & CD, FR\,II  & \\
 16 & Coma A                         & 12:54:12.01 & $+$27:37:34.0 & 0.08534 &768 &384 & FD, FR\,II  & 3C 277.3, 4C $+$27.22 \\
 17 &                                & 12:54:12.26 & $+$26:46:24.7 &         & 24 & 12 & BT  & \\
 18 & WISEA J125413.96$+$264153.8    & 12:54:11.78 & $+$26:42:14.4 & 0.00007 & 24 & 12 & A  & \\
 19 & SDSS J125416.43$+$271556.4     & 12:54:16.42 & $+$27:15:56.1 & 0.29825 & 36 & 18 & HT, WAT  & \\
 20 & WISEA J125432.27$+$263453.4    & 12:54:34.33 & $+$26:35:43.7 &         & 36 & 18 & D, Tr  & \\
 21 &                                & 12:54:50.88 & $+$27:56:31.3 &         & 48 & 48 & CD  & \\
 22 & WISEA J125458.24$+$271513.5    & 12:54:58.21 & $+$27:15:12.8 & 0.25733 & 36 & 18 & A, D  & \\
 23 & WISEA J125501.22$+$270222.8    & 12:55:02.00 & $+$27:02:23.8 & 0.45838 & 24 & 12 & HT, WAT  & \\
 24 & 5C 04.025$^*$                  & 12:55:03.50 & $+$27:54:13.7 & 0.40920 & 48 & 48 & CD. FR\,II  & \\
 25 &                                & 12:55:10.55 & $+$27:35:43.1 &         & 36 & 48 & CD  & \\
 26 &                                & 12:55:12.70 & $+$26:59:21.7 &         & 24 & 18 & A  & \\
 27 & WISEA J125520.65$+$271408.1$^*$& 12:55:20.62 & $+$27:14:09.1 & 1.76500 & 36 & 24 & CD, FR\,II  & \\
 28 & WISEA J125519.16$+$272305.3    & 12:55:21.03 & $+$27:22:31.4 &         & 24 & 36 & CD, FR\,II  & \\
 29 & SDSS J125523.89$+$265640.9     & 12:55:24.72 & $+$26:56:40.2 & 0.43852 & 24 & 24 & BT  & \\
 30 & MCG\,$+$05$-$31$-$007          & 12:55:25.05 & $+$27:47:53.0 &         & 36 & 48 & CD, FR\,II  & \\
 31 & AGC 228097                     & 12:55:25.70 & $+$26:51:04.6 &         & 24 & 12 & CD, FR\,II  & LEDA 5060293 \\
 32 &                                & 12:55:26.75 & $+$26:21:28.1 &         & 48 & 12 & CD, FR\,II  & \\
 33 & SDSS J125526.52$+$273804.7$^*$ & 12:55:27.62 & $+$27:38:07.8 & 0.10645 & 24 & 24 & BT  & \\
 34 & WISEA J125537.60$+$262540.4    & 12:55:36.80 & $+$26:25:45.7 & 0.42347 & 36 & 12 & A, D  & \\
 35 &                                & 12:55:37.21 & $+$26:29:10.3 &         & 24 & 12 & CD  & \\
 36 &                                & 12:55:46.32 & $+$27:08:30.2 &         & 24 & 12 & A, D  & \\
 37 &                                & 12:55:46.72 & $+$27:03:51.0 &         & 24 & 18 & CD  & \\
 38 &                                & 12:55:48.22 & $+$27:17:02.4 &         & 24 & 18 & CD, FR\,II  & \\
 39 &                                & 12:55:55.48 & $+$27:41:27.2 &         & 96 & 96 & A, D  & \\
 40 & SDSS J125556.79$+$270957.1     & 12:55:56.26 & $+$27:09:58.7 & 0.53605 & 24 & 12 & A, D  & \\
 41 & WISEA J125601.82$+$274525.3$^*$& 12:56:00.96 & $+$27:45:15.3 & 2.05500 & 24 & 24 & A, D  & \\
 42 &                                & 12:56:01.81 & $+$27:32:05.3 &         & 24 & 12 & CD  & \\
 43 & WISEA J12560986$+$2750393      & 12:56:09.86 & $+$27:50:39.3 &         & 24 & 48 & CD  & \\
 44 &                                & 12:56:19.34 & $+$28:10:41.4 &         & 48 &    & CD  & \\
 45 & WISEA J125619.70$+$271940.2$^*$& 12:56:19.49 & $+$27:19:33.9 & 0.07003 & 24 & 12 & BT, CD  & \\
 46 &                                & 12:56:24.40 & $+$27:02:36.1 &         & 24 & 48 & W  & \\
 47 &                                & 12:56:38.25 & $+$27:29:02.7 &         & 24 & 12 & Z  & \\
 48 &                                & 12:56:50.38 & $+$26:43:53.4 &         & 48 & 18 & CD  & \\
 49 &                                & 12:56:52.88 & $+$26:39:43.0 &         & 48 & 48 & CD  & \\
 50 & WISEA J125657.08$+$264742.6    & 12:56:57.08 & $+$26:47:43.1 & 0.63894 & 24 & 18 & A, D  & \\
 51 & WISEA J125708.10$+$274853.4$^*$& 12:57:06.17 & $+$27:49:03.9 & 0.23997 & 24 & 12 & Tr, D  & \\
 52 &                                & 12:57:08.06 & $+$27:52:51.9 &         & 24 & 12 & BT, FR\,I  & \\
 53 &                                & 12:57:25.97 & $+$27:44:40.3 &         & 24 & 12 & CD, Tr  & \\
 54 &                                & 12:57:28.95 & $+$27:56:31.3 &         & 18 &    & CD  & \\
 55 & WISEA J125740.06$+$282701.9    & 12:57:40.00 & $+$28:27:00.5 & 0.41283 & 24 &    & BT  & \\
 56 & SDSS J125739.29$+$282012.3     & 12:57:40.29 & $+$28:20:25.0 & 0.55437 & 36 &    & CD  & \\
 57 & WISEA J125752.93$+$280609.4    & 12:57:52.76 & $+$28:06:07.6 & 0.45360 & 48 & 48 & CD  & \\
 58 &                                & 12:57:54.73 & $+$26:59:10.0 &         & 24 & 12 & BT, FR\,II  & \\
 59 & SDSS J125803.78$+$275601.7     & 12:58:03.31 & $+$27:56:11.0 & 0.26110 & 24 & 24 & Tr  & \\
 60 & WISEA J125801.49$+$273006.0    & 12:58:04.45 & $+$27:30:45.3 &         & 24 & 12 & Tr, FR\,II  & \\
 61 &                                & 12:58:04.51 & $+$27:50:23.8 &         & 24 & 18 & A, D  & \\
 62 & WISEA J125812.36$+$273534.7$^*$& 12:58:12.33 & $+$27:35:33.6 & 0.44673 & 24 & 24 & CD, FR\,II  & \\
\hline
\end{tabular}
\end{center}
\tablecomments{Notes on our interpretation of the radio morphologies for these extended radio sources in the Coma cluster field. \\
Column~1: source name as identified in the NED. \\
Columns~2 and 3: the right ascension and declination (J2000), in increasing right ascension.  Source position is given by the position of the probable radio core. \\
Column~4: spectroscopic redshift; `$*$' depicts photometric redshift of the source. \\
Columns~5 and 6: the local \textsc{rms} noise in near vicinity of source for 250--500 MHz (band-3) and 550--850 MHz (band-4). \\
Column~7: the radio source morphology classification following \citet{Rudnick2021}: A = amorphous; BT = bend-tail; HT = head-tail; WAT = wide-angle-tail; T = tailed source; CD = classical double; FR\,II: FR type II source; Tr: Triple; NAT: narrow-angle-tailed; D: diffuse; W: Winged (or X-shaped); and Z: Z symmetric. \\
Column~8: alternate name(s), as reported in NED, `$**$': \citet{NHM2009}.}
\end{table*}

\setcounter{table}{0}
\tabletypesize{\scriptsize}
\def\arraystretch{0.5}
\begin{table*}[ht]
\begin{center}
\tablewidth{0pt}
%\caption{A summary of the nature of radio sources in the Coma cluster field using the uGMRT.}
\begin{tabular}{rlllcccll}
\hline \hline
 & Source\_ID & \multicolumn{2}{c}{Radio core position} & \multicolumn{1}{c}{$z$} &\multicolumn{2}{c}{\textsc{rms}} & Radio & Alternate name(s) \\
\cline {1-2}  & &\multicolumn{2}{c}{R.A. (400 MHz) Dec.}             &  & \multicolumn{1}{c}{B-3} & \multicolumn{1}{c}{B-4} & morphology & \\
  &     & \multicolumn{2}{c}{(J2000)}  &   & \multicolumn{3}{l}{($\mu$Jy~beam$^{-1}$)} & \\
  & (1) & \multicolumn{1}{c}{(2)} & \multicolumn{1}{c}{(3)} & (4) & (5) & \multicolumn{1}{c}{(6)} & \multicolumn{1}{c}{(7)} & \multicolumn{1}{c}{(8)} \\
\hline\noalign{\smallskip}
 63 &                                & 12:58:21.50 & $+$27:22:01.1 &         & 24 & 18 & A, BT  & \\
 64 & 5C 04.064                      & 12:58:21.69 & $+$27:01:38.0 &         & 36 & 24 & CD  & \\
 65 & 5C 04.065                      & 12:58:23.12 & $+$26:56:12.7 &         & 48 & 48 & CD  & 7C 1255$+$2712 \\
 66 &                                & 12:58:28.08 & $+$27:00:43.4 &         & 24 & 18 & CD  & \\
 67 & WISEA J125829.96$+$275818.0$^*$& 12:58:29.97 & $+$27:58:17.3 & 1.68500 & 24 & 12 & Tr  & \\
 68 & 2MASX J12583190$+$2802587      & 12:58:31.91 & $+$28:02:58.6 &         & 12 & 12 & A, D  & \\
 69 & WISEA J125838.15$+$274921.2    & 12:58:34.55 & $+$27:49:17.2 &         & 24 & 12 & Tr  & \\
 70 &                                & 12:58:37.14 & $+$28:27:55.0 &         & 24 &    & A  & \\
 71 & SDSS J125838.32$+$272258.2$^*$ & 12:58:38.10 & $+$27:23:03.6 & 1.39500 & 24 & 36 & CD  & \\
 72 & 2MASX J12583860$+$2756356      & 12:58:38.44 & $+$27:56:29.3 & 0.15680 & 12 & 12 & A, D  & \\
 73 & WISEA J125838.69$+$270047.6    & 12:58:40.54 & $+$27:01:12.9 &         & 24 & 36 & BT  & \\
 74 & WISEA J125840.28$+$283425.5    & 12:58:40.75 & $+$28:35:03.2 &         & 24 &    & Tr  & \\
 75 &                                & 12:58:41.87 & $+$27:54:09.2 &         & 48 & 36 & FD  & \\
 76 & WISEA J125846.09$+$273009.2    & 12:58:46.01 & $+$27:30:07.2 & 0.28481 & 24 & 24 & WAT  & \\
 77 &                                & 12:58:51.30 & $+$26:49:40.9 &         & 36 &    & Tr  & \\
 78 & WISEA J125856.74$+$274057.7$^*$& 12:58:55.62 & $+$27:41:05.4 & 0.28578 & 24 & 24 & CD, FR\,II  & \\
 79 &                                & 12:58:57.00 & $+$27:25:41.0 &         & 24 & 48 & CD  & \\
 80 & WISEA J125859.32$+$274644.9    & 12:58:59.29 & $+$27:46:44.3 &         & 96 & 24 & Tr, FR\,II  & \\
 81 & WISEA J125903.24$+$281200.2    & 12:59:03.24 & $+$28:12:00.2 &         & 36 & 18 & CD, FR\,II  & \\
 82 & 2MASX J12592846$+$2805078      & 12:59:28.46 & $+$28:05:07.4 & 0.01241 & 18 & 12 & CD  & \\
 83 & 5C 04.084                      & 12:59:30.64 & $+$27:01:45.2 &         & 36 &    & A  & C3B-217$^{**}$ \\
 84 &                                & 12:59:40.05 & $+$28:44:33.0 &         & 24 &    & CD  & \\
 85 & WISEA J125945.22$+$272703.6    & 12:59:42.18 & $+$27:27:20.8 &         & 24 & 24 & CD  & \\
 86 & WISEA J125945.22$+$272703.6    & 12:59:44.11 & $+$27:26:41.7 &         & 24 & 24 & CD  & \\
 87 &                                & 12:59:49.41 & $+$28:24:32.8 &         & 24 & 12 & CD  & \\
 88 & WISEA J130010.42$+$273540.9    & 13:00:13.91 & $+$27:35:53.2 &         & 24 & 18 & CD, FR\,II  & \\
 89 & WISEA J130016.16$+$274838.8    & 13:00:15.10 & $+$27:48:39.7 & 0.22373 & 24 & 12 & CD, FR\,II & \\
 90 &                                & 13:00:24.60 & $+$27:11:09.8 &         & 96 &    & A, D  & \\
 91 & WISEA J130024.14$+$280927.8    & 13:00:26.28 & $+$28:09:23.4 & 2.17935 & 48 & 24 & A, BT  & 5C\,04.102 \\
 92 & IC 4033                        & 13:00:28.39 & $+$27:58:20.7 &         & 18 & 12 & A, D  & \\
 93 & 5C 04.104                      & 13:00:30.36 & $+$27:11:27.6 &         & 48 &    & A  & \\
 94 & WISEA J130032.97$+$272225.3    & 13:00:31.10 & $+$27:22:37.0 & 0.72957 & 36 &    & FD  & \\
 95 & SSTSL2J130031.07$+$283300.0    & 13:00:31.43 & $+$28:33:00.5 & 0.22103 & 24 &    & Tr  & \\
 96 & 5C 04.114                      & 13:00:50.92 & $+$28:08:03.8 &         & 96 & 48 & FD  & \\
 97 & WISEA J130106.38$+$281813.0    & 13:01:06.35 & $+$28:18:09.1 & 0.37339 & 24 & 36 & CD, FR\,II  & \\
 98 & WISEA J130114.14$+$272859.1    & 13:01:12.45 & $+$27:28:35.5 &         & 18 &    & CD  & \\
 99 & WISEA J130113.14$+$281429.5    & 13:01:13.32 & $+$28:14:27.6 & 0.28795 & 24 & 24 & Z  & \\
100 &                                & 13:01:14.03 & $+$27:51:41.7 &         & 18 & 18 & CD  & \\
101 & WISEA J130121.84$+$280726.7    & 13:01:21.25 & $+$28:07:22.6 & 2.26500 & 36 & 48 & FD  & \\
102 &                                & 13:01:22.28 & $+$27:58:50.9 &         & 24 & 18 & Tr  & \\
103 & WISEA J130131.35$+$280334.8$^*$& 13:01:31.44 & $+$28:03:35.9 & 0.58188 & 24 & 24 & CD  & \\
104 &                                & 13:01:43.02 & $+$27:48:11.6 &         & 24 &    & CD  & \\
105 & 2MASX J13015023$+$2753367      & 13:01:50.22 & $+$27:53:36.9 &         & 12 &    & A, D  & \\
106 & WISEA J130153.52$+$273604.3    & 13:01:52.32 & $+$27:36:06.7 & 0.15856 & 12 &    & CD  & \\
107 &                                & 13:01:58.86 & $+$28:29:27.1 &         & 24 &    & A, D  & \\
108 & WISEA J130220.42$+$275159.8    & 13:02:20.94 & $+$27:52:10.7 & 0.25082 & 18 &    & A, D  & \\
109 & WISEA J130232.28$+$280117.0    & 13:02:32.74 & $+$28:01:19.0 & 0.16059 & 24 &    & CD  & \\
110 & TT 50                          & 13:02:41.52 & $+$28:08:10.7 & 0.16446 & 36 &    & HT, NAT  & \\
111 & NGC 4931                       & 13:03:00.88 & $+$28:01:56.9 & 0.01816 & 24 &    & CD, BT  & UGC 08154, CGCG 160-118 \\
\hline
\end{tabular}
\end{center}
\tablecomments{Continued.}
\end{table*}

\begin{figure*}[ht]
\begin{center}
\begin{tabular}{c}
\includegraphics[width=17cm]{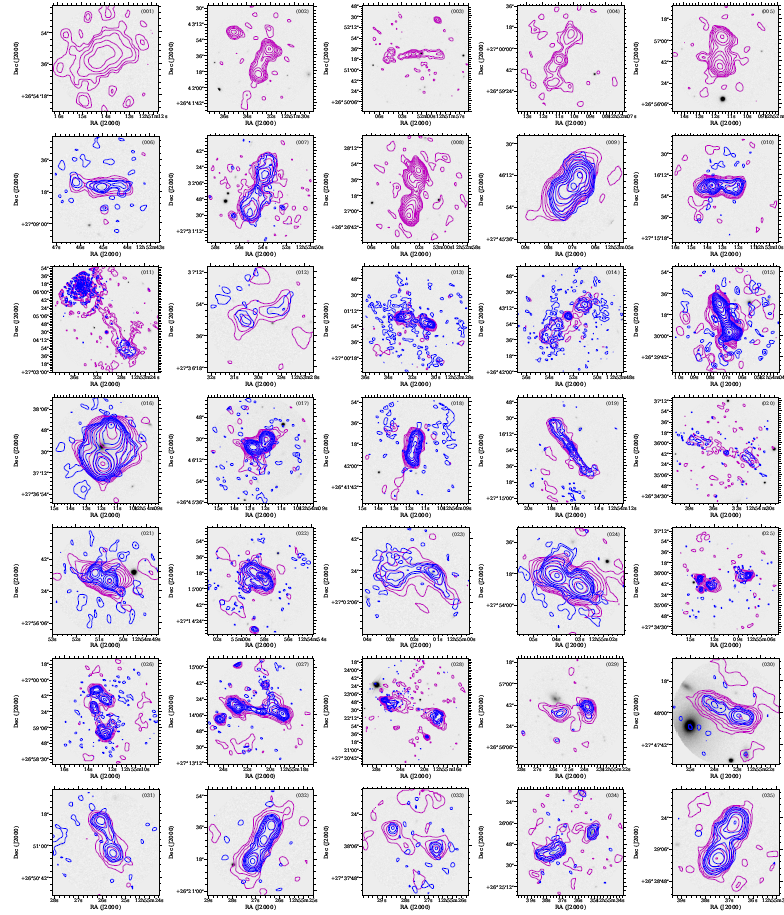}
\end{tabular}
\end{center}
\caption{Images of extended ($\simeq$12.6~kpc) radio sources in the Coma cluster field that are either foreground/background galaxies or do not have redshift data.  The source\_ID as tabulated in the Table~\ref{tab:app-tab1} is labelled in the upper-right corner of each image.  These radio sources are in the order presented in Table~\ref{tab:app-tab1}, from left to right and top to bottom.  The magenta and blue colour radio contours correspond to 250--500 MHz (band 3) and 550--850 MHz (band 4) at the angular resolutions of 6\farcs1 and 3\farcs7, respectively of the uGMRT that are are overlaid on the grayscale SDSS (DR12) $r$-band image.  The lowest radio contour plotted is three times the local \textsc{rms} noise and increasing by factors of 2. The local \textsc{rms} noise are tabulated in Table~\ref{tab:app-tab1} for 250--500 MHz band (Col.~5) and 550--850 MHz band (Col.~6) of the uGMRT.  The grayscale images are displayed in logarithmic scales to emphasize the optical hosts associated with the respective radio galaxies.}
\label{fig:app1}
\end{figure*}

\setcounter{figure}{1}
\begin{figure*}[ht]
\begin{center}
\begin{tabular}{c}
\includegraphics[width=17cm]{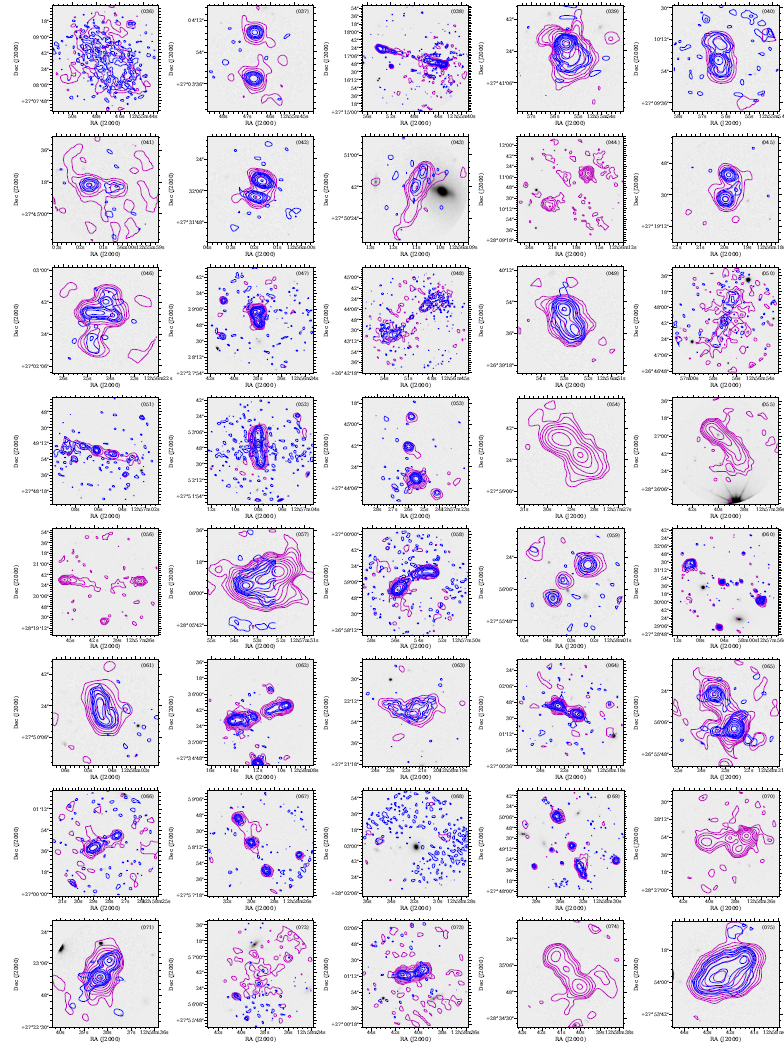}
\end{tabular}
\end{center}
\caption{Continued}
\end{figure*}

\setcounter{figure}{1}
\begin{figure*}[ht]
\begin{center}
\begin{tabular}{c}
\includegraphics[width=17cm]{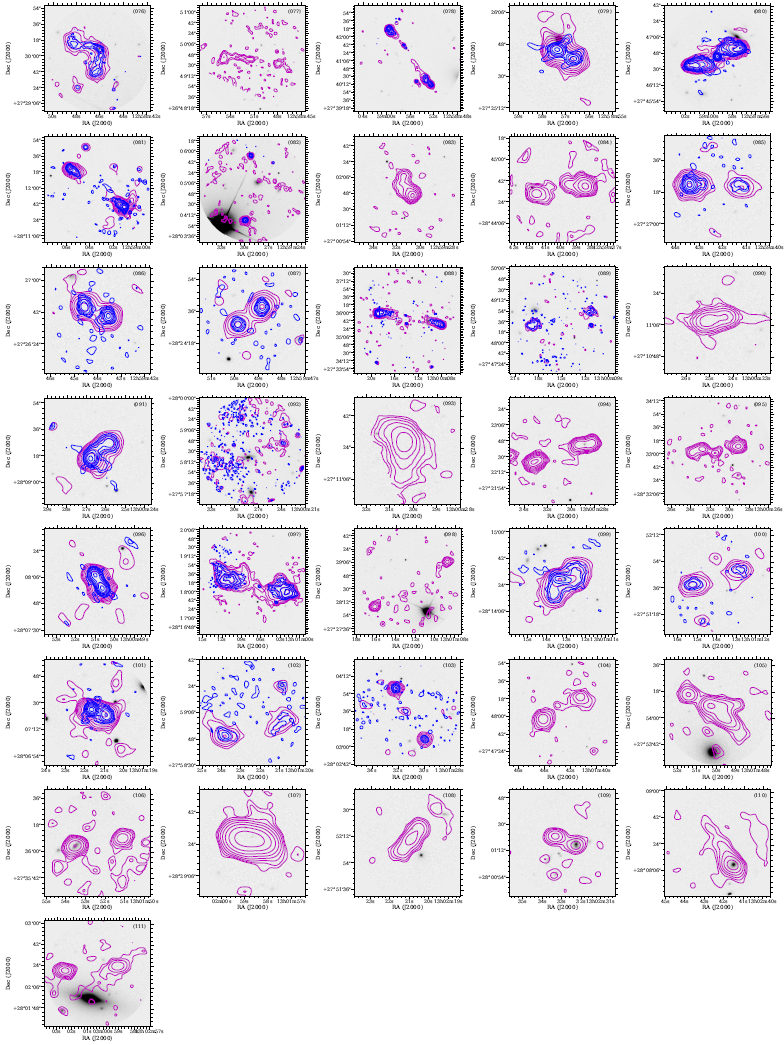}
\end{tabular}
\end{center}
\caption{Continued}
\end{figure*}


\begin{thebibliography}{}

\bibitem[Abell (1958)]{Abell1958} Abell, G. O. 1958, ApJS, 3, 211

\bibitem[Abell, Corwin \& Olowin (1989)]{Abell1989} Abell, G. O., Corwin, H. G., Jr., Olowin, R. P. 1989, ApJS, 70, 1

\bibitem[Adam et~al.(2021)]{Adametal}  Adam, R., Goksu, H., Brown, S., Rudnick, L., Ferrari, C. 2021, A\&A, 648, A60

% \bibitem[Arnaud(1996)]{Arnaud1996} Arnaud K. A., 1996, in ASP Conf. Ser. 101, Astronomical Data Analysis Software and Systems, ed. G. H. Jacoby, \& J. Barnes J. (San Francisco, CA: ASP), 17

% \bibitem[Arnaud et~al.(2001)]{Arnaudetal2001} Arnaud, M., Aghanim, N., Gastaud, R., et~al. 2001, A\&A, 365, L67

\bibitem[Baghmanyan et~al.(2021)]{Baghmanyanetal} Baghmanyan, V., Zargaryan, D., Aharonian, F., et~al. 2022, arXiv:2110.00309 (in press)

\bibitem[Baier, Fritze \& Tiersch(1990)]{BaierTiersch1990} Baier, F. W., Fritze, K., \& Tiersch. H. 1990, AN, 311, 89

\bibitem[\protect\citeauthoryear{Basu et al.}{2016}]{2016A&A...591A.142B} Basu K., Vazza F., Erler J., Sommer M., 2016, A\&A, 591, A142 %. doi:10.1051/0004-6361/201527726

\bibitem[Beck \& Krause(2005)]{2005AN....326..414B} Beck, R., \& Krause, M. 2005, AN, 326, 414

\bibitem[Bonafede et~al.(2022)]{Bonafedeetal2022} Bonafede, A., Brunetti, G., Rudnick, L., et al., 2022, ApJ (submitted)


\bibitem[Bonafede et~al.(2021)]{Bonafedeetal2021} Bonafede, A., Brunetti, G., Vazza, F., et al., 2021, ApJ, 907, 32

\bibitem[Boselli et~al.(2016)]{Bosellietal2016} Boselli A., Roehlly, Y., Fossati, M., et al. 2016, A\&A, 596, A11

\bibitem[Boselli, Fossati \& Sun(2021)]{Bosellietal} Boselli, A., Fossati, M., \& Sun, M. 2021, in press

\bibitem[Bothun \& Dressler(1986)]{BothunandDressler} Bothun, G. D., \& Dressler, A. 1986, ApJ, 301, 57

\bibitem[Bravo-Alfaro et~al.(2000)]{BravoAlfaro} Bravo-Alfaro H., Cayatte V., van Gorkom J. H., \& Balkowski C. 2000, AJ, 119, 580

\bibitem[Brown \& Rudnick(2011)]{BrownRudnick2011} Brown, S., \& Rudnick, L. 2011, MNRAS, 412, 2

\bibitem[Brunetti, Setti \& Comastri(1997)]{BSC1997} Brunetti, G., Setti, G., \& Comastri, A. 1997, A\&A, 325, 898

\bibitem[Brunetti, \& Lazarian(2011)]{BL2011} Brunetti, G., \& Lazarian, A. 2011, MNRAS, 410, 127

\bibitem[Brunetti \& Jones(2014)]{BrunettiJones} Brunetti, G., \& Jones, T. W. 2014, Int. Jl. of Modern Physics D, 23

\bibitem[Brunetti, \& Lazarian(2016)]{BL2016} Brunetti, G., \& Lazarian, A. 2016, MNRAS, 458, 2584

\bibitem[Burns(1990)]{Burns1990} Burns, J. O. 1990, AJ, 99, 14

%\bibitem[Bykov et al.(2015)]{bykov2015} Bykov, A. M., Churazov, E. M., Ferrari, C., et~al. 2015, \ssr, 188, 141

\bibitem[Bykov et al.(2019)]{bykov2019} Bykov, A.~M., Vazza, F., Kropotina, J.~A., et al.\ 2019, \ssr, 215, 14

\bibitem[Capetti et~al.(2000)]{2000A&A...362..871C} Capetti, A., de Ruiter, H. R., Fanti, R., et~al. 2000, A\&A, 362, 871

\bibitem[Chen et~al.(2020)]{Chenetal} Chen, H., Sun, M., Yagi, M., et~al. 2020, MNRAS, 496, 4654

\bibitem[Churazov et~al.(2003)]{Churazovetal2003} Churazov, E., Forman, W. R., Jones, C., \& B\"ohringer, H. 2003, ApJ, 590, 225

\bibitem[Churazov et~al.(2012)]{Churazovetal2012} Churazov, E., Vikhlinin, A., Zhuravleva, I., et~al. 2012, MNRAS, 421, 1123

\bibitem[Churazov et~al.(2021)]{Churazovetal2021} Churazov, E., Khabibullin, I., Lyskova, N., Sunyaev, R., \& Bykov, A. M. 2021, A\&A, 651, A41

\bibitem[Crowl et~al.(2005)]{Crowletal} Crowl H. H., Kenney J. D. P., van Gorkom J. H., \& Vollmer B. 2005, AJ, 130, 65

\bibitem[Curtis(1918)]{Curtis1918} Curtis, H. D. 1918, Publ. of Lick Obs., 13, 9

% \bibitem[Dallacasa et~al.(1989)]{Dallacasaetal1989} Dallacasa, D., Feretti, L., Giovannini, G. \& Venturi, T.  1999, A\&ASS, 79, 391

% \bibitem[Dehghan et~al.(2014)]{Dehghanetal} Dehghan, S., Johnston-Hollitt, M., Franze, T. M. O., et~al. 2014, AJ, 148, 75

% \bibitem[de~Young(1986)]{deYoung1986} De Young, D. S. 1986, ApJ, 307, 62

% \bibitem[de~Young(1996)]{deYoung1996} De Young D. S. 1996, in ASP Conf. Ser. 100, Energy Transport in Radio Galaxies and Quasars, ed. P. E. Hardee, A. H. Bridle, \& J. A. Zensus J. A. (San Francisco CA: ASP), 261

% \bibitem[Dickey \& Lockman(1990)]{DickeyLockman} Dickey, J. M., \& Lockman, F. J. 1990, ARA\&A, 28, 215

\bibitem[Fanti et~al.(1990)]{Fantietal1990} Fanti, R., Fanti, C., Schilizzi, R. T., et~al. 1990, A\&A, 231, 333

\bibitem[Feretti et~al.(1990)]{Ferettietal1990} Feretti, L., Dallacasa, D., Giovannini, G., \& Venturi, T. 1990, A\&A, 232, 337

\bibitem[Feretti \& Venturi(2002)]{FerettiVenturi} Feretti, L., \& Venturi, T. 2002, in Merging Processes in Galaxy Clusters 272, Astrophysics and Space Science Library, ed. L. Feretti, I. M. Gioia \& G. Giovannini  (Dordrecht), 163

\bibitem[Gavazzi(1989)]{Gavazzi1989} Gavazzi, G. 1989, AJ, 346, 59

% \bibitem[Gendron-Marsolais et~al.(2020)]{Gendron-Marsolaisetal} Gendron-Marsolais, M., Hlavacek-Larrondo, J., van Weeren, R. J., et~al. 2020, MNRAS, 499, 5791

\bibitem[Giovannini, Feretti \& Stanghellini(1991)]{Giovanninietal1990} Giovannini, G., Feretti, L, \& Stanghellini, C. 1991, A\&A, 252, 528

\bibitem[Giovannini et~al.(1993)]{1993ApJ...406..399G} Giovannini, G., Feretti, L., Venturi, T., Kim, K. -T., \& Kronberg, P. P.  1993, ApJ, 406, 399

% \bibitem[Grevesse \& Sauval(1998)]{GrevesseSauval} Grevesse, N., \& Sauval, A. J., 1998, Space Sci. Rev., 85, 161

\bibitem[Grishin et~al.(2021)]{Grishinetal}  Grishin, K. A., Chilingarian, I., Afanasiev, A. V., et~al. 2021, Nat. Astron. 5, 1308

\bibitem[Gunn \& Gott(1972)]{GunnGott} Gunn, J. E., \& Gott, J. R. III 1972, ApJ, 176, 1

\bibitem[Gupta et~al.(2017)]{Guptaetal2017} Gupta, Y., Ajithkumar, B., Kale, H., et~al. 2017, Current Science, 113, 707

% \bibitem[The H.E.S.S. Collaboration(2020)]{HESScollaboration} The H.E.S.S. Collaboration 2020, Nature, 582, 356

% \bibitem[Hardcastle et~al.(2002)]{Hardcastleteal2002} Hardcastle, M. J., Worrall, D. M., Birkinshaw, M., et~al. 2002, MNRAS, 334, 182

\bibitem[Herschel(1785)]{Herschel1785} Herschel, W. 1785, Phil. Trans. of the Roy. Soc. of London, 75, 213

\bibitem[Huchra et~al.(1990)]{Huchra1990} Huchra, J. P., Geller, M. J., de~Lapparent, V., \& Corwin, H. G. Jr. 1990 ApJS 72, 433

\bibitem[Jaffe \& Perola(1973)]{JaffePerola1973} Jaffe, W. J., \& Perola, G. C. 1973, A\&A, 26, 423

\bibitem[Joye \& Mandel(2003)]{joye_new_2003} Joye, W.~A., \& Mandel, E. 2003, in ASP Conf. Ser., 295, Astronomical Data Analysis Software and Systems, ed. H. E. Payne, R. I. Jedrzejewski, \& R. N. Hook (San Francisco, CA: ASP), 489.

\bibitem[Kenney, Abramson \& Bravo-Alfaro (2015)]{Kenney2015} Kenney, J. D. P., Abramson, A., \& Bravo-Alfaro, H. 2015  AJ 150, 59

\bibitem[Kent \& Gunn(1982)]{KentandGunn} Kent, S. M., \& Gunn, J. E. 1982 AJ 87, 945

\bibitem[Kim et~al.(1989)]{Kimetal1990} Kim, K.-T., Kronberg, P. P., Giovannini, G., \& Venturi, T. 1989, Nat., 341, 720

\bibitem[Kim et~al.(1994)]{Kimetal1994} Kim, K.-T., Kronberg, P. P., Dewdney, P. E., \& Landecker, T. L. 1994, ApJS, 105, 385

\bibitem[Koopmann \& Kenney(2004)]{KoopmannKenney2004} Koopmann R. A., \& Kenney J. D. P. 2004, ApJ, 613, 866

\bibitem[Lal \& Rao(2004)]{LalandRao2004} Lal, D. V., \&  Rao, A. P. 2004, A\&A, 420, 491

% \bibitem[Lal et~al.(2013)]{Laletal2013} Lal, D. V., Kraft, R. P., Randall, S. W., et~al. 2013, ApJ, 764, 83

\bibitem[Lal(2020a)]{Lal2020a} Lal, D. V. 2020a, ApJS, 250, 22

\bibitem[Lal(2020b)]{Lal2020b} Lal, D. V. 2020b, AJ, 160, 161

% \bibitem[Lal(2021)]{Lal2021} Lal, D. V. 2021, ApJ, 915, 126

\bibitem[Lyskova et~al.(2019)]{Lyskovaetal2018} Lyskova, N. Churazov, E., Zhang, C., et~al. 2019, MNRAS, 485, 2922

% \bibitem[Markevitch \& Vikhlinin(2007)]{MaximAlexey} Markevitch, M., \& Vikhlinin, A. 2007, Physics Reports, 443, 1

\bibitem[Mazzarella \& Balzano(1986)]{MazzarellaandBalzano} Mazzarella, J. M., \& Balzano V. A. 1986, ApJS, 62, 751

% \bibitem[McBride \& McCourt(2014)]{McBrideMcCourt} McBride, J., \& McCourt, M. 2014, MNRAS, 442, 838

\bibitem[McMullin et~al.(2007)]{mcmullin_casa_2007} McMullin, J.~P., Waters, B., Schiebel, D., Young, W., \& Golap, K. 2007, in ASP Conf. Ser., 376, Astronomical Data Analysis Software and Systems, ed. R. A. Shaw, F. Hill, \& D. J. Bell (San Francisco, CA: ASP), 127

\bibitem[Miley(1980)]{Miley1980} Miley, G. K. 1980, ARA\&A, 18, 165

% \bibitem[Miley et~al.(1972)]{Mileyetal1972} Miley, G. K., Perola, G. C., van der Kruit, P. C. \& van der Laan, H. 1972, Nature, 237, 269

\bibitem[Miller \& Owen(2002)]{MillerOwen2002} Miller, N. A., \& Owen, F. N. 2002, AJ, 124, 2453

\bibitem[Miller, Hornschemeier \& Mobasher(2009)]{NHM2009} Miller, N. A., Hornschemeier, A. E., Mobasher, B. 2009, AJ, 137, 4436

% \bibitem[Missaglia et~al.(2019)]{Missagliaetal}Missaglia, V., Massaro, F., Capetti, A., et~al. 2019, A\&A, 626, A8

\bibitem[Neumann et~al.(2001)]{2001A&A...365L..74N}  Neumann, D. M., Arnaud, M., Gastaud, R., et~al. 2001, A\&A, 365, L74
 
\bibitem[Nulsen (1982)]{Nulsen1982} Nulsen P. E. J. 1982, MNRAS, 198, 1007

\bibitem[O'Dea(1998)]{ODea1998}, O'Dea, C. 1998, PASP, 110, 493

\bibitem[Oemler(1976)]{Oemler1976} Oemler, A. 1976, ApJ, 209, 693

\bibitem[Ozawa et~al.(2015)]{Ozawaetal2014} Ozawa, T., Nakanishi, H., Akahori, T., et~al. 2015, PASJ, 67, 110

\bibitem[Perley \& Butler(2017)]{PerleyButler} Perley, R. A., \& Butler, B. J. 2017, ApJS 230, 7

\bibitem[Pinzke, Oh \& Pfrommer(2017)]{Pinzkeetal2017} Pinzke, A., Oh, S. P., \& Pfrommer, C. 2017, MNRAS, 465, 4800

\bibitem[Planck Collaboration(2013)]{Planck-collaboration} Planck Collaboration 2013, A\&A, 554, A140

\bibitem[Roberts \& Parker(2020)]{RobertsParker} Roberts, I. D., \& Parker, L. C. 2020, MNRAS, 495, 554

\bibitem[Roberts et~al.(2021)]{Robertsetal2021}  Roberts, I. D., van~Weeren, R. J., McGee, S. L., et~al. 2021, A\&A, 650, A111

\bibitem[Robitaille et~al.(2013)]{robitaille_astropy_2013} Robitaille, T.~P., Tollerud, E.~J., Greenfield, P., {et~al.} 2013, A\&A, 558, A33

\bibitem[Rudnick(2021)]{Rudnick2021} Rudnick, L. 2021, Galaxies, 9, 85

% \bibitem[Sanders et~al.(2013)]{2013Sci...341.1365S} Sanders, J. S., Fabian, A. C., Churazov, E., et~al. 2013, Science, 341, 1365

\bibitem[Sanders et~al.(2014)]{Sandersetal} Sanders J. S., Fabian A. C., Sun M., et~al. 2014, MNRAS, 439, 1182

\bibitem[Sheardown et~al.(2019)]{Sheardown2019} Sheardown, A.,  Fish, T. M.,  Roediger, E., et~al. 2019, ApJ, 874, 112

%\bibitem[\protect\citeauthoryear{Simionescu et al.}{2013}]{2013ApJ...775....4S} Simionescu A., Werner N., Urban O., Allen S.~W., Fabian A.~C., Mantz A., Matsushita K., et al., 2013, ApJ, 775, 4. doi:10.1088/0004-637X/775/1/4

\bibitem[\protect\citeauthoryear{Simionescu et al.}{2013}]{2013ApJ...775....4S} Simionescu A., Werner N., Urban O., et al., 2013, ApJ, 775, 4 %. doi:10.1088/0004-637X/775/1/4

% \bibitem[Smith et~al.(2004)]{2004AJ....128.1558S} Smith, R. J., Hudson, M. J., Nelan, J. E., et~al. 2004, AJ, 128, 1558

\bibitem[Smith et~al.(2010)]{Smith2010} Smith R. J., Lucey, J. R., Hammer, D., et al. 2010, MNRAS 408, 1417

% \bibitem[Sun, Jerius \& Jones(2005)]{Sunetal2005} Sun, M., Jerius, D. \& Jones, C. 2005, ApJ, 633, 165

\bibitem[Sun et~al.(2005)]{Sunetal2005} Sun, M., Vikhlinin, A., Forman, W., Jones, C., \& Murray, S. S. 2005, ApJ, 619, 169

\bibitem[Sun et~al.(2007)]{Sunetal2007} Sun, M., Jones, C., Forman, W., et~al. 2007, ApJ, 657, 197

\bibitem[Sun et~al.(2021)]{Sunetal2021} Sun, M., Ge, C., Luo, R., et~al. 2021, Nat. Astron., 243

\bibitem[Swarup et~al.(1991)]{Swarupetal1991} Swarup, G., Ananthakrishnan, S., Kapahi, V. K., et~al. 1991, Current Science, 60, 95

\bibitem[Vikhlinin et~al.(2001)]{Vikhlininetal} Vikhlinin, A., Markevitch, M. M., Forman, W. R., \& Jones, C. 2001, ApJ, 555, L87

\bibitem[Vikhlinin et~al.(2005)]{Vikhlininetal2005} Vikhlinin, A., Markevitch, M. M., Murray, S. S., et~al. 2005, ApJ, 628, 655

\bibitem[Venturi, Feretti \& Giovannini(1989)]{Venturietal1989} Venturi, T., Feretti, L., \& Giovannini, G. 1989, A\&A, 213, 49

\bibitem[Venturi, Giovannini \& Feretti(1990)]{1990AJ.....99.1381V} Venturi, T., Giovannini, G., \& Feretti, L. 1990, AJ, 99, 1381

\bibitem[Venturi et~al.(2022)]{Venturietal2022}  Venturi, T., Giacintucci, S., Merluzzi, P., et~al. 2022, A\&A, in press

\bibitem[van~Moorsel, Kemball \& Greisen(1996)]{van_moorsel_aips_1996} van~Moorsel, G., Kemball, A., \& Greisen, E. 1996, in ASP Conf. Ser. 101, Astronomical Data Analysis Software and Systems, ed. G. H. Jacoby \& J. Barnes J. (San Francisco, CA: ASP), 37

\bibitem[van~Weeren et~al.(2009)]{vanWeerenetal2009} van~Weeren, R. J., R\"ottgering, H. J. A., Bagchi, J., et~al. 2009, A\&A, 506, 1083 
\bibitem[van~Weeren et~al.(2017)]{vanWeeren2017} van Weeren, R. J., Andrade-Santos, F., Dawson, W. A., et~al. 2017, Nat. Astron. 1, 5

\bibitem[van~Weeren et~al.(2019)]{vanWeerenetal2019} van Weeren, R. J., de Gasperin, F., Akamatsu, H., et~al. 2019, Sp. Sc. Rev. 215, 16

\bibitem[Wagner et~al.(1999)]{Wagneretal} Wegner, G., Colless, M., Saglia, R. P., et~al. 1999, MNRAS, 305, 259

\bibitem[Willson(1970)]{Willson1970} Willson, M. A. G. 1970, MNRAS, 151, 1

\bibitem[Yagi et~al.(2010)]{Yagi2010} Yagi M., Yoshida, M., Komiyama, Y., et al. 2010 AJ 140, 1814

\bibitem[Zhang et~al.(2019)]{Zhangetal} Zhang, C., Churazov, E.,  Forman, W. R., \& Lyskova, N. 2019, MNRAS, 488, 5259

\bibitem[Zwicky(1933)]{Zwicky1933} Zwicky, F. 1933, Helvetica Physica Acta, 6, 110

\end{thebibliography}
\end{document}